\documentclass[preprint,trackchanges]{aastex62}

\usepackage{amsmath}
\usepackage{natbib}
\usepackage{multirow} 
\hypersetup{linkcolor=magenta, citecolor=cyan, filecolor=yellow, urlcolor=blue}

\received{2018 October 5}
\revised{2019 March 22}
\accepted{2019 March 24}
\published{2019 May 31}

\shorttitle{MultiPulse {\it Fermi} Gamma-Ray Bursts}
\shortauthors{Li.}

\begin{document}
\title{MultiPulse {\it Fermi} Gamma-Ray Bursts. I. \\Evidence of the Transition from Fireball to Poynting-flux-dominated Outflow}

\correspondingauthor{Liang Li}
\email{liang.li@icranet.org}

\author[0000-0002-1343-3089]{Liang Li}
\affiliation{ICRANet, Piazza della Repubblica 10, I-65122 Pescara, Italy}

\begin{abstract}

The composition of a jet is still an unsolved problem in gamma-ray bursts (GRBs). Several previous studies have suggested that the prompt emission spectrum of GRBs is likely to consist of a few components that may arise from different jet compositions. Here we present a systematic analysis to search for the GRBs that seem to show the transition from a fireball to the Poynting-flux-dominated outflow between well-separated pulses within a single burst, like the GRB 160626B, using the Gamma-ray Burst Monitor data of the \emph{Fermi} satellite. We obtain 43 GRBs with clear multiple pulses and find that 9/43 (21\%) bursts may exhibit such a transition based on the time-integrated spectral analysis. We then select a further four bursts with the data of adequate quality to perform a detailed time-resolved spectral analysis, and we find that in three bursts the thermal-like pulse is a precursor. Furthermore, based on the time-resolved spectra, we constrain the outflow properties for those thermal pulses and find them consistent with the typical properties of the photosphere emission. Also, the later pulses with the softer low-energy spectral index are compatible with the optically thin synchrotron emission model. Our analysis indicates that a good fraction of the multipulse {\it Fermi} bursts may obtain the transition from fireball to the Poynting-flux-dominated outflow.

\end{abstract}


\section{Introduction} \label{sec:Introduction}

Nearly 50 yr after the discovery of gamma-ray bursts (GRBs), the prompt emission of GRBs is still puzzling in several fundamental perspectives \cite[e.g.,][]{1999ApJ...518..356P, 1999A&A...350..334R, 2000A&A...359..855R, 2002ApJ...579..706D, 2006ApJ...643L..87G, 2007ApJ...661.1025L, 2011CRPhy..12..206Z, 2011ApJ...726...90Z, 2013ApJS..207...23X, 2015PhR...561....1K, 2015AdAst2015E..22P, 2017NewAR..79....1L}. 
A quite fundamental question is the jet composition (baryonic-dominated fireball or Poynting-flux-dominated outflow),  which determines or at least strongly affects the energy dissipation mechanism (the shock or the magnetic reconnection), the particle acceleration mechanism (thermally driven or magnetically driven), and especially the radiation mechanism (synchrotron, \citealt{1994ApJ...422..248K, 1994ApJ...432..181M, 1996ApJ...473..204S, 1998MNRAS.296..275D, 2000ApJ...543..722L, 2008MNRAS.384...33K, 2011A&A...526A.110D, 2011ApJ...726...90Z, 2013ApJ...769...69B, 2014MNRAS.445.3892B}; or Comptonization of quasi-thermal emission from the photosphere, \citealt{1986ApJ...308L..47G, 1994MNRAS.270..480T, 2005ApJ...628..847R, 2006A&A...457..763G, 2010MNRAS.407.1033B, 2010ApJ...725.1137L, 2012MNRAS.422.3092G, 2013ApJ...767..139B, 2014IJMPD..2330003V, 2017IJMPD..2630018P}). Several scenarios of the jet composition have been proposed in the literature \cite[e.g.,][]{2009ApJ...702.1211R, 2012MNRAS.420..468P, 2018MNRAS.475.1708A} based on the components in the observed spectrum: (i) a quasi-thermal component indicating a hot fireball origin \citep{2010ApJ...709L.172R}; (ii) a nonthermal component from the synchrotron radiation of the Poynting-flux-dominated outflow \cite[e.g.,][]{1998A&A...338L..87P, 2002astro.ph..4080P}; and (iii) a combination of the quasi-thermal component and nonthermal component from a hybrid jet, consisting of a hot fireball and a cold Poynting flux \citep{2013MNRAS.433.2739I, 2015MNRAS.450.1651I,2015ApJ...801..103G}.

As mentioned above, the synchrotron radiation and the photosphere emission in the observed spectrum are crucial to judging the jet composition.  
A criterion that could distinguish the synchrotron and photosphere emission is the observed low-energy spectral index $\alpha$. Generally speaking, the photosphere model predicts a much harder value of $\alpha$.
For the fast-cooling synchrotron emission (the dynamical timescale is greater than the synchrotron cooling timescale) the predicted $\alpha$ is -1.5 \citep{1998ApJ...497L..17S, 1999ApJ...511L..93G}, much softer than the typical value of -0.8 \citep{1998ApJ...506L..23P,2000ApJS..126...19P, 2011ApJ...741...24B, 2011MNRAS.415.3153N, 2011CRPhy..12..206Z, 2014ApJ...784...17B, 2013ApJS..208...21G}. 
The slow-cooling synchrotron emission predicts a much harder value of $\alpha$=-2/3, the so-called line of the death of synchrotron emission \citep{1998ApJ...506L..23P}, more consistent with that typical value. 
Noteworthily, the electrons in the slow-cooling regime are unexpected, 
due to the usually strong local magnetic field in the prompt emission region \citep{1999ApJ...511L..93G, 2014NatPh..10..351U}. 
However, reconciling the slow-cooling spectra, as required by observations, with high efficiency for the prompt emission (as obtained by fast-cooling electrons) is possible to do naturally if synchrotron cooling is balanced by a continuous source of heating \citep{2018MNRAS.476.1785B}. 
Also, the specific question of electron cooling in magnetic GRB jets has been extensively discussed in \cite{2014MNRAS.445.3892B}. 
Besides, recent broad time-dependent spectral analysis of GRBs, going beyond the simple Band function (a smoothly joint broken power law, \citealt{1993ApJ...413..281B}), suggests that the synchrotron spectrum is consistent with observations \citep[e.g.,][]{2017ApJ...846..137O, 2018arXiv181006965B}.
For the photosphere model, pure blackbody (BB) emission in the Rayleigh-Jeans regime predicts $\alpha \sim$ +1.0, while with the curvature effect (equal-arrival-time surface effect) $\alpha \simeq$ +0.4 \cite[e.g.,][]{2010MNRAS.407.1033B, 2014ApJ...785..112D}. In reality, a pure BB emission is in general never expected from the photosphere, since geometry \citep{2008ApJ...682..463P} and dissipation \citep{2010MNRAS.407.1033B} will always broaden the spectrum a bit, but it can be used as an approximation.
It is worth noting that the deviation from the straightforward prediction of both models for $\alpha$ can be modified with reasonable assumptions \citep[e.g.,][]{2000ApJ...543..722L,2000ApJ...530..292M, 2011MNRAS.415.1663T, 2013MNRAS.428.2430L, 2018ApJS..234....3G, 2018ApJ...860...72M}.

Photosphere emission and synchrotron radiation widely exist in the observed spectrum of GRBs, indicating the different kinds of jet composition. A hot fireball, the Poynting-flux-dominated outflow, or a hybrid jet with both these components in a single burst are the common types of jet composition. But whether the composition can change between different pulses in a single burst is an interesting issue, which seems to be confirmed in \cite{2018NatAs...2...69Z}. 
In that work, the spectral property of the bright burst GRB 160625B \citep[see also][]{2017ApJ...836...81W, 2017ApJ...834L..13W}, with three sub-bursts, is found to change very noticeably between the different sub-bursts, suggesting a different physical origin.
Thus, \cite{2018NatAs...2...69Z} for the first time suggest, from early precursor to late main sub-bursts, the transition from fireball to the Poynting-flux-dominated outflow.
\textcolor{orange}{To go a step further, it is interesting to address the question, what is the proportion of GRBs having observed evidence of the transition from fireball to the Poynting-flux-dominated outflow within a single burst?}

In this paper, we have systematically searched for the bright {\it Fermi} GRBs to perform the time-integrated and the detailed time-resolved spectral analysis.
To know whether there is apparent spectral evolution among the multiple pulses\footnote{In the following discussions of the paper, the first pulse (or sub-burst), the second pulse (or sub-burst), the third pulse (or sub-burst), and the fourth pulse (or sub-burst) will be denoted as 1st, 2nd, 3rd, and 4th, respectively.} (or sub-bursts\footnote{The sub-bursts are defined as different pulses that are separated with clear quiescent intervals.}) within a single GRB, we are particularly concerned with those GRBs with clear multiple pulses (or sub-bursts) from precursor to the main burst, or extended emission.
The explicit goal of this task is to investigate whether there is evidence of the transition from fireball to the Poynting-flux-dominated outflow of GRBs.

The paper is organized as follows. 
In \S 2, we perform the spectral analysis and then discuss the fitting parameters of different empirical models.
In \S 3, we derive the fireball parameters for the early thermal pulse and check the synchrotron origin for the later nonthermal pulse.
The discussions and conclusions are presented in \S 4.  
Throughout the paper, a concordance cosmology with parameters $H_{0}= 71$ ${\rm km s^{-1}}$ ${\rm Mpc^{-1}}$, $\Omega_{M}=0.30$, and $\Omega_{\Lambda}=0.70$ is adopted. The convention $Q=10^{x}Q_{x}$ is adopted in cgs units.

\section{Data Analysis} \label{sec:Data Analysis}

This task was carried out by working with the data from the Gamma-ray Burst Monitor (GBM; \citealt{2009ApJ...702..791M}) on board the {\it Fermi Gamma-ray Space Telescope}. 
More than $2200$ GRBs have been detected by the GBM, from 2008 July until 2018 March.
The GBM covered the energy range from 8 keV to 40 MeV. 
It carries two sets of detectors: sodium iodide (NaI) scintillation with 12 detectors, and bismuth germanate (BGO) scintillation with two detectors. 
All 14 GBM detectors point to different directions and collected the signals of the photon by a Central Data Processing Unit.
The 12 NaI detectors play the same role but with a different direction, and they provide an energy range of observations from the 8 keV to 1 MeV. 
The two BGO detectors provide coverage from 200 keV to 40 MeV. 
This overlaps with the energy range at the lower end of NaI detectors and lower energies of the Large Area Telescope (LAT), allowing for cross-calibration of the detectors.
The BGO detectors are located on opposite sides of the spacecraft to ensure that at least one BGO detector gets illuminated for each possible source location. 
Three data types of science data that are available in GBM are 128-channel resolution (CSPEC), 8-channel resolution (CTIME), and the time-tagged event (TTE) data. The CSPEC and CTIME are the binned data events. CTIME contains data collected from each detector with 8-channel pulse height resolution and provides data with continuous high time resolution, with a temporal resolution of 0.064 s after triggered time and 0.256 s before triggered time, with the time range from --1000s to 1000 s. CSPEC contains data collected from each detector with 128-channel pulse height resolution and provides data with continuous high spectral resolution, with a temporal resolution of 1.024 s after triggered time and 4.096 s before triggered time, with the time range from --4000 s to 4000 s. The TTE data event is an unbinned data type, consisting of individually digitalized pulse height events from the GBM detector during the event of the burst. It has an energy resolution of 128 channels, which is similar to CSPEC, and records the time interval of photons from the --20s to 300 s.

The TTE data and the standard response files are used as provided by the GBM team.
We selected the data from all the NaI detectors that are triggered by GBM (most cases are one to three) and the brightest BGO detector.

\subsection{Data Reduction and Sample Selection}

Two main criteria are adopted for a rough sample selection:
(i) To perform a detailed time-resolved spectral analysis, brighter bursts\footnote{Peak flux $>$ 40 photons cm$^{-2}$ s$^{-1}$ or fluence $>$ 1$\times$ 10$^{4}$ erg cm$^{-2}$.} are required. Therefore, we initially selected all bright bursts from the {\it Fermi} GBM catalog;
(ii) Since it is interesting to investigate whether we have an apparent spectral evolution among the multipulses (or the sub-bursts) within a single GRB, we then obtained all those GRBs that have a clear couple (or more) of distinct pulses (sub-bursts) in our sample, particularly in the GRBs that do not overlap each other significantly. 
With those selected criteria, we obtain 43 bursts from all 2281 GRBs detected by the GBM until 2018 March 31. In total, 118 pulses/sub-bursts are obtained from these 43 GRBs. In addition to the 43 bursts showing at least two clear pulses/sub-bursts (including 22 bursts that only contain two clear pulses), we also identified 12 bursts having three distinct pulses/sub-bursts, 7 bursts exhibiting four different pulses/sub-bursts, and 2 bursts displaying five pulses/sub-bursts within one single burst.

Observationally, {\it Fermi} Gamma-ray Observatory has revealed three elemental spectral features of GRBs \citep{2011ApJ...730..141Z}: 
(i) The Band spectral component, Band function \citep{1993ApJ...413..281B}, which is written as
\begin{eqnarray}
\label{n}  N_{E}=A \left\{ \begin{array}{ll}
(\frac{E}{E_{\rm piv}})^{\alpha} e^{-E/E_{0}}, & (\alpha-\beta)E_{0}\geq E \\
(\frac{(\alpha-\beta)E_{0}}{E_{\rm piv}})^{(\alpha-\beta)} e^{(\beta-\alpha)}(\frac{E}{E_{\rm piv}})^{\beta}, & (\alpha-\beta)E_{0}\leq E \\
\end{array} \right.
\end{eqnarray}
where $N_{E}$ is the photon flux ($\rm ph/cm^{2}/keV/s$), $A$ is the normalization for the spectral fit, $E_{\rm piv}$ is the pivot energy fixed at 100 keV, $E$ is the photon energy, and $E_{0}$ is the break energy.
The function consists of two power laws that are smoothly separated by a peak energy $E_{\rm pk}$ $\sim$ 250 keV \cite[e.g.,][]{2006ApJS..166..298K, 2011MNRAS.415.3153N}, described by the low-energy photon index $\alpha$ $\sim$ -1.0, 
\begin{equation}
N_{E} \propto E^{\alpha} e^{-E/E_{\rm pk}},
\label{Eq:CPL}
\end{equation}
and the high-energy photon index $\beta$ $\sim$ -2.1,
\begin{equation}
N_{E} \propto E^{\beta}.
\label{???}
\end{equation}
Equation (\ref{Eq:CPL}) is also called the CPL function; it can be used when the high-energy photon index $\beta$ of Band is unconstrained.
(ii) A quasi-thermal component \citep{2010ApJ...709L.172R, 2011ApJ...727L..33G, 2015ApJ...807..148G, 2012ApJ...757L..31A}. 
The thermal energy of the photon will be emitted as BB emission when the optical depth goes below unity; the BB emission can be modified by the Planck spectrum, which is given by the photon flux
\begin{equation}
N_{E}(E,t)=A(t)\frac{E^{2}}{exp[E/kT(t)]-1},
\end{equation}
where $E$ is the photon energy and $k$ is the Boltzmann's constant. 
It has two free parameters: $T=T(t)$ and the normalization $A(t)$ of the photon flux\footnote{In reality, photospheric emission is not expected to be a pure Planck function, since several effects \citep{2008ApJ...682..463P, 2010ApJ...709L.172R} need to be taken into account (e.g. the observed BB temperature and the optical depth are the angle dependent). Physically, the photospheric emission therefore should be a multicolor BB (mBB; e.g., \citealt{2010ApJ...709L.172R}, \citealt{2018ApJ...866...13H}) instead of a single Planck function.}.
And (iii) A power-law component extending to high energies \citep{2010ApJ...712..558A, 2011ApJ...730..141Z}
\begin{equation}
N_{E} \propto E^{s},
\label{PL}
\end{equation}
where $s$ is the power-law index.

Of these, only the BB component indicates a clear physical meaning, which explains as the thermal emission from the photosphere of the outflow. The other two functions do not have any direct physical meaning, although the Band function can be interpreted as being the result of a radiation mechanism (e.g., synchrotron emission), and the power-law component can be a valid approximation of various emission processes over a limited energy range.

\subsection{Spectral Fitting and Model Comparison}

When performing spectral fits, we need to know several important things:
(i) How can we derive the best-fit model parameters from data?
(ii) How well does our model fit our data? 
(iii) How can we choose the best model from given models?

\subsubsection{Spectral Fitting}

The above-mentioned questions involve two operations that require statistics in spectral fitting: one is parameter estimation, which includes finding the best-fit parameters and its uncertainties for a given model; another is testing whether the model and its best-fit parameters actually match the data, which is usually referred to as determining the goodness of fit (GOF).

For the former operation (parameter estimation), the purpose of performing a spectral fit is first to find the ``best-fit'' parameters that make a given model optimally consistent with data. On the other hand, the equally important thing is also to find the true likelihood\footnote{The maximum likelihood is the standard statistic used for parameter estimation. This is because the best values of the parameters are those that maximize the probability of the observed data given the model. In other words, finding the maximum likelihood means finding the set of model parameters that maximize the likelihood function.} map in the entire parameter space, with which how the model parameters can be really constrained in a globally confident manner. 
We adopt the Markov chain Monte Carlo (MCMC) technique to conduct the time-resolved spectral fitting for the following reasons: (i) traditional fitting algorithms (e.g., Levenberg-Marquardt) typically fail to map the multidimensional likelihood; (ii) the theoretical models always involve in multiple model parameters, which may be affected by correlation and multimodality of the likelihood in their parameter space; (iii) the MCMC method can effectively solve multidimensional problems.
We therefore perform the MCMC fitting technique based on the Bayesian statistic by using the 3ML tool (see \S 2.3) to carry out the parameter estimation of data. The priors distribution is used and multiplied by the likelihood that combines the model and the observed data, yielding a posterior distribution of the parameters.
Similar techniques have been widely and successfully applied to GRB modeling (e.g., \citealt{2012A&A...538A.134X, 2015ApJ...799....3R, 2015ApJ...806...15Z}).
The parameters of our model in the Monte Carlo fit are allowed in the following range---index (PL model): [-5, 1]; $KT$ (keV, BB model): [1, 10$^{3}$]; $\alpha$ (CPL model): [-5, 1]; $E_{\rm c}$ (keV, CPL model): [1, 10$^{4}$]; $\alpha$ (Band model): [-5, 1]; $\beta$ (Band model): [-10, 1]; $E_{\rm pk}$ (keV, Band model): [1, 10$^{4}$].

For the latter operation (test the model and its best-fit parameters), the reduced $\chi^{2}$ is often used as a measure of fit quality. Indeed, most of the time it is unreliable and incorrect to use \citep{2010arXiv1012.3754A}. Instead, we can almost always use the bootstrap method to estimate the quality of a Maximum Likelihood Estimate (MLE) analysis; the statistic to choose for the MLE is a Poisson--Gaussian profile likelihood \citep{1979ApJ...228..939C}, the so-called Pgstat\footnote{Which statistics should be used depends on the probability distributions underlying the data. Nearly all astronomical data are drawn from two main distributions: Gaussian (normal) or Poisson distributions, e.g., a Gaussian-distributed data with a background (Chi), or a Poisson-distributed data with a Poisson background (Cstat), or a Poisson-distributed data with a Gaussian background (Pgstat). A more detailed discussion can be found in \cite{2011hxra.book.....A} or at the official XSPEC website: \url{https://heasarc.gsfc.nasa.gov/xanadu/xspec/}.}.
The free parameter $\vec{\theta}$ of the likelihood function $L(\vec{\theta})$ usually tends to be very small; therefore, numerically it is more tractable to deal with the logarithmic form of the likelihood function. In practice, the 3ML provides an algorithm to find the minimum of the $-$$L(\vec{\theta})$ function, which is equivalent to finding the maximum of the $L(\vec{\theta})$ function. 
To perform the analysis, either the MLE analysis -ln(likelihood) or the Bayesian analysis -ln(posterior) can be used.
In this paper, we use the minimum logarithm of likelihood-based statistics based on both the MLE analysis (time-integrated spectral fits) and the Bayesian analysis (time-resolved spectral fits).
When our data are fitted, we can assess the GOF for the MLE analysis by using --ln(likelihood) and for the Bayesian analysis via simulating synthetic data sets, --ln(posterior). 

\subsubsection{Model Comparison} 

Since we often need to know which of a suite of models best represents the data, one question is how to choose between different models. 
The Akaike Information Criterion (AIC; \citealt{1974ITAC...19..716A}), defined as AIC=-2ln$L(\vec{\theta})$+2k, is preferred to discretely select (non-nested, e.g., BB vs. CPL, BB vs. Band, CPL vs. Band) models from a set of models that contains the true data-generating process, under the assumption that the model errors or disturbances are normally distributed, where $L$ is the maximized value of the likelihood function for the estimated model, and $k$ is the number of free parameters to be estimated.
The Bayesian Information Criterion (BIC; \citealt{1978AnSta...6..461S}) is better for comparing nested models (e.g., BB vs. PL+BB, BB vs. CPL+BB, BB vs. Band+BB), which is defined as BIC=-2ln$L(\vec{\theta})$+kln(n), where $n$ is the number of observations\footnote{Note that if the sample size is large, BIC is superior to AIC.} (or the sample size).
In the Bayesian statistics MCMC case, another criterion, namely, the deviance information criterion (DIC), should be invoked, defined as DIC=-2log[$p$(data$\mid\hat{\theta}$)]+2$p_{\rm DIC}$, where $\hat{\theta}$ is the posterior mean of the parameters and $p_{\rm DIC}$ is the effective number of parameters. In this paper, we adopt the corresponding criteria for different model comparison scenarios. Given any two estimated models, the preferred model is the one that provides the minimum AIC, BIC, or DIC scores, which are often compared as $\Delta$AIC, $\Delta$BIC, or DIC scores (the difference between the best model and each model).
For example, we can use $\Delta$BIC to describe the evidence against a candidate model as the best model in the nested model comparisons. If $\Delta$BIC is greater than 10, the evidence against the candidate model is very strong \citep{kass1995bayes}.
Figure \ref{BIC} presents the time evolution of the $\Delta$AIC/BIC/DIC in comparison with two different empirical models. For the first pulse/sub-burst (Fig.\ref{BIC}a), we compared the PL+BB model with the CPL model, the PL+BB model with the single BB model, and the CPL model with the single BB model.  For the rest pulses/sub-bursts (Fig.\ref{BIC}b), the case displays the CPL model against the  Band model. For instance, a majority of $\Delta$BIC  have negative values and much less than -10 for the case of the PL+BB model against the single BB model, indicating that the PL+BB is the preferred model in these two models.

Another common method is to do hypothesis testing.
We can use a likelihood-ratio test to compare the GOF of two statistical models: a null model against an alternative model. This test is based on the likelihood ratio, which expresses how many times more likely the data are under one model than the other.
The steps in hypothesis testing are like this:
First, set up two possible exclusive hypotheses: the null hypothesis (e.g. BB), which is usually formulated to be rejected, and the alternative hypothesis, which is the research hypothesis (e.g. CPL).
Second, specify a priori the significance level $\hat{a}$ (a typical value of 0.05 is usually set). We can reject the null model if the test yields a value of the statistics whose probability of occurrence under the null model is $p<\hat{a}$ (small probability events occurrence). The null model is usually set as a simpler model, while the alternative model is configured as a more complex model. 

\subsection{Time-integrated Spectral Fit Results}

In this work, we adopted the Bayesian approach analysis package, namely, the Multi-Mission Maximum Likelihood Framework (3ML; \citealt{2015arXiv150708343V}), as the main tool to carry out all the temporal and the spectral analyses. We also used the RMFIT (version 3.3pr7) package\footnote{\url{https://fermi.gsfc.nasa.gov/ssc/data/analysis/user/}} and XSPEC (version 12.9.0) package\footnote{\url{https://heasarc.gsfc.nasa.gov/xanadu/xspec/}} \citep{1996ASPC..101...17A} to ensure consistency of the results across various fitting tools. In this section, we will perform a detailed temporal and spectral analyses to find the character and commonality for our sample.

Observationally, a single power-law component cannot well fit the data for most cases, whereas the Band (or a power-law plus an exponential cutoff, hereafter CPL) component favors the majority of bursts. The interplay among three elemental components indeed can explain a variety of observed GRB spectra \citep{2015ApJ...807..148G}.
Therefore, we first perform time-integrated spectral fits by the Band model to each pulse/sub-burst for individual burst.
If the values of $\beta$ are not well constrained (have fairly large values and large uncertainties), we adopt the CPL model to possibly obtain equally good fits for $\alpha$ and $E_{\rm c}$. 
In reality, we find that the high-energy index $\beta$ indeed cannot be well constrained by the Band model fitting in many time-integrated spectra. Also, as suggested in \cite{2016A&A...588A.135Y}, the CPL model is probably the best model for a majority of bursts. 
We therefore use the CPL model for the time-integrated spectral analyses, and mainly focus on the properties of $\alpha$, and $E_{\rm c}$. Since the $\alpha$ value is more important for comparing the physical models, the CPL model is a better choice than the Band model for a rough sample selection when we perform spectral fits.

We select two background intervals (before and after a burst, marked with ``Bkg.Selections'' in Figure \ref{LCs}) and fit the light curve with order 0-4 polynomials for one of the triggered NaI detectors. The 0-order of the polynomial can optimally describe a majority of bursts. A source selection is carried out by which we visually inspected the TTE light curves from each of the 43 GRBs (labelled ``Selection" in Figure \ref{LCs}). The maximum-likelihood-based statistics are used, the so-called Pgstat, given by a Poisson (observation; \citealt{1979ApJ...228..939C})---gaussian (background) profile likelihood.

We reported the properties of all 43 GRBs in Table 1, together with their results of the time-integrated spectral fits, which include the GBM-triggered number, the each time slice (time interval) of the pulses/sub-bursts, the temporal sequence of the pulses/sub-bursts, the low-energy index $\alpha$, and the peak energy $E_{\rm pk(c)}$ of the Band or CPL model; we used the GBM-triggered detectors, the selection criteria (peak flux or fluence), and the -ln(L)/dof (degrees of freedom).
The light curve of prompt emission, along with the source selection, the background selections, and the best background fitting, is shown in Figure \ref{LCs}.

We note that here -ln($L$)\footnote{Here -ln(L) needs to be multiplied by 2.}/dof should approach 1 if the statistic follows a chi-square distribution\footnote{Chi-square distribution is defined as if all $n$ independent random variables are subject to the standard normal distribution, and then the sum of the squares of these $n$ random variables obeying the standard normal distribution constitutes a new random variable, whose distribution rule is called the $\chi^{2}_{\nu}$ distribution, where the parameter $\nu$ is called the degree of freedom. The chi-square distribution is a new distribution formed by a normal distribution. When the degree of freedom $\nu$ is large, the chi-square distribution is approximately normal.}. We use the Pgstat\footnote{Pgstat = -2 ln($L$).} rather than the chi-square in this paper; this is why -2ln($L$)/dof cannot be expected to be close to 1 for most of our cases (see Table 1). A detailed discussion of why Pgstat is preferred can be found in \cite{2016ApJ...827L..38G}. The main reason that we are using different statistics like Pgstat and Cstat rather than chi-square is that, in the {\it Fermi} data analysis here, counts per bin are not large and each bin therefore cannot be considered as an observation from a normal distribution. So when counts per bin become large enough, the Poisson distribution can be well approximated by a Gaussian, then Pgstat or Cstat will approach chi-square behaviour, and then in that case -2ln($L$)/dof can be expected to be close to 1. Otherwise, it is not necessary to be close to 1. 

We next compare the low-energy index $\alpha$ and peak energy $E_{\rm pk(c)}$ between pulses/sub-bursts to investigate how $\alpha$ and $E_{\rm pk(c)}$ evolve in different pulses/sub-bursts within one single burst. If $\alpha$ and $E_{\rm pk(c)}$ values are very different, it indicates a different physical process. 
In Figure \ref{SpectralRelations}, we present $\alpha^{\rm P}$-$\alpha^{\rm F}$ (Fig.\ref{SpectralRelations}a) and $E^{\rm P}_{\rm pk(c)}$-$E^{\rm F}_{\rm pk(c)}$ (Fig.\ref{SpectralRelations}b) relations\footnote{The symbols `P' and `F' represent previous and following pulses (or sub-bursts), respectively.} between different multi-pulses/sub-bursts based on time sequence and find that the relations in later pulses show a more discrete behavior compared with previous ones. This result is confirmed with the quantitative test by using the probabilities of the nonparametric Kendall coefficient\footnote{For the Band model $\alpha^{\rm P}$-$\alpha^{\rm F}$ relations, we have $\tau$=0.45 (1st to 2nd), $\tau$=0.41 (2nd to 3rd), and $\tau$=-0.33 (3rd to 4th). For the Band model $E^{\rm P}_{\rm pk}$-$E^{\rm F}_{\rm pk}$ relations, one has $\tau$=0.46 (1st to 2nd), $\tau$=0.25 (2nd to 3rd), and $\tau$=0.24 (3rd to 4th).}.
This indicates that the spectral evolution prominently occurs in the later pulses. On the other hand, we compare the $E_{\rm pk(c)}-\alpha$ relation of different pulses/sub-bursts (Figure \ref{EpAlpha}) and find that the $E_{\rm pk(c)}-\alpha$ relation generally presents two different behaviors for two different $\alpha$ distributions, which show a positive correlation for $\alpha \lesssim$ -1.0 while show a negative correlation for $\alpha \gtrsim$ -1.0 in the log-linear plots. 
Moreover, in comparing with the different empirical models, the slope obtained from the Band model is significantly steeper than that of the CPL model for $\alpha \lesssim$ -1.0, while interestingly present the same for $\alpha \gtrsim$ -1.0 (Figure \ref{EpAlpha}). 
In general, the global behavior of $E_{\rm pk(c)}$ presents a hard-to-soft evolution \cite[e.g.,][]{2012ApJ...756..112L}.

\begin{figure*}
\includegraphics[angle=0, scale=0.45]{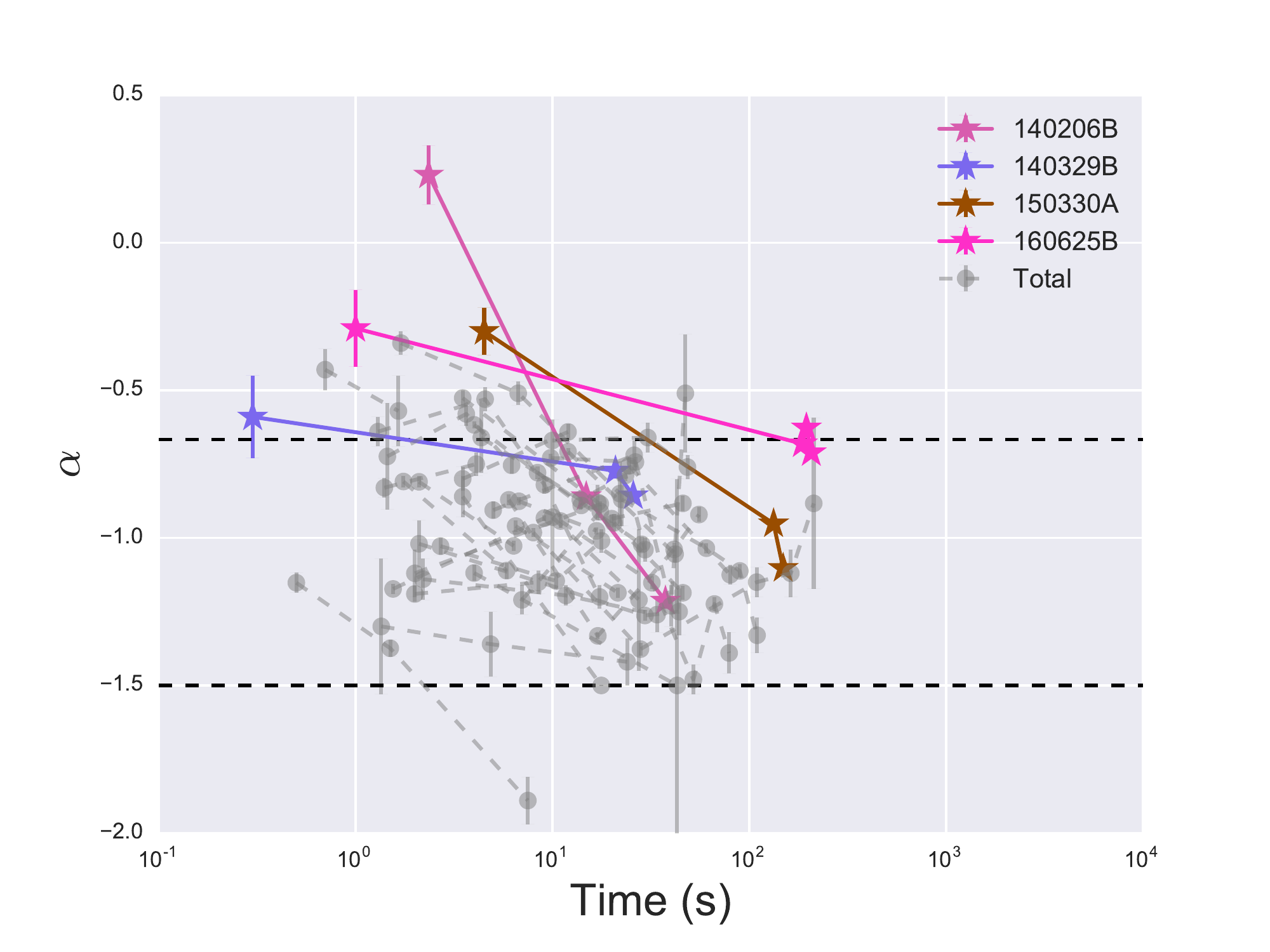}
\includegraphics[angle=0, scale=0.45]{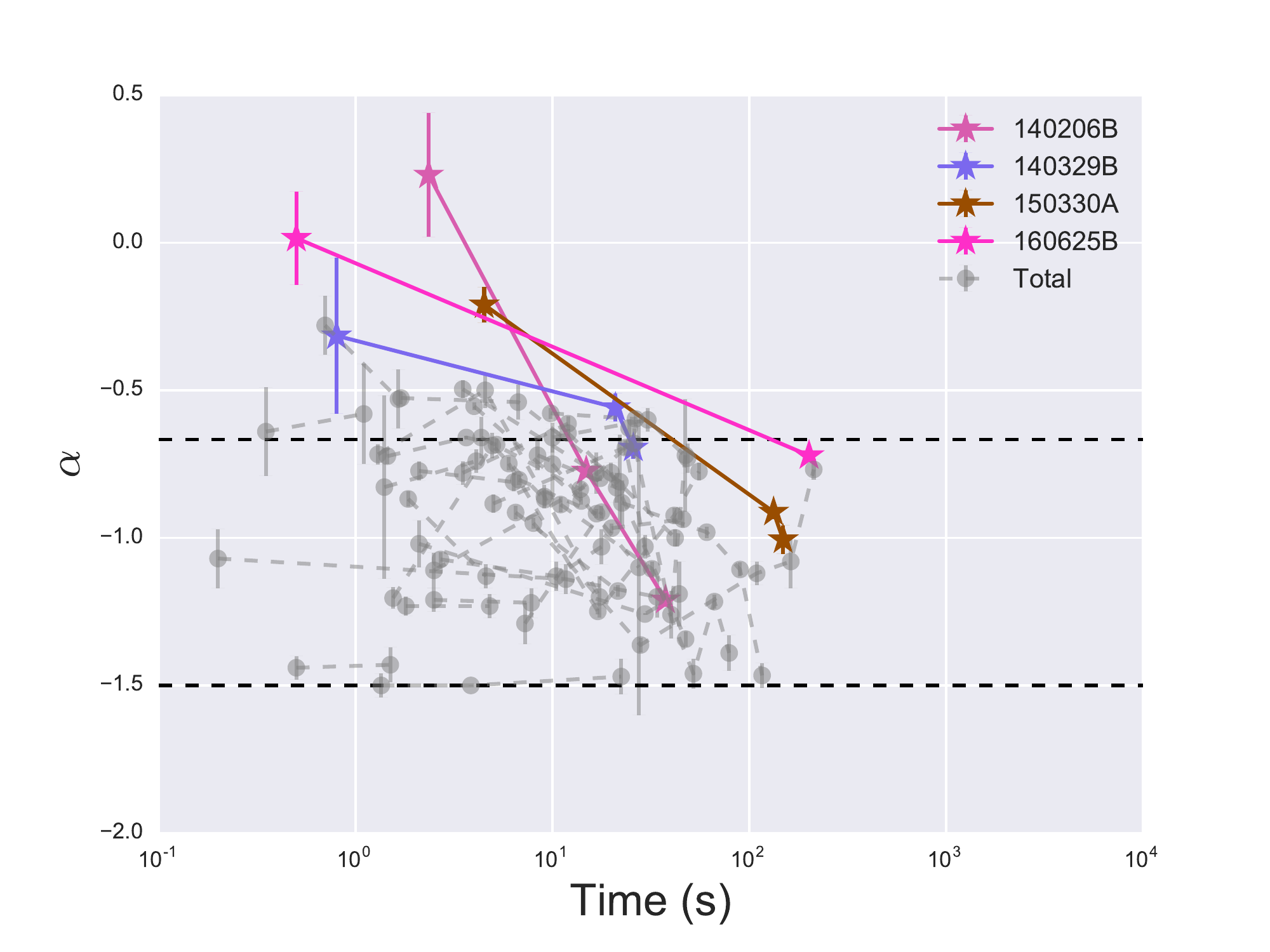}
\caption{Temporal evolution of $\alpha$ based on the time-integrated spectra for each pulse/sub-burst for our global sample. Each data point represents one pulse/sub-burst, and they are connected by solid lines within one single burst. Two horizontal dashed lines represent the limiting values of $\alpha$=-2/3 and $\alpha$=-3/2 for electrons in the slow- and fast-cooling regimes, respectively. Left panel: for the CPL model; right panel: for the Band model.}\label{AlphaTime}
\end{figure*}

We select the bursts with such features in our sample: the low-energy index $\alpha$ above -2/3 (much harder and close to 0) for the first pulse/sub-burst, and dropping below -2/3 (generally range from -3/2 to -2/3) for the second pulse/sub-burst. 
In total, based on the time-integrated spectra fitting results, we find that 9 out of 43 bursts in our sample satisfy this criterion significantly. These bursts are GRB 090926181 \citep{2011ApJ...729..114A}, GRB 100719989, GRB 140206275 \citep{2014MNRAS.444.2776G}, GRB140329295, GRB 140523129, GRB 150330828, GRB 160625945 \citep{2018NatAs...2...69Z}, GRB 160820259 \citep{2018ApJ...862..154C}, and GRB 171102107.
Such a transition that across the line of the death of synchrotron emission in different pulses/sub-bursts within a single burst could provide a good clue for us to study the jet composition.
Based on this, we finally obtain four interesting bursts with adequate data for a further time-resolved spectral analysis and model discussion: GRB 140206B, GRB 140329B, GRB 150330A, and GRB 160625B\footnote{Corresponding to the {\it Fermi} GBM-triggered burst ID: GRB 140206B (140206275), GRB 140329B (140329295), GRB 150330A (150330828), and GRB 160625B (160625945).}.
The lower-energy power-law index\footnote{Time-integrated spectral fit results for the first pulses give 0.23$\pm$0.21 (Band) and 0.23$\pm$0.10 (CPL) for GRB 140206B; -0.32$\pm$0.26 (Band) and -0.59$\pm$0.14 (CPL) for GRB 140329B; -0.21$\pm$0.06 (Band) and -0.30$\pm$0.08 (CPL) for GRB 150330A; and 0.02$\pm$0.16 (Band) and -0.29$\pm$0.13 (CPL) for GRB 160625B.}, $\alpha$, evolves with time in different pulses/sub-bursts for those bursts and is shown in Figure \ref{AlphaTime}.

\subsection{Time Bin Techniques}

To perform a time-resolved spectral analysis, how to effectively time-bin the data is a crucial issue, since time-resolved spectral analysis is the main method of extracting information from the data. \cite{2014MNRAS.445.2589B} investigated several techniques for temporal binning of GRB spectra---constant cadence (CC), Bayesian blocks (BBlocks, \citealt{2013ApJ...764..167S}), signal-to-noise ratio (S/N), and Knuth bins (KB)---and concluded that the S/N and the BBs are two most effective binning methods. Yet both these two techniques have their own disadvantages. The traditional S/N method ensures enough photons to perform the spectral fits, but sometimes it could also destroy the physical structure. BBlocks refer to the technique with the following features \citep{2014MNRAS.445.2589B}:
(i) each time bin conforms to a constant Poisson rate;
(ii) time bins selection is conducted by algorithmically subdividing the flux history of the GRB light curve;
(iii) comparing the likelihood of the distribution of the count rate of each bin to being piecewise constant or constant; 
(iv) possess a variable width and variable S/N;
(v) the selection of the time bins demonstrates the true variability of the data.
Nevertheless, the technique does not guarantee adequate signal in the bins to carry out an accurate spectral fit.
 
Therefore, to ensure all the advantages from both these two methods, we first use the BBlocks binning method to time-bin the data, then calculate the S/N value for each individual bin, and obtain the bins that have suitable S/N values (we applied S/N greater than 20 in this paper). Time bins selected in this case thus could be more reasonable.

In Figure \ref{SNRTime}, we take GRB 140206B as an example to show temporal evolution of the S/N based on the BBlcoks technique, along with its light curve of the prompt emission.  Time evolution of the S/N and flux track each other for all time bins.

\subsection{Time-resolved Spectral Fit Results}

We rebin the TTE data by using the BBlcoks method with false alarm probability p0 = 0.01 to the TTE light curves of one of the brightest NaI detectors for each burst, and other trigged detectors follow the same bin time information.
In total, we obtained 13 pulses/sub-bursts and 132 time-resolved spectra\footnote{The number of time-resolved spectra for each burst and their temporal sequence are as follows: 140206B, 3 (1st), 12 (2nd), 8 (3rd); 140329B, 4 (1st), 6 (2nd), 11 (3rd); 150330A, 6 (1st), 21 (2nd), 9 (3rd); 160625B, 7 (1st), 38 (2nd).} from these four bursts following the BBlocks.
All these four bursts in our sample are long GRBs (type II GRBs), with $T_{90}$ being 93.6s (GRB 140206B), 21.5s (GRB 140329B), 153s (GRB 150330A), and 460s (GRB 160625B), respectively.
Three bursts (all except GRB 140206B) show a clear short precursor \citep{2014ApJ...789..145H} characteristics before the main emission episode. The first pulses/sub-bursts for all three bursts in our analysis are consistent with being a precursor of the main burst. This interesting finding is inconsistent with some previous findings \cite[e.g.,][]{2010ApJ...723.1711T, 2014ApJ...789..145H}, which claimed that no thermal precursor is found in early time. The reason for this could be that early short precursors are typically so weak that they cannot ensure enough photons to perform precise spectral fits. This is why we select the bright bursts in our sample. There is also another possibility. If precursors have low S/N values\footnote{Note that we also keep these spectra that have lower S/N values ($<$20) in the analysis for precursors, since early short precursors are always weak, and a majority of time-resolved spectra of precursors have low S/N values ($<$20). This indicates that fewer spectra can be obtained and it is difficult to study their evolution properties. On the other hand, we also discard the time bins before triggered time, which are indicated by ellipsis dots in the tables for each analyzed bursts.}, the underlying spectrum may not be reliably obtained. This is because it is difficult to do the model selection, and the finding of the thermal component is not convincing.

\begin{figure*}
\includegraphics[angle=0, scale=0.45]{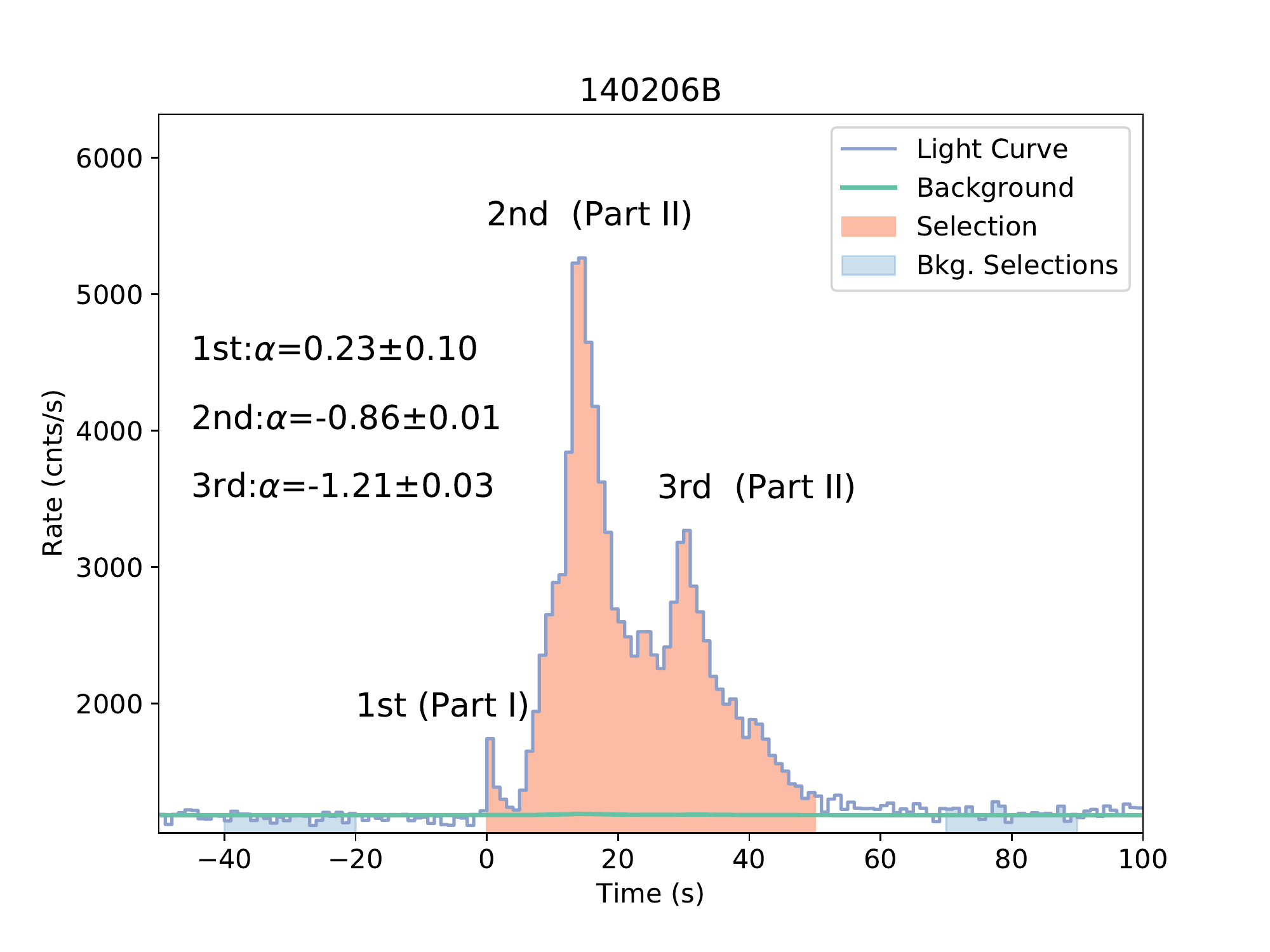}
\includegraphics[angle=0, scale=0.50]{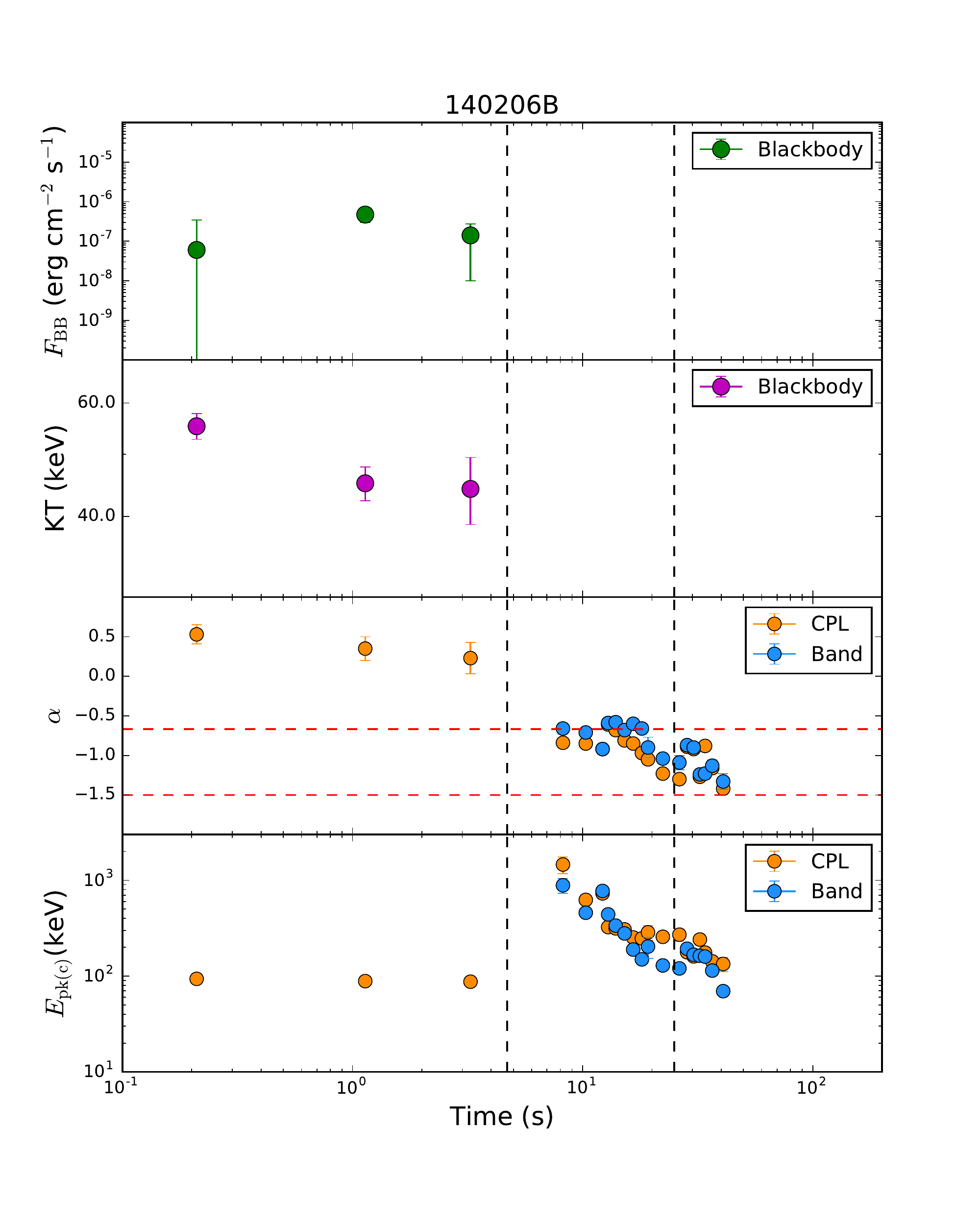}
\includegraphics[angle=0, scale=0.45]{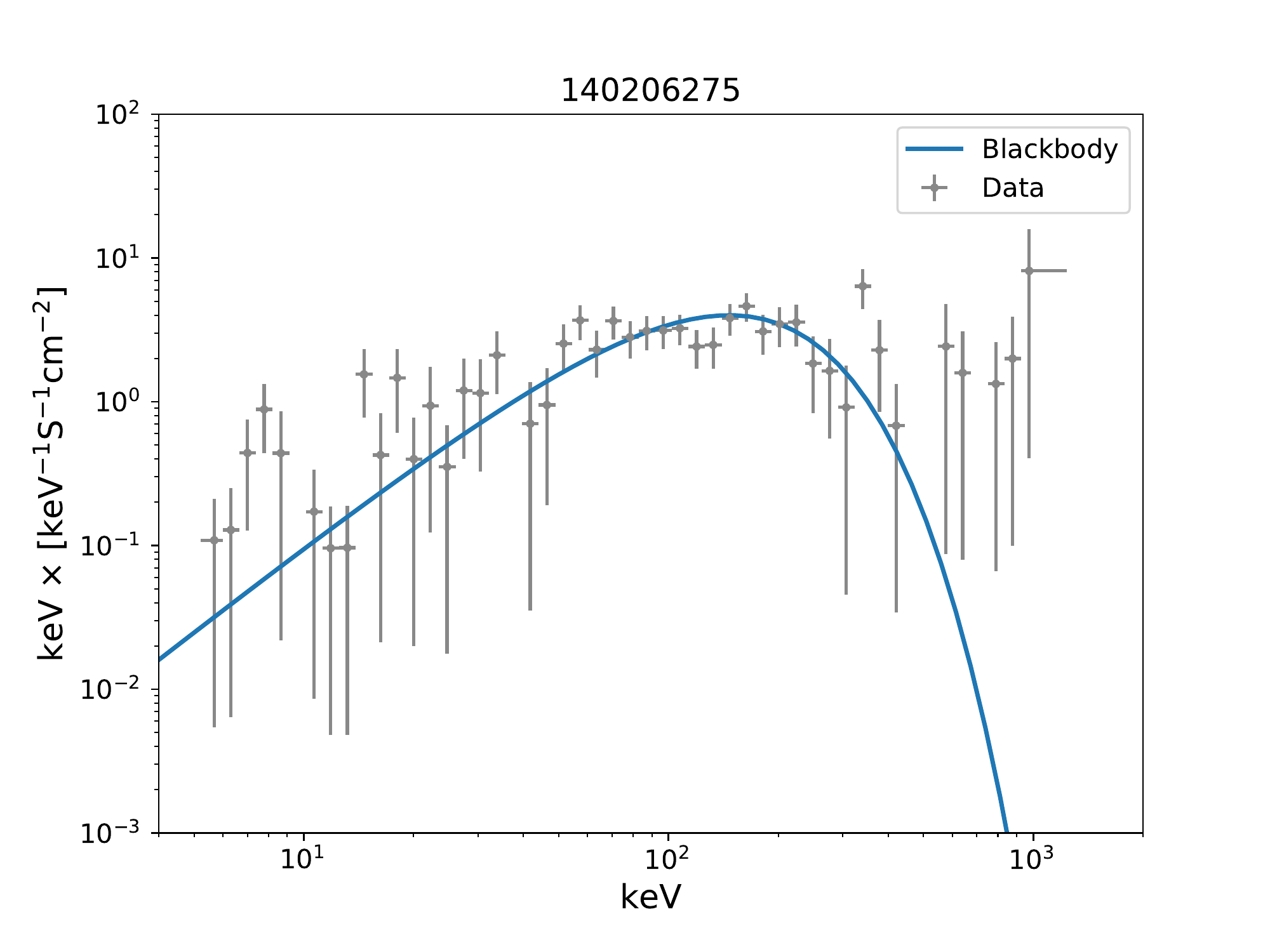}
\includegraphics[angle=0, scale=0.45]{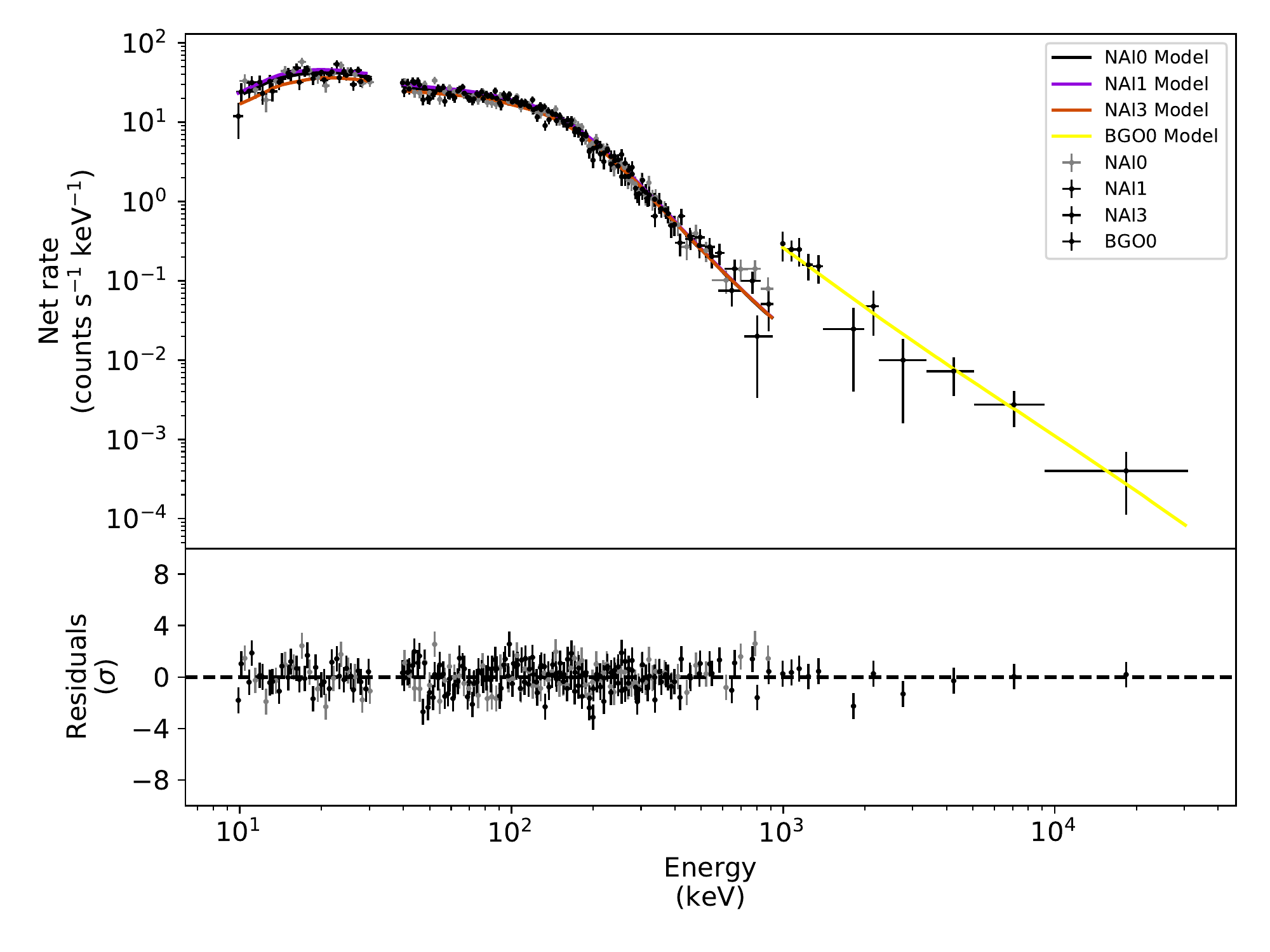}
\caption{Top left panel: light curve of the prompt emission of GRB 140206B, along with the source selection, the background selections, and the best background fits. Top right panel: temporal evolution of $\alpha$ (Band and CPL), $E_{\rm pk(c)}$ (Band and CPL), BB temperature, and BB flux $F_{\rm BB}$ based on the time-resolved spectrum for GRB 140206B. Three pulses/sub-bursts are divided by two vertical black dashed lines. Two horizontal red dashed lines represent the limiting values of $\alpha$=-2/3 and $\alpha$=-3/2 for electrons in the slow- and fast-cooling regimes, respectively. Bottom left panel: typical spectral fits using the brightest time bin (-0.04, 0.46) for Part I. Bottom right panel: typical spectral fits using the brightest time bin (13.24, 14.60) for Part II.}\label{GRB140206275}
\end{figure*}

Similar to the time-integrated spectral fits, we also find that many cases show an unconstrained $\beta$ in the time-resolved spectra by the Band model fitting. On the other hand, the CPL model is the best model for a majority of bursts in the GBM GRB time-resolved spectral catalog as suggested in \cite{2016A&A...588A.135Y}. 
We therefore use the CPL to perform time-resolved fits for our sample. 
One interesting question is the difference of the $\alpha$ values comparing  the CPL and the Band model (see \S \ref{sec:Comparison}). 
To clarify this, we first take GRB 140206B as an example to study such questions by performing the detailed spectral analysis. 
GRB 140206B was triggered by GBM (8 keV--40 MeV) on board the {\it Fermi}, and had a fluence of (9.93$\pm$0.05)$\times$10$^{-5}$erg cm$^{-2}$ in the energy range 10 keV--1000 keV.
The light curve of the prompt emission of GRB 140206B shows three clear distinct pulses (see Figure \ref{GRB140206275}).
Before the main pulse, we find a small pulse with a duration of $\sim$ 5 s. Such a case cannot be defined as a precursor since it does not have a long quiescent interval between it and the main pulse.

Time evolution of $\alpha$ (both the Band and the CPL), $E_{\rm pk(c)}$ (both the Band and the CPL), BB temperature, and BB flux $F_{\rm BB}$ based on the time-resolved spectrum for GRB 140206B, together with its light curve of prompt emission and typical spectral fits to each sub-burst, are shown in Figure \ref{GRB140206275}.
The best-fit parameters and their uncertainties, the -ln(posterior), the degree of freedom of data, and the statistical parameters AIC, BIC, and DIC, are presented in Table \ref{Table:140206B}.

We obtain four spectra for the first pulse. We first fit the data with the CPL model and find that all $\alpha$ values are above 0, which are much higher than what is typically expected in the synchrotron emission model.
We are then motivated to fit the observed spectrum with the single BB model, and we find that it can be fitted well by such a model (an example is shown in Figure \ref{MCFitBB}).
Also, comparing with the CPL model, the single BB has lower AIC, BIC, and DIC scores, indicating that the single BB is a preferred model to the data.
We also try to add an additional component to the BB component and find that adding a CPL component to the BB neither improves of the GOF nor leads to constraints on the parameters of the new component. However, we find that adding a PL component to the BB can obtain a constraint on the parameters of the new component and an acceptable result of the GOF.

Next, we perform the time-resolved spectral fits for the second and the third pulses by both the Band and the CPL models. A total of 10 spectra in 2nd and 7 spectra in 3rd are obtained\footnote{Three spectra are discarded since we require an S/N value $>$20.}. We find that the $\beta$ parameter can not be well constrained for a majority of spectra, ranging within [3.0-8.0]. However, we almost can obtain a good fit result for all the spectra by the CPL model. The result is consistent with the finding in \cite{2016A&A...588A.135Y}.
An example fit to the data in a one time bin of GRB 140206B comparing the Band model and the CPL model is shown in Figure \ref{MCFitBandCPL1} for the case of an unconstrained parameter $\beta$ and in Figure \ref{MCFitBandCPL2} for the case of a constrained parameter $\beta$. 

As discussed above, we adopt the following steps for the time-resolved spectral analysis for the remaining bursts (GRB 140329B, GRB 150330A, and GRB 160625B). 
GRB 140329B, GRB 150330A, and GRB 160625B were all triggered by GBM (8 keV-40 MeV) on board {\it Fermi}. 
GRB 140329B also has {\it Fermi}/IPN burst and LAT detection, and the intense high-energy photon flux of GRB 160625B also triggered the Large Area Telescope (LAT) on board {\it Fermi}, Konus-{\it Wind}, INTEGRAL/SPI-ACS, {\it RHESSI} and {\it CALET} \citep{2018NatAs...2...69Z}. 
Both GRB 140329B and GRB 150330A have two sub-bursts that include three well-separated pulses. The fluence in the energy range 10 keV-40 MeV for 
GRB 140329B is (6.70$\pm$0.03) $\times$ 10$^{-5}$ erg cm$^{-2}$ and for GRB 150330A is (1.44$\pm$0.06) $\times$ 10$^{-4}$ erg cm$^{-2}$. GRB 160625B is one of the few extremely bright bursts, composed of three sub-bursts separated by two quiescent times, and has a fluence of (6.43$\pm$0.02) $\times$ 10$^{-4}$ erg cm$^{-2}$ in the energy range 10 keV-40 MeV.
All three bursts show a short precursor before their main emission episodes, but the duration of precursors and the quiescent interval are quite different. GRB 140329B has a duration of the short precursor of $\sim$ 1.6 s, and a quiescent interval of $\sim$ 18 s, separated between the short precursor and the main burst. GRB 150330A has the longest duration of a precursor, with $\sim$ 10 s. After the precursor, a very long quiescent interval is found before the main bursts arrival, with $\sim$ 110 s.
Similarly, a very long quiescent interval is also found in GRB 160625B, with $\sim$ 180 s. However, the duration of precursor for GRB 160625B is short\footnote{The short duration of the precursor with such a long quiescent interval (e.g., GRB 160625B) gives it the appearance of a traditional short GRB, which is undoubtedly related to the main event. To confirm this,  the properties between different sub-bursts need to be tested \citep{2018NatAs...2...69Z}.}, with $\sim$ 1 s, which is different from GRB 150330A but similar to GRB 140329B. 
 
Since the first pulses/sub-bursts (hereinafter Part I) in our sample always are involved in a quasi-thermal spectral component, we therefore use all kinds of models that contain the BB component (e.g, single BB, PL+BB, CPL+BB, Band+BB) to perform the time-resolved spectral analysis and choose the best one via the hypothesis testing (an example of the null hypothesis testing between the single BB and the PL+BB model is reported in Table \ref{Table:Hypothesis}) and the AIC/BIC/DIC judgment (Figure \ref{BIC}). 
To choose the best model, we compare all the used models (together with the CPL model) by analyzing the statistical parameters for the brightest time bins (the highest S/N value) for each burst. We present these statistical parameters in Table \ref{Table:Comparison}, which includes the -ln(posterior) for MCMC based on the Bayesian analysis, the AIC/BIC/DIC scores, and the degree of freedom of data, dof. Based on the statistical analysis, we also list the best model suggested for each burst in the table.
Observationally, GRB spectra are typically fitted with standard empirical models containing a single Band function with or without an additional BB \citep{2015MNRAS.450.1651I}. However, we find that the Band+BB components to fit the data show unconstrained fit parameters for all the cases. 
Besides, in order to check the consistency of the results with the time-integrated spectral fitting, we also redo the spectral fits with the CPL model and find that $\alpha$ values are still very hard (above -2/3) for almost all the cases. 

After the first pulses/sub-bursts, the subsequent pulses/sub-bursts (e.g. 2nd, 3rd, 4th, 5th, hereinafter Part II) in our sample typically show a featureless nonthermal spectrum; we thus apply the CPL model to perform the spectral fits. Similarly, we also redo the spectral fits with a BB component if we still find a case with a hard $\alpha$ index (above -2/3). However, we find that almost all the resolved spectra have a softer $\alpha$ index. Also, the CPL model can fit the data well for nearly all cases. The typical spectral fits (the BB for Part I and the CPL for Part II) for our sample are presented in Table \ref{Table:140206B}, Table \ref{Table:140329B}, Table \ref{Table:150330A}, and Table \ref{Table:160625B}.

As argued above, we find that the best models of each sub-burst for time-resolved spectral fits for our sample are as follows: 
\begin{itemize}
\item GRB 140206B: single BB (part I), CPL (part II);
\item GRB 140329B: PL+BB (part I), CPL (part II); 
\item GRB 150330A: CPL+BB (part I), CPL (part II);  
\item GRB 160625B: PL+BB (part I), CPL (part II).
\end{itemize}

All the best models in the final choice for both Part I and Part II are based on the likelihood-ratio test, namely, hypothesis testing (see in Table \ref{Table:Hypothesis} and \S 2.2.2). We apply the suggested best models to derive the characteristic temperature of the photosphere emission, which is used to derive the physical parameters of the photosphere (see \S 3).
The detailed comparison between the single BB model and the PL plus BB model is based on the likelihood-ratio test; see Table \ref{Table:Comparison}.

\begin{figure*}
\includegraphics[angle=0, scale=0.45]{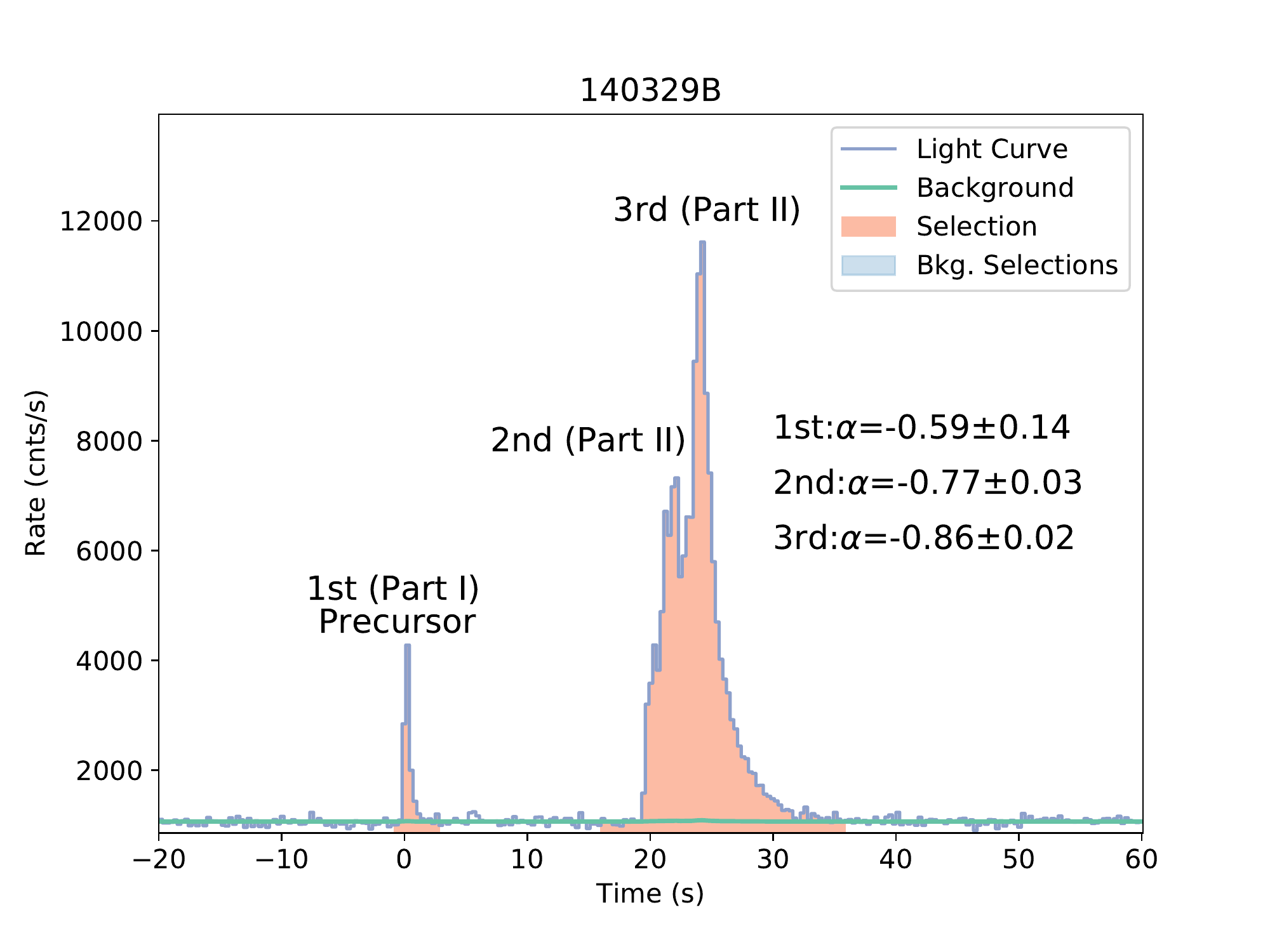}
\includegraphics[angle=0, scale=0.50]{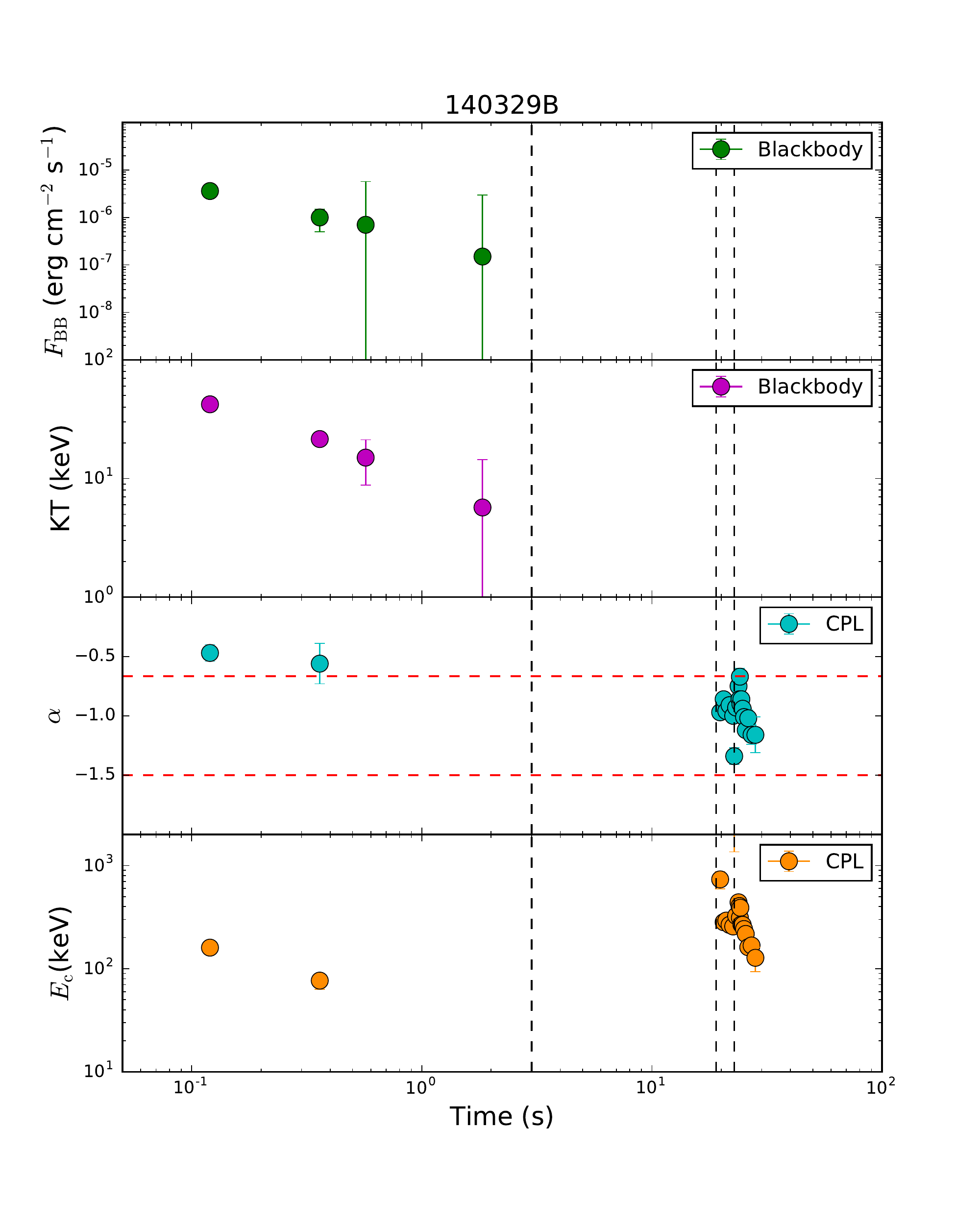}
\includegraphics[angle=0, scale=0.45]{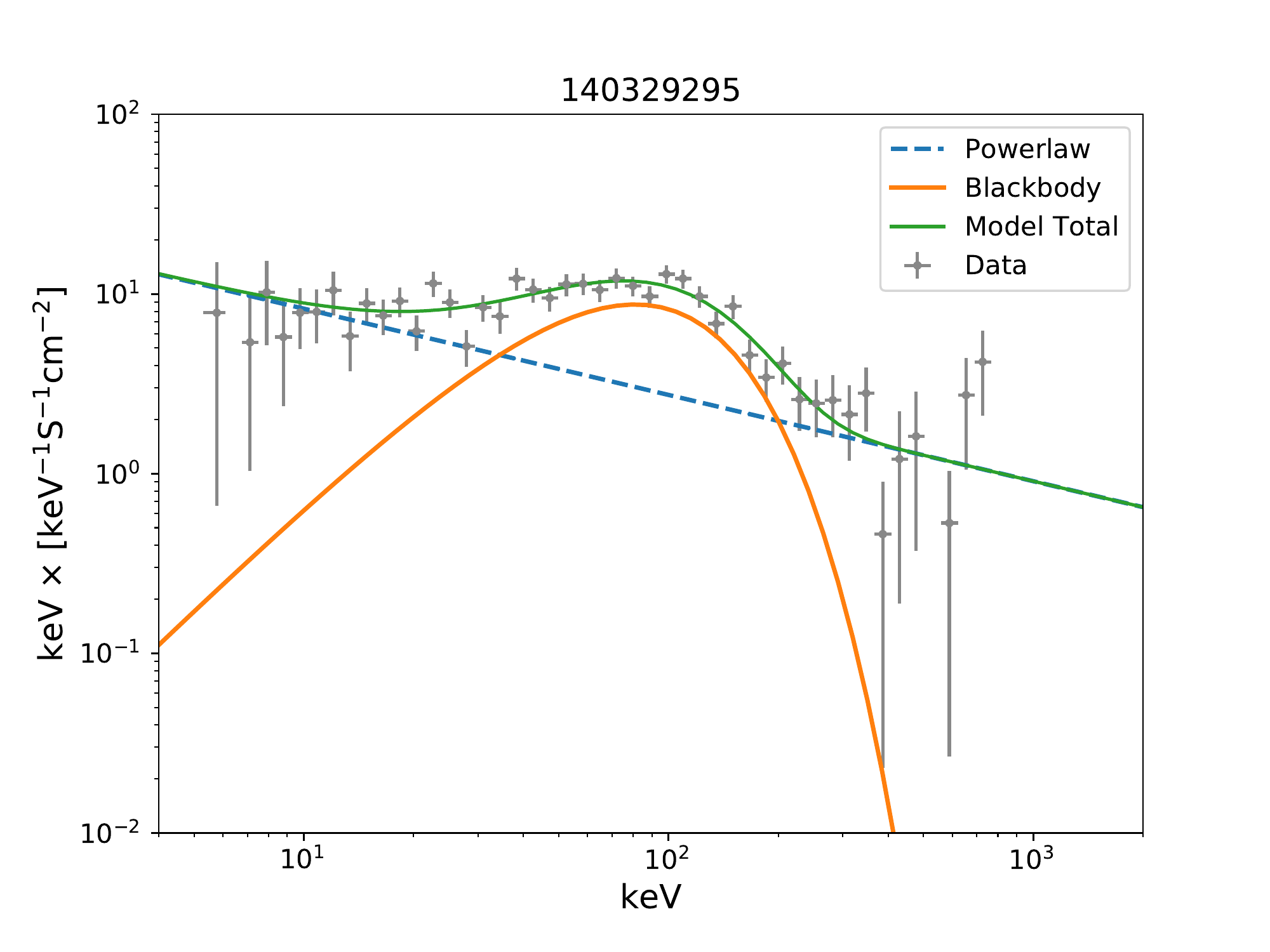}
\includegraphics[angle=0, scale=0.45]{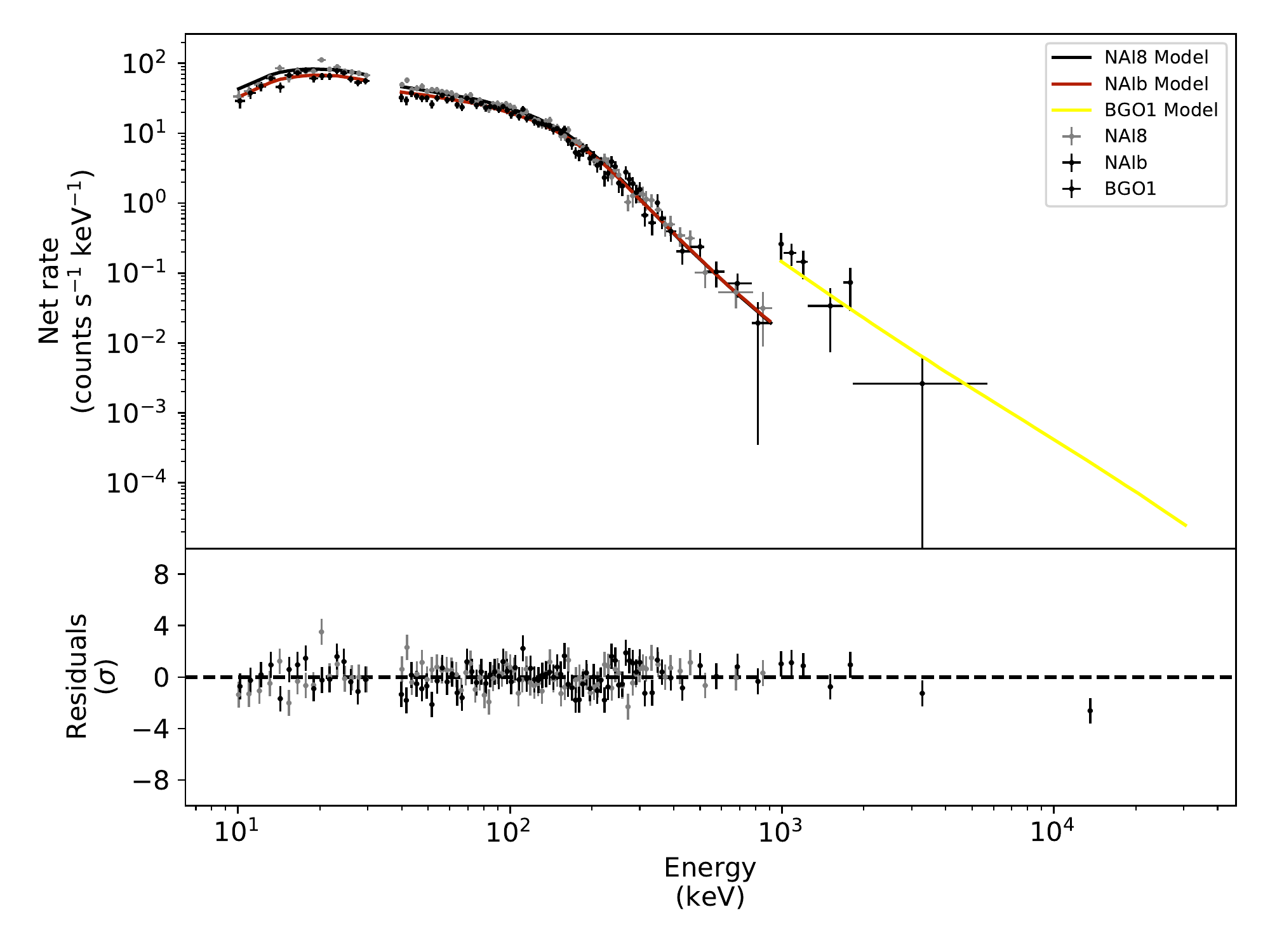}
\caption{Same as Figure \ref{GRB140206275} but for GRB 140329B.
 The brightest time bins (highest S/N values) for Part I (-0.01-0.25) and for Part II (21.17-22.77)  are used.}\label{GRB140329295}
\end{figure*}

\begin{figure*}
\includegraphics[angle=0, scale=0.45]{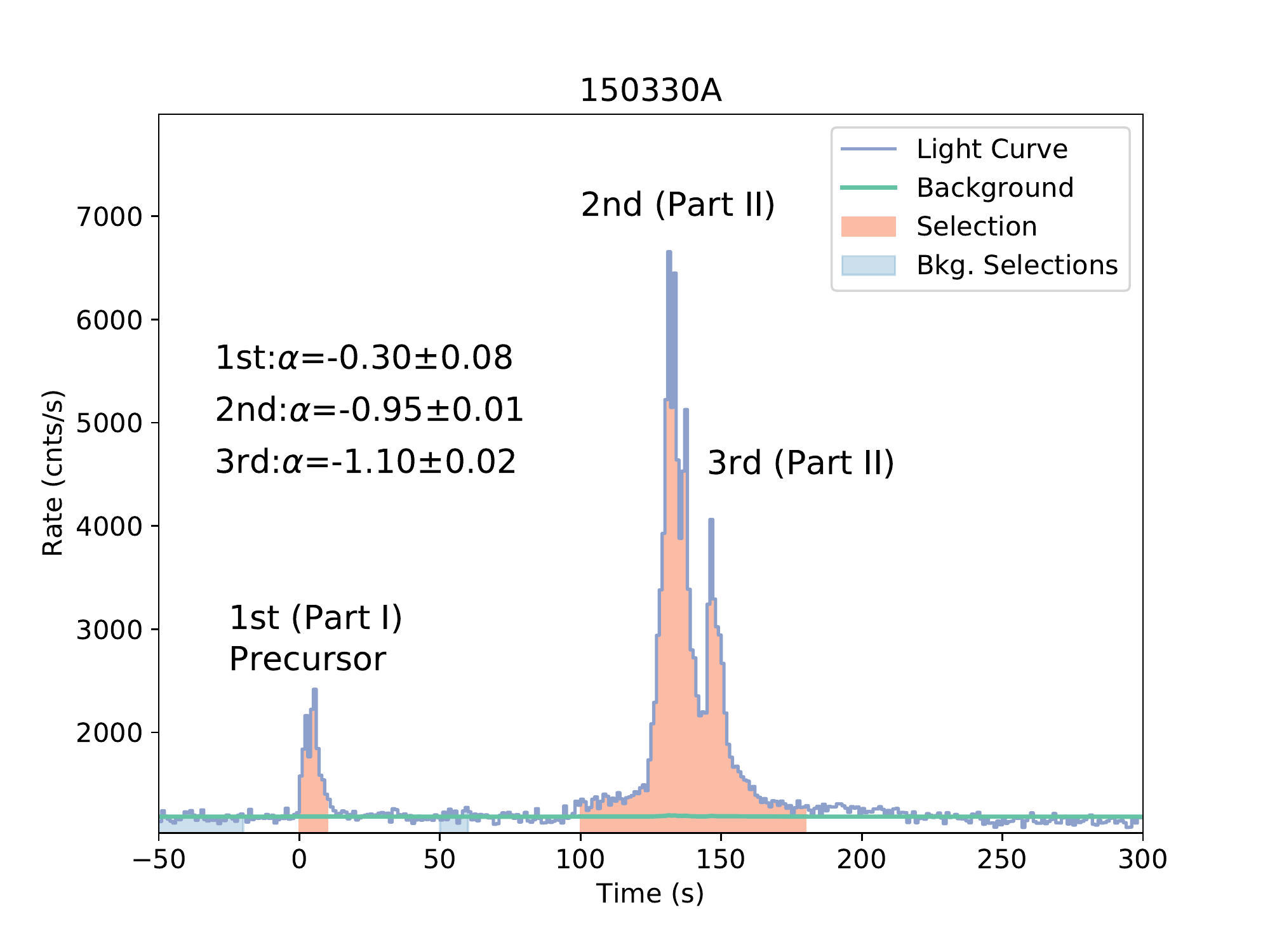}
\includegraphics[angle=0, scale=0.50]{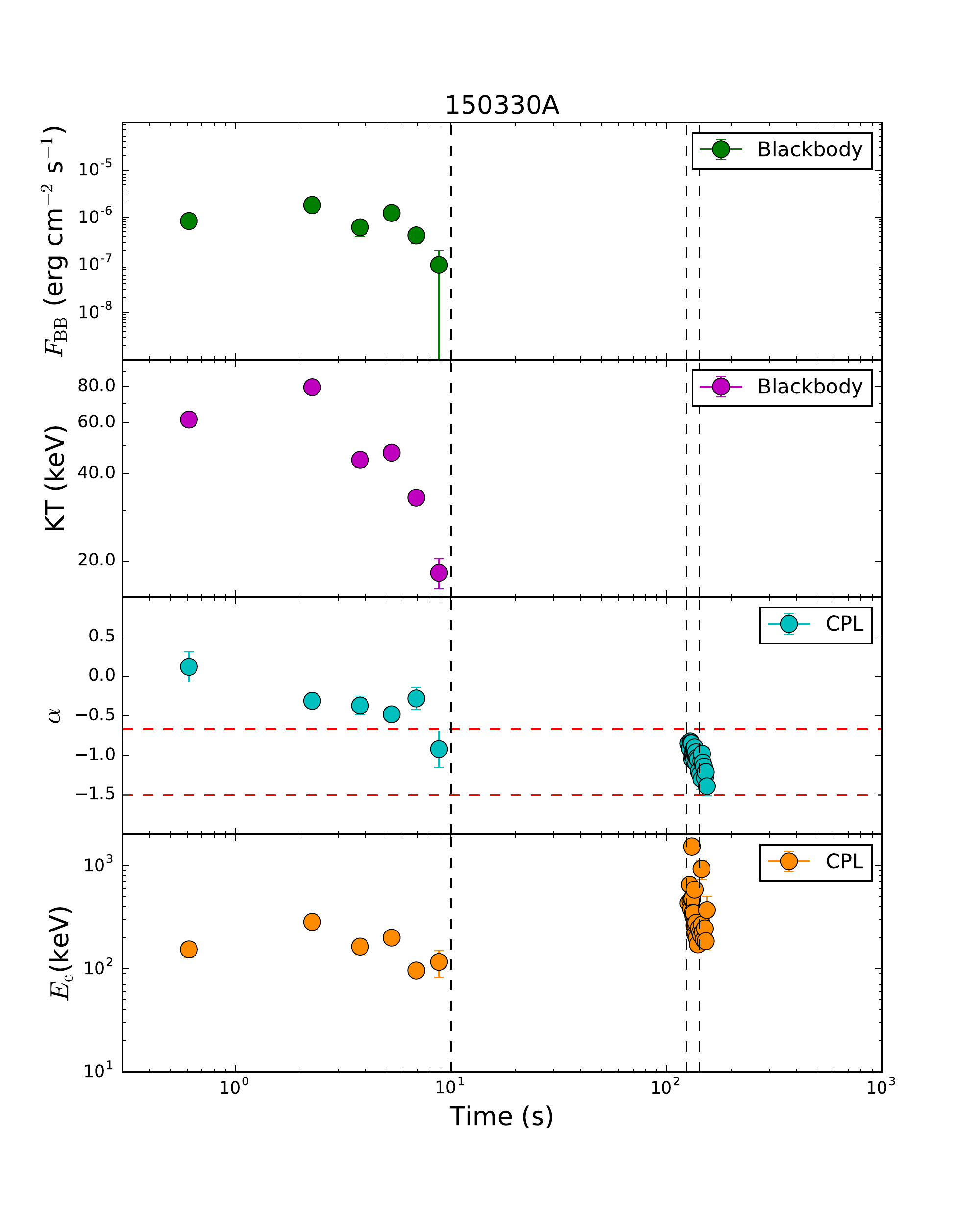}
\includegraphics[angle=0, scale=0.45]{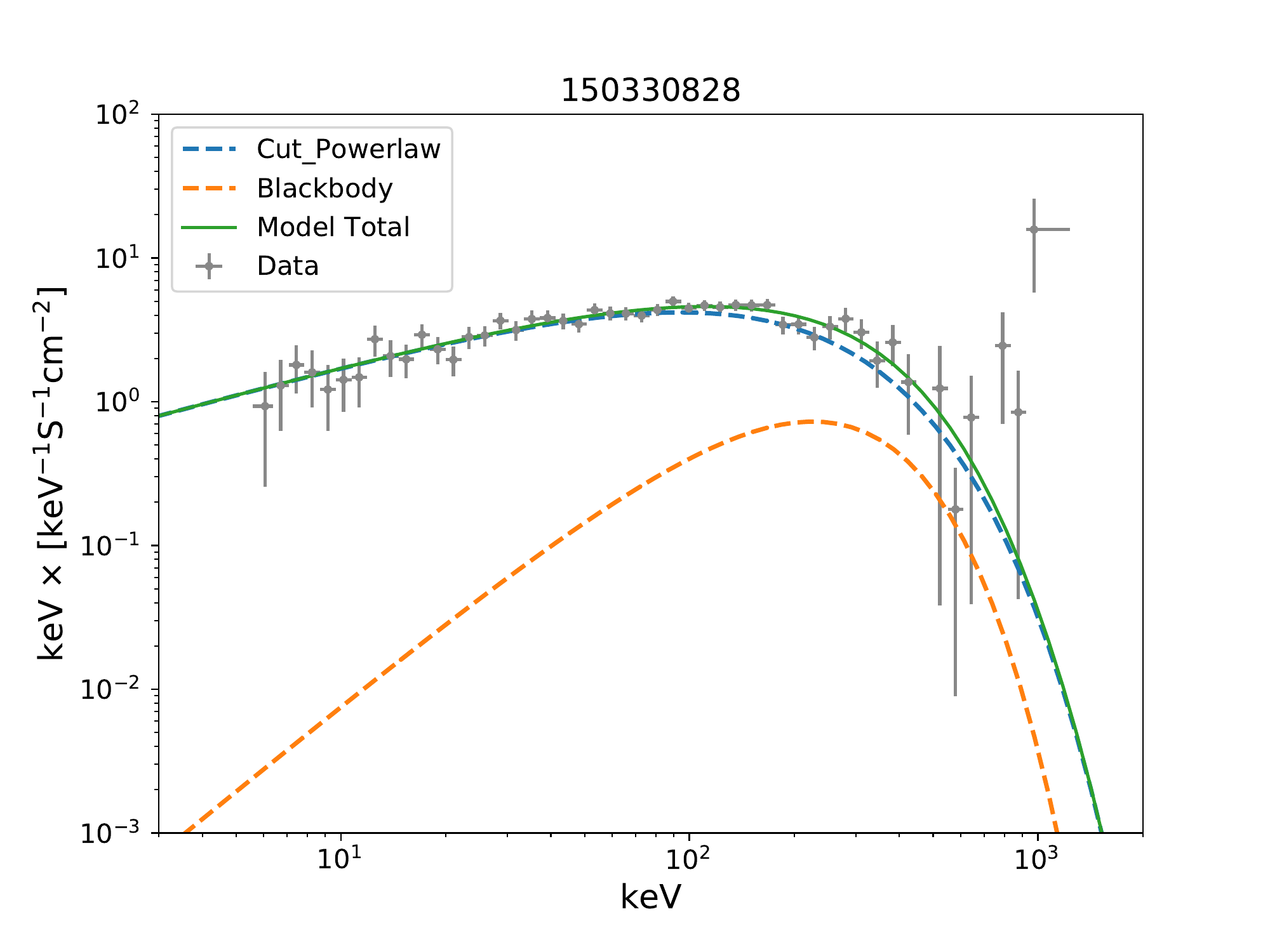}
\includegraphics[angle=0, scale=0.45]{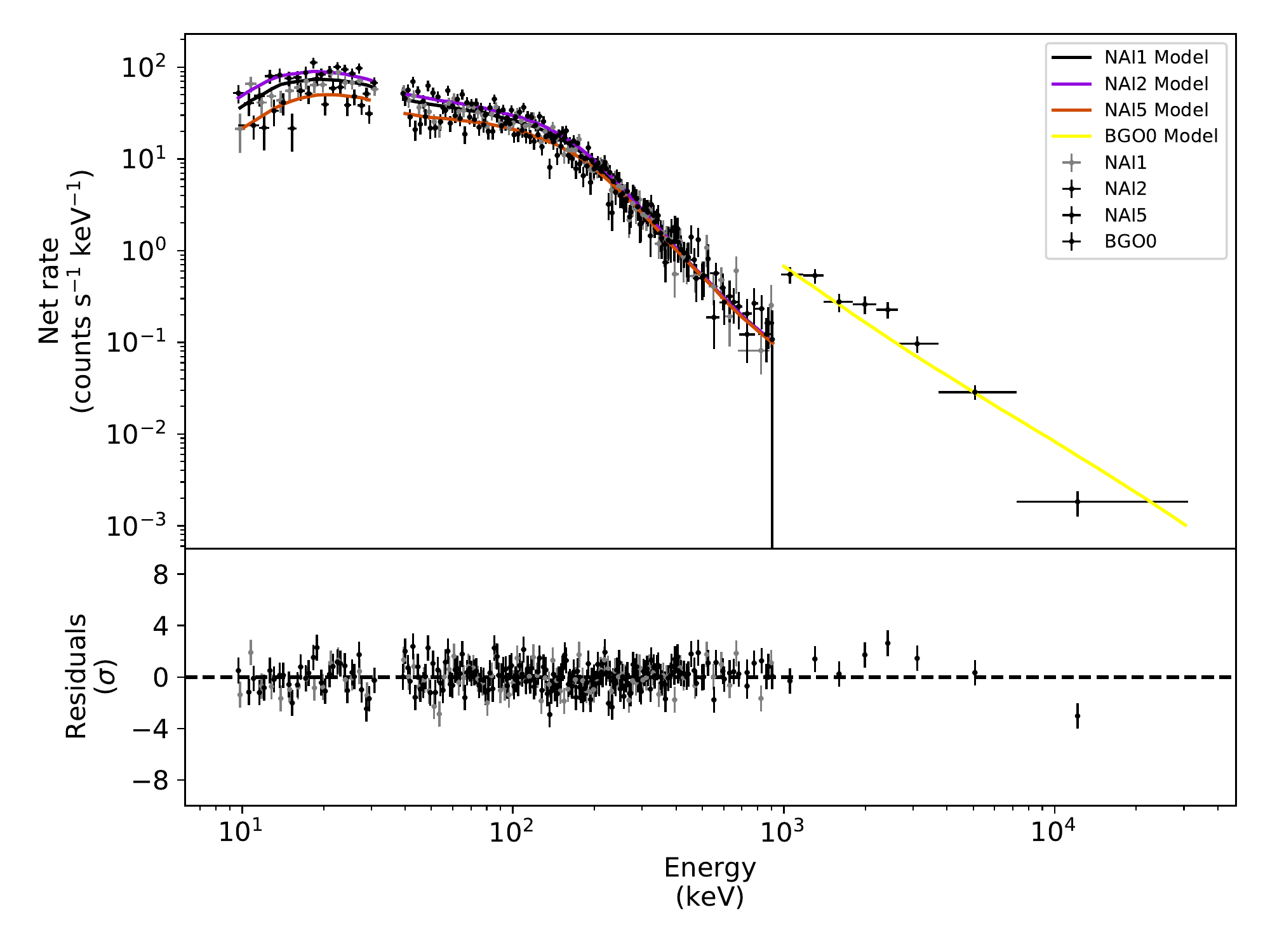}
\caption{Same as Figure \ref{GRB140206275} but for GRB 150330A. The brightest time bins for Part I (4.43-6.20) and for Part II (131.01-131.52) are used.}\label{GRB150330828}
\end{figure*}

\begin{figure*}
\includegraphics[angle=0, scale=0.45]{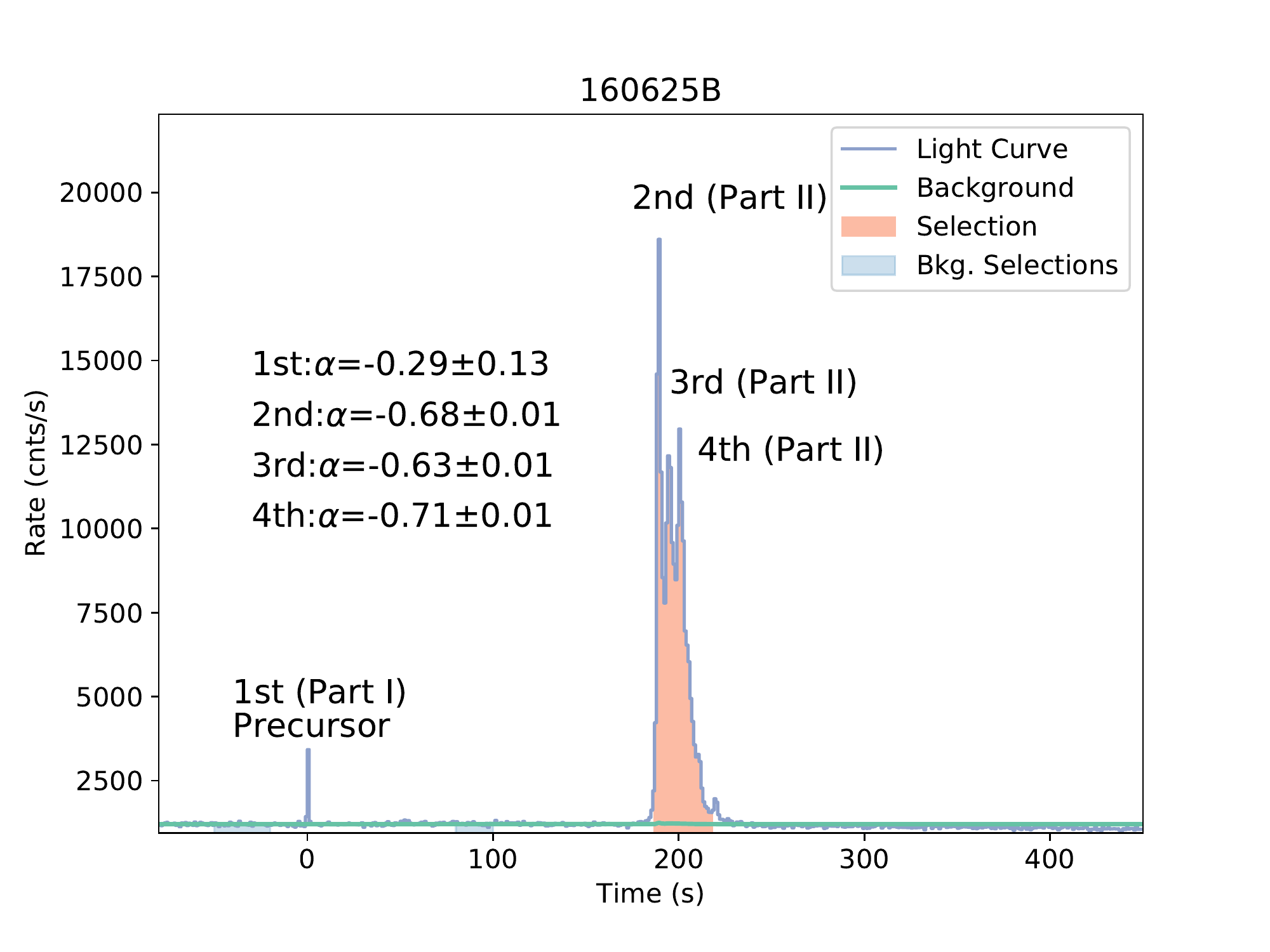}
\includegraphics[angle=0, scale=0.45]{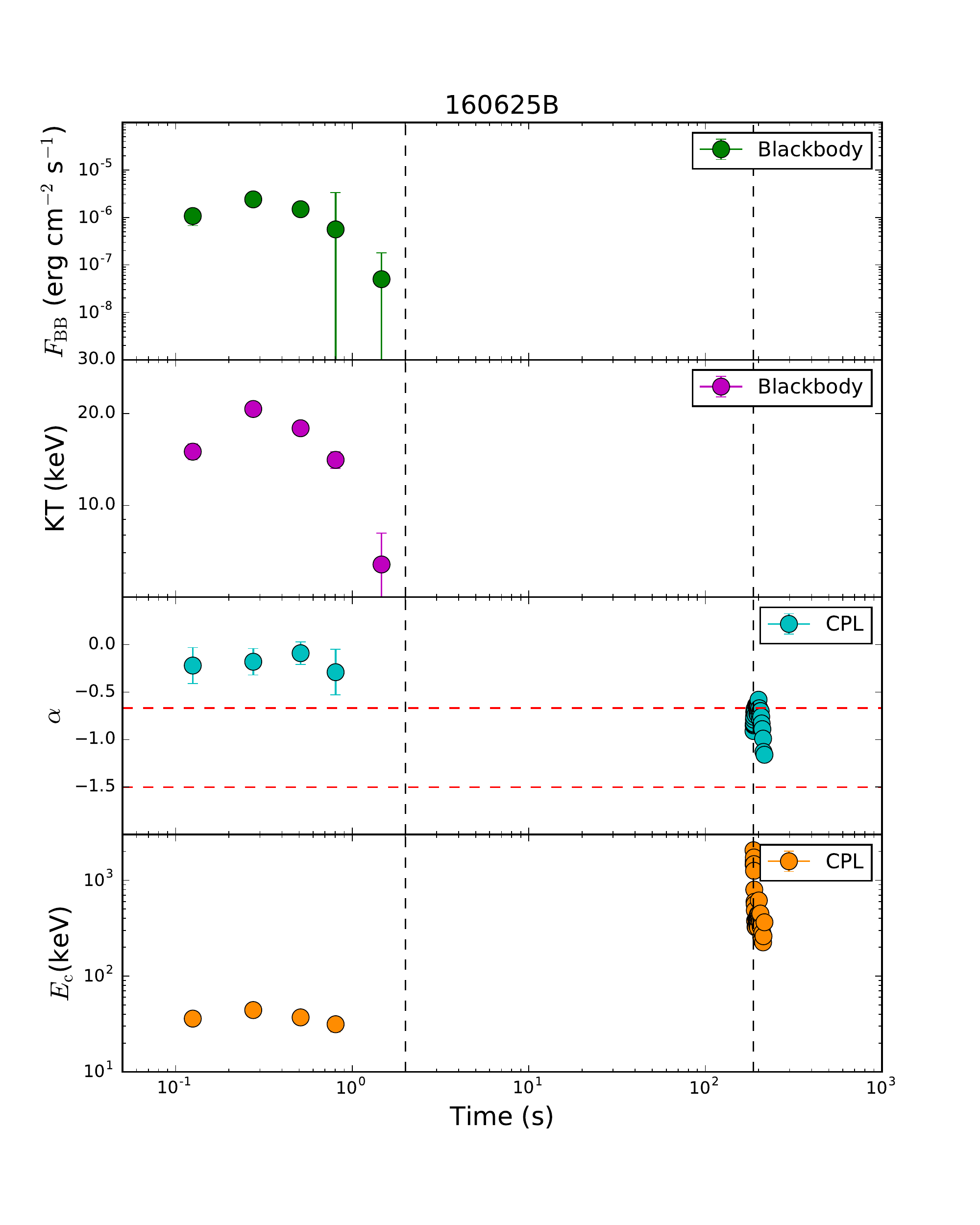}
\includegraphics[angle=0, scale=0.45]{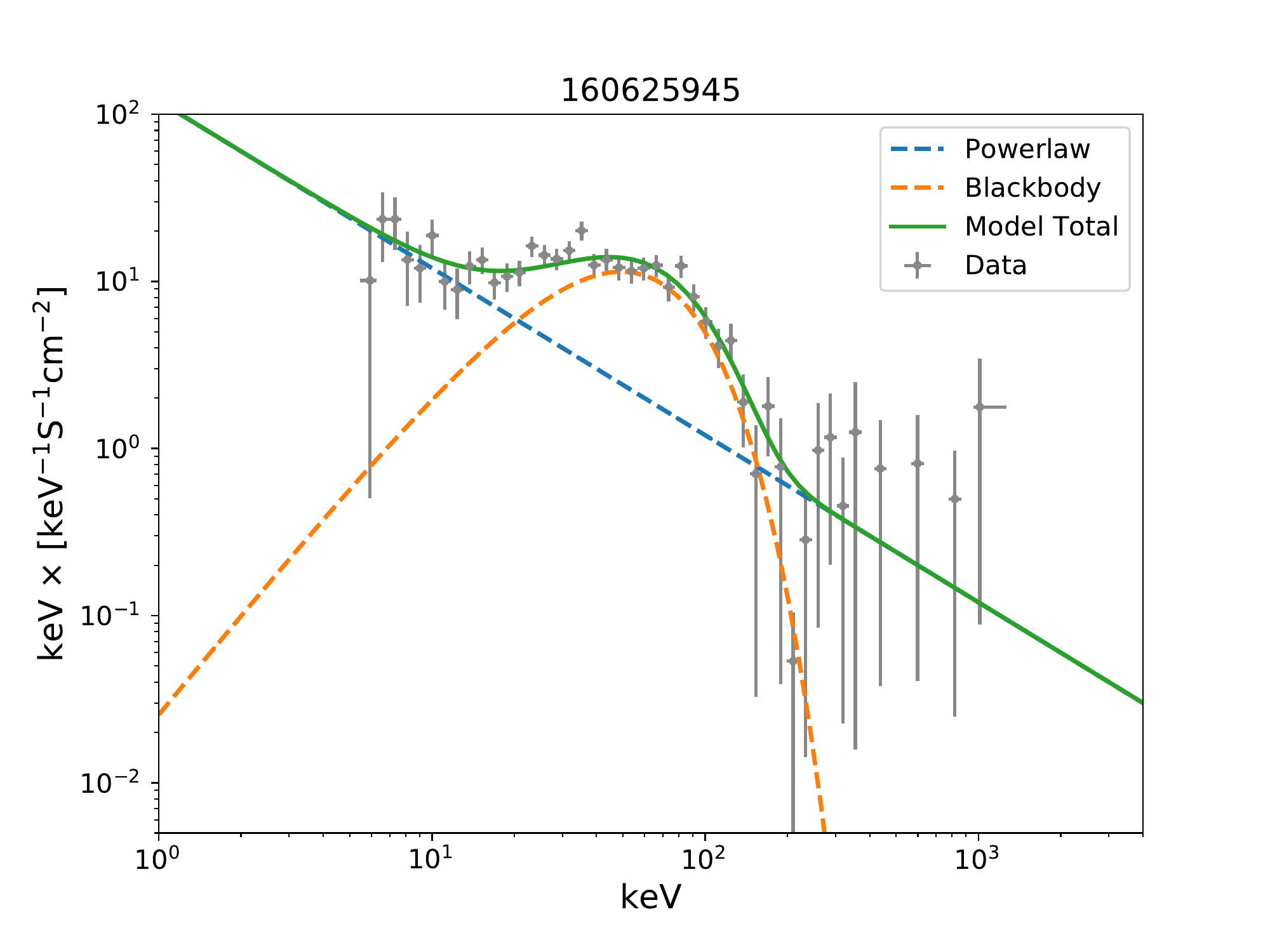}
\includegraphics[angle=0, scale=0.45]{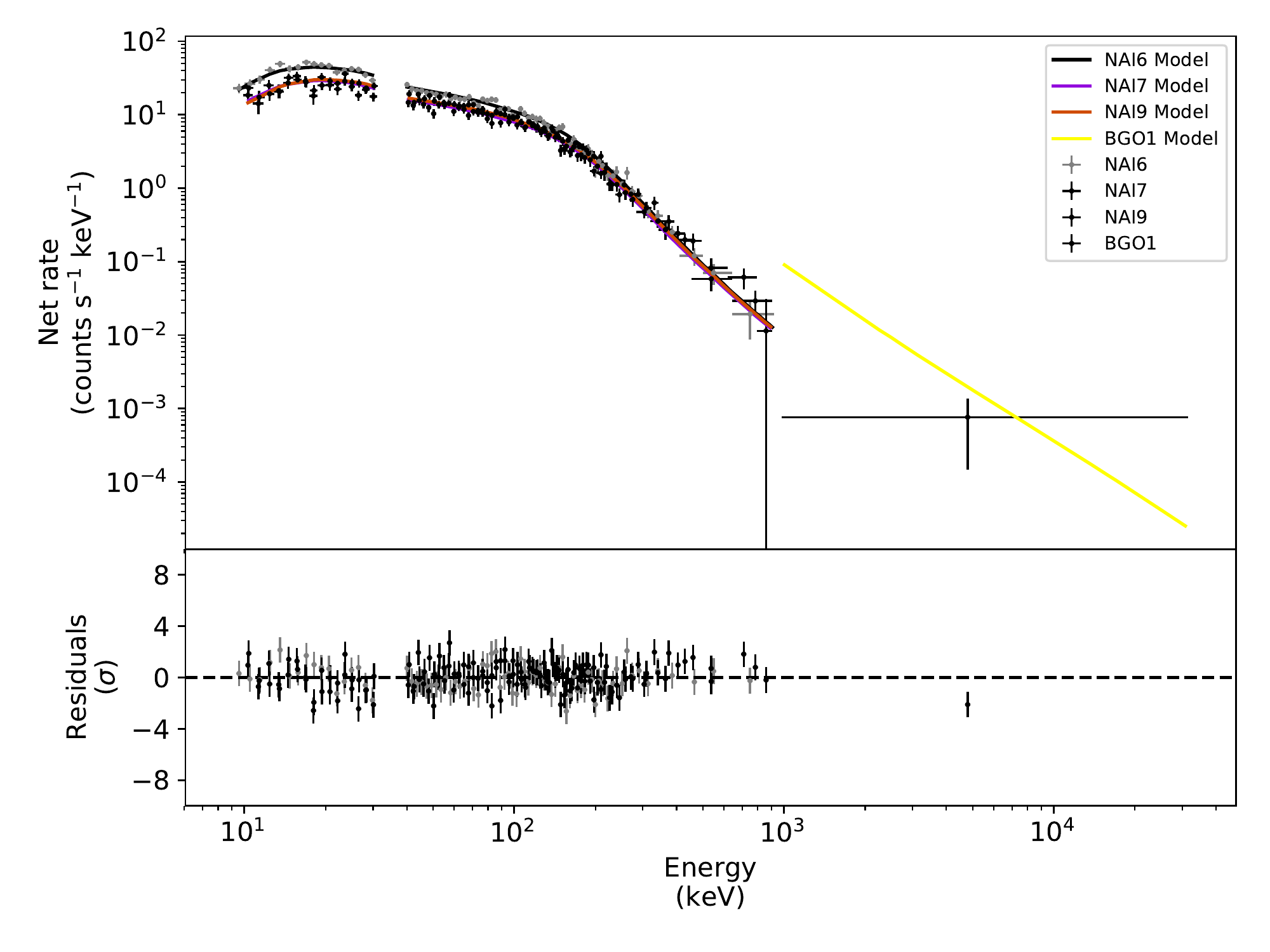}
\caption{Same as Figure \ref{GRB140206275} but for GRB 160625B. Time bins for Part I (0.34-0.68) and for Part II (208.65-211.64) are used.}\label{GRB160625945}
\end{figure*}

Figures (\ref{GRB140329295}-\ref{GRB160625945}) present time evolution of $\alpha$ (CPL), $E_{\rm c}$ (CPL), temperature (BB), and BB flux $F_{\rm BB}$ based on the time-resolved spectrum for the remaining sample (GRB 140329B, GRB 150330A, GRB 160625B), together with their light curves of prompt emission and typical spectral fits to each sub-burst. The evolution of the observed BB temperature $kT$ either a power-law (e.g. GRB 140206B and GRB 140329B) or a broken power-law decay (e.g. GRB 150330A and GRB 160625B).
Initially, the temperature is constant with time; later on, it presents a power-law decay after a break, which is consistent with the typical evolution characteristics of $kT$, found in \cite{2009ApJ...702.1211R}.
It is interesting to find that the temporal evolution of the observed BB flux $F_{\rm BB}$, and the BB temperature track each other for all four bursts in our sample\footnote{Note that the lower S/N values for some spectra in Part I would cause an unconstrained spectral fitting and an unrealistic evolution picture, such as time bins (0.47-0.67) and (0.67-3.00) for GRB 140329B and (0.93-2.00) for GRB 160625B. Therefore, we need to be cautious when giving the physical interpretation of the data.} \citep{1997ApJ...479L..39C, 2009ApJ...702.1211R}. 

After obtaining the observed BB temperature $kT$, the observed BB flux $F_{\rm BB}$, and the observed total flux $F_{\rm tot}$ (thermal + nonthermal) from the spectral fits, following \cite{2007ApJ...664L...1P}, we can derive the relevant photosphere properties (see \S 3 for a more detailed discussion): the isotropic equivalent luminosity of the thermal component $L_{\rm BB}$, the Lorentz factor of the bulk motion of the flow at the photospheric radius $\eta$, and the physical size at the base of the flow $r_{0}$. 

\subsection{Comparison of Spectral Characteristics with Different Scenarios} \label{sec:Comparison}

Before physics, the key point we need to address from data analysis is whether different phenomenological models (Band vs. CPL), types of spectrum (time-integrated vs. time-resolved), energy channels (e.g., one NaI-triggered detector vs. all), and fitting techniques (frequentist statistic by using the MLE vs. Bayesian statistic based on the MCMC) affect the results.

\begin{figure*}
\includegraphics[angle=0, scale=0.45]{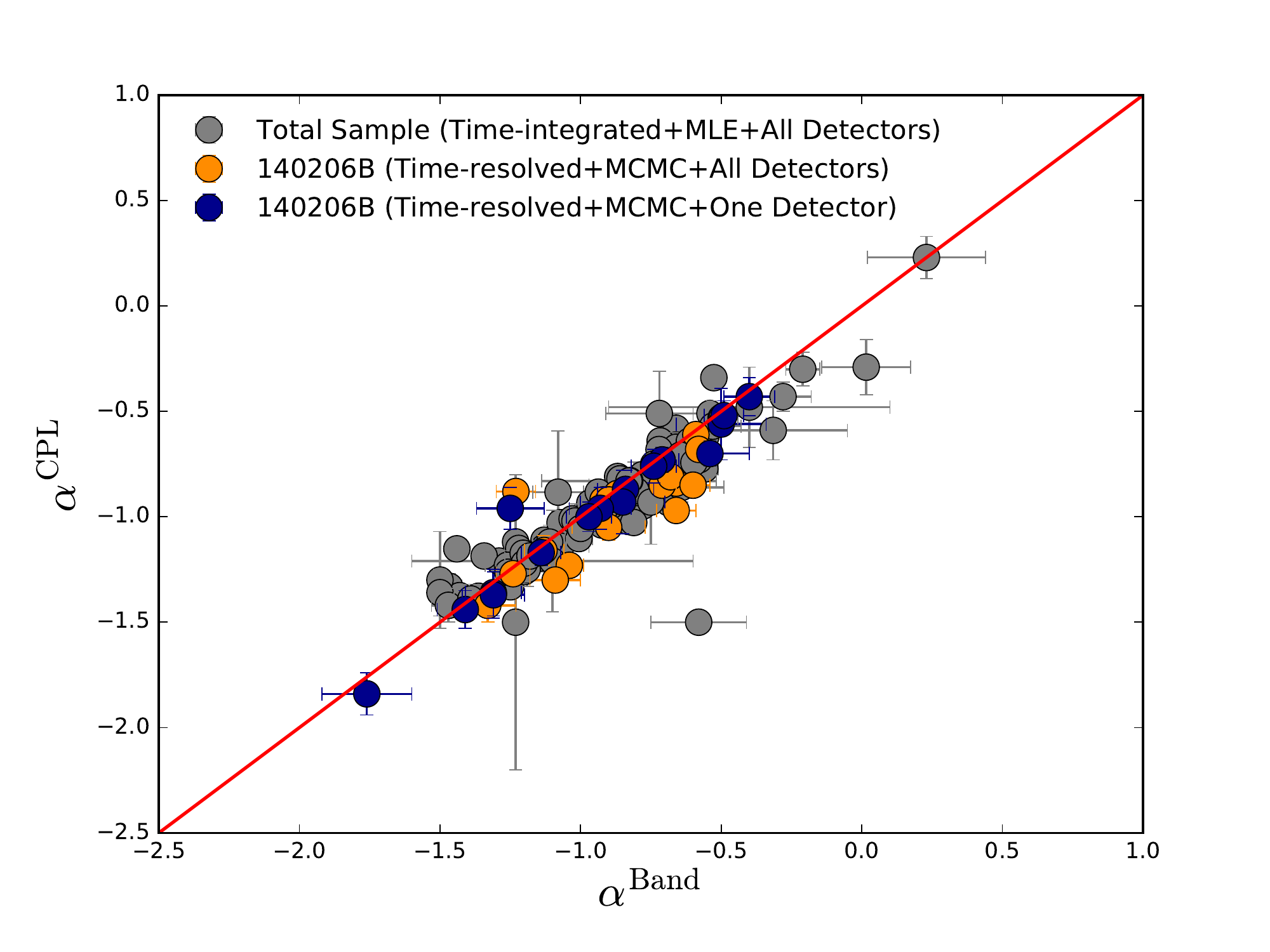}
\includegraphics[angle=0, scale=0.45]{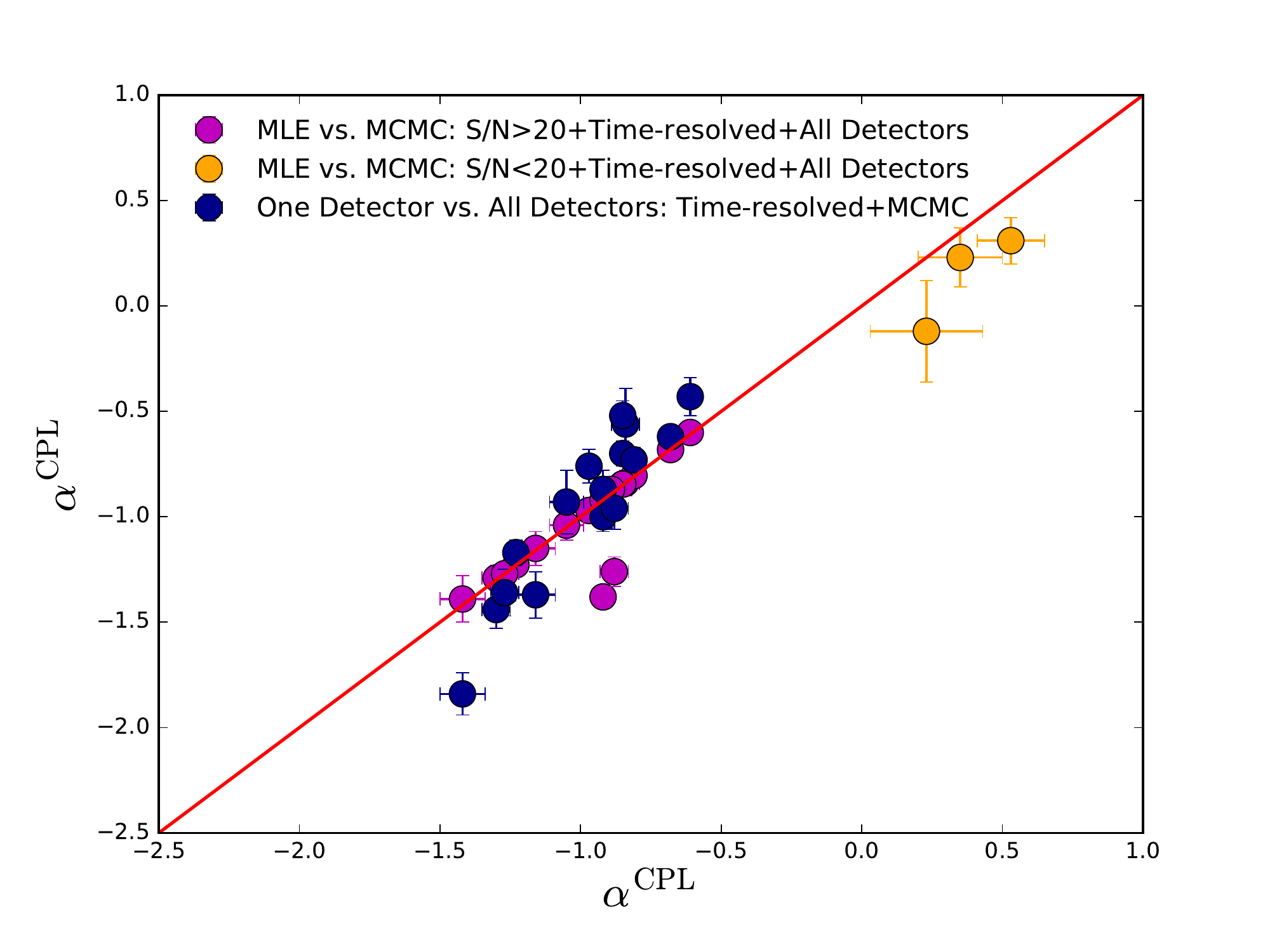}
\caption{Correlation analysis of $\alpha$ for different cases. The left panel investigates the difference of $\alpha$ between the Band and the CPL models with three different cases. 
The right panel compares $\alpha$ also with three different scenarios but based on the same CPL model.}\label{BandCPLRelation}
\end{figure*}

To account for this, we investigate the difference of $\alpha$ (or $E_{\rm pk(c)}$) values based on these different scenarios (Figure \ref{BandCPLRelation}). 
For different empirical model comparison (Band vs. CPL), we consider the investigation with the following three cases (Figure \ref{BandCPLRelation}a): 
\begin{itemize}
\item Case I: the time-integrated spectrum by using the total sample, and the MLE method (based on frequentist statistic), and all the triggered detectors.
\item Case II: the time-resolved spectrum by using one single GRB (140206B), and the MCMC technique (based on the Bayesian statistic), and all the triggered detectors.
\item Case III: the time-resolved spectrum by using one single GRB (140206B), and the MCMC technique (based on the Bayesian statistic), and one NaI-triggered detector.
\end{itemize}

We find that $\alpha$ values for the CPL model are slightly softer than the ones for the Band model (Case I), whereas the $E_{\rm pk(c)}$, in general, present the same. Equivalently, the same result is also found in their time-integrated spectrum (Case II). Note that our sample meets our selection criteria whether by the Band model or by the CPL model fitting.
For a more detailed discussion of a comparison with the Band function fits, see \cite{2018arXiv181007313Y}. 

Besides, we also check the consistency between different energy channels by using one triggered detector (Case III) to compare with all triggered detectors (Case II). One interesting finding is that the $\alpha$ values are more consistent with each other between these two models if only one NaI-triggered detector is used (marked with orange in Figure \ref{BandCPLRelation}a). The time-resolved spectral fitting results of GRB 140206B are reported in Table \ref{Table:140206B}.

For different fitting techniques (MLE based on the frequentist statistic vs. MCMC based on the Bayesian statistic), we compare the values of $\alpha$ based on the same CPL model with the following two groups of S/N range (Figure \ref{BandCPLRelation}b):
\begin{itemize}
\item Group I: the time-resolved spectrum by using one single GRB (140206B) with S/N$\geq$20 and using all the triggered detectors.
\item Group II: the time-resolved spectrum by using one single GRB (140206B) with S/N$<$20 and using all the triggered detectors.
\end{itemize}

We find that a majority of data points are consistent with each other (tightly distributed around the equal line) for S/N$\geq$20 (Group I), and only two data points are exceptions.
For S/N$<$20 (Group II), we find that the data points present significant deviation, and with a larger error. This is because lower S/N values cannot ensure that there are enough photons to conduct an accurate spectral fitting. This is why we apply for the criterion of S/N$\geq$20.
 
Likewise, based on the same CPL model, we also compare $\alpha$ values between different energy channels (using one NaI detector versus all, marked with blue in Figure \ref{BandCPLRelation}b); this exhibits an apparent nonmonotonous relation. This implies that the result could be prominently affected by the selection effect.

\begin{figure*}
\includegraphics[angle=0, scale=0.5]{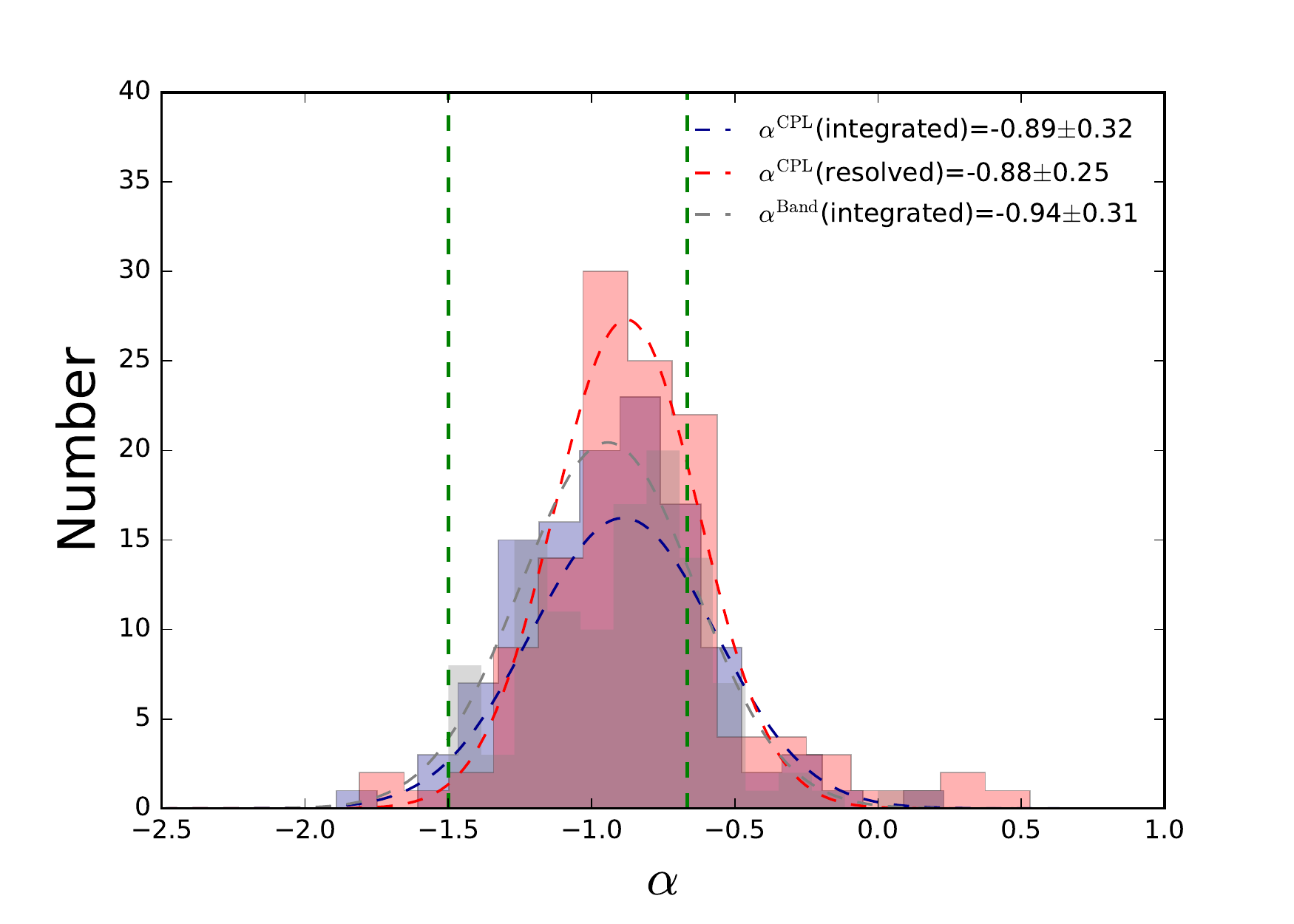}
\includegraphics[angle=0, scale=0.5]{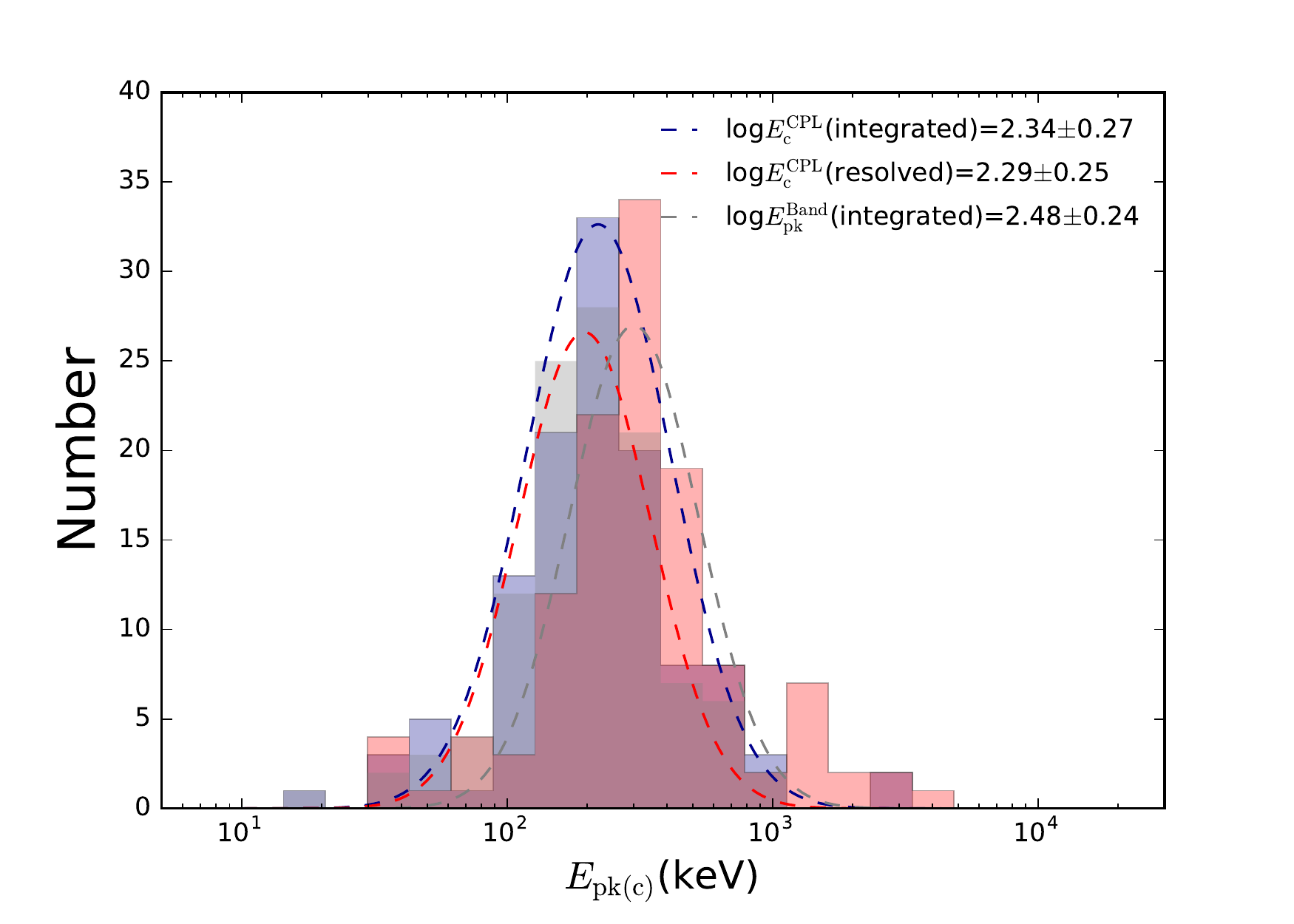}
\caption{Comparisons of $\alpha$-distributions (left panel) and $E_{\rm pk(c)}$-distributions (right panel) in different scenarios (Band or CPL, time-resolved or time-integrated). 
Based on the CPL model and the time-integrated spectral fit results, the best Gaussian fits give $\alpha^{\rm CPL}$=0.89$\pm$0.32 and log$E^{\rm CPL}_{\rm c}$=2.34$\pm$0.27; based on the CPL model and the time-resolved spectral fit results, we have $\alpha^{\rm CPL}$=0.88$\pm$0.25 and log$E^{\rm CPL}_{\rm c}$=2.29$\pm$0.25; based on the Band model and the time-integrated spectral fit results, we have $\alpha^{\rm Band}$=0.94$\pm$0.31 and log$E^{\rm Band}_{\rm pk}$=2.48$\pm$0.24. Two vertical green dashed lines represent the limiting values of $\alpha$=-2/3 and $\alpha$=-3/2 for electrons in the slow- and fast-cooling regimes, respectively.}\label{alphaEpDis}
\end{figure*}

On the other hand, we also compare the distributions of $\alpha$ \cite[e.g.,][]{2013ApJ...764...75G}, and $E_{\rm pk(c)}$ \cite[e.g.,][]{2003ApJ...583L..71S, 2016ApJ...821...12P} between the Band and the CPL models and between the time-integrated and the time-resolved spectral fitting results, as presented in Figure \ref{alphaEpDis}.
They are all well fitted with Gaussian distributions for each sample. 
For $\alpha$-distribution by the CPL model, we have $\alpha$=-0.89$\pm$0.32 based on the time-integrated spectral fits and $\alpha$=-0.88$\pm$0.25 based on the time-resolved spectral fits.
For $\alpha$-distribution but by the Band model, we get $\alpha$=-0.94$\pm$0.31 based on the time-integrated spectral fits (Fig.\ref{alphaEpDis}a).
Similarly, by the same analysis for $E_{\rm c}$-distribution, we find log$E_{\rm c}$=2.34$\pm$0.27 based on the time-integrated spectral fits by the CPL model, and log$E_{\rm c}$=2.29$\pm$0.25 based on the time-resolved spectral fits also by the CPL model; and log$E_{\rm pk}$=2.48$\pm$0.25 based on the time-integrated spectral fits but by the Band model (Fig.\ref{alphaEpDis}b).

The results indicate that, compared with the distributions based on the time-integrated spectral fits between the Band and the CPL models, the distributions based on the CPL model between the time-integrated and the time-resolved spectral fits are more consistent with each other, both in the $E_{\rm pk(c)}$-distribution and in the $\alpha$-distribution.

\section{Physical Scenario}

There are two main mechanisms for the acceleration of a GRB jet: thermally driven or magnetically driven. The former one is relevant for a hot fireball and proceeds very rapidly, whereas the latter is relevant for a Poynting-flux-dominated outflow and proceeds relatively more slowly \citep{2015ApJ...801..103G}.
In this section, we constrain the outflow properties for those thermal pulses (Part I) to check the photosphere emission origin and confirm the optically thin synchrotron emission for the later pulses (Part II) based on some empirical relations.

\subsection{Photosphere Emission Component and Determination of the Outflow Properties}

In the early picture of the fireball model, the fireball is composed of the thermal photon and electron-positron pair \citep{1986ApJ...308L..47G, 1986ApJ...308L..43P}. As the fireball expands and the optical depth goes below unity (at the photosphere radius $r_{\rm ph}$), the thermal energy of the photon will be emitted as BB emission and thus produce the photosphere emission. 
Later, it is found that a small amount of the baryon should be included in the fireball, namely, a baryonic fireball rather than the pure radiative fireball \citep{1990ApJ...365L..55S}. 
Since the photon luminosity of the fireball is much larger than the Eddington luminosity, in other words, the radiation pressure exceeds self-gravity, the fireball must expand \citep{1993ApJ...405..278M, 1993MNRAS.263..861P}. As the fireball expands rapidly, the thermal energy of photons will be converted into the kinetic energy of the baryons from an initial nozzle radius $r_{0}$.  
According to the conservation of energy and entropy, the bulk Lorentz factor of the outflow increases with radius as $\Gamma \propto r$ while the comoving temperature of BB decreases as $T^{'}\propto r^{-1}$, before the saturation radius $r_{\rm s}=\eta r_{0}$, where the Lorentz factor reaches the maximum value $\eta \equiv L_{w}/\dot{M}c^{2}$. Here $L_{w}$ is the isotropic equivalent luminosity, $\dot{M}$ is the ejection mass, and $c$ is the speed of light. Above the saturation radius $r_{s}$, the Lorentz factor keeps being constant.

The dissipation mechanism, e.g., internal shocks \citep{1994ApJ...427..708P, 1994ApJ...430L..93R, 2017ApJ...844...56T}, magnetic reconnection \citep{2005A&A...430....1G, 2006A&A...457..763G}, or external shocks \citep{1993ApJ...405..278M, 1999ApJ...513L...5D} for the nonthermal emission is still uncertain. 
However, it is clear that the thermal component originates from the photosphere. Therefore, once identifying a BB component in the observed spectrum, as following we can derive the physical parameters of the outflow as follows (considering the case without subphotospheric dissipation): $r_{0}$, $r_{\rm s}$, $\eta$, and $r_{\rm ph}$.

\cite{2007ApJ...664L...1P} developed a method to determine the initial size of $r_{0}$, and Lorentz factor $\Gamma$ of GRB fireballs using a thermal emission component. The method in \cite{2007ApJ...664L...1P} can be applied for the pure fireball model (thermal $\gamma$ photons+e$^{\pm}$ pair).
Following \cite{2007ApJ...664L...1P}, once we know all the three observed quantities, the observed BB temperature $T_{\rm obs}$, the observed BB flux $F_{\rm BB}$, and the observed total flux $F_{\rm tot}$ (thermal+nonthermal), we can infer the values of the isotropic equivalent luminosity of the thermal component $L_{\rm BB}$, the Lorentz factor of the bulk motion of the flow at the photospheric radius $\eta$, and the physical size at the base of the flow $r_{0}$. This is because in the pure fireball model there are three unknowns, $L_{\rm BB}$, $\eta$, and $r_{0}$.
For the bursts with known redshift and measured thermal flux, $L_{\rm BB}$ can be directly measured. The calculations of $\eta$ and $r_{0}$ can follow Equation (4) and Equation (5) in \cite{2007ApJ...664L...1P} if we know all these three observed quantities ($F_{\rm BB}$, $T_{\rm obs}$ and $F_{\rm tot}$).

\subsubsection{Flux Ratio, $F_{\rm BB}/F_{\rm tot}$}

The observed BB flux, $F_{\rm BB}$, is calculated by integrating the intensity over the emitting surface
\begin{equation}
F_{\rm BB}=\frac{2\pi}{d^{2}_{\rm L}}\int d\mu \mu r^{2}_{\rm ph} D^{4} (\sigma T^{'4}/\pi),
\end{equation}
where $T^{'}$ is the comoving temperature at the photospheric radius; $D=D(\theta)$ is the Doppler factor, $D=(\Gamma(1-\beta \mu))^{-1}$; and so the observed BB temperature is $T=DT^{'}=(\Gamma(1-\beta \mu))^{-1}T^{'}$, where $\theta$ is the angle to the line of sight, $\mu\equiv cos \theta$, and $\Gamma=(1-\beta^{2})^{-1/2}$ is the outflow Lorentz factor.

The observed ratio of $F_{\rm BB}/F_{\rm tot}$ is shown in the left hand panel of Figure \ref{RatioReTime} for bursts with the best models of PL+BB or CPL+BB, where $F_{\rm tot}$ is the total flux.
The evolution properties for this ratio vary from burst to burst.
GRB 140329B nearly displays a constant ratio throughout the first sub-pulse with a thermal flux of about 30 per cent, while GRB 150330A show a single power-law decay with a thermal flux initially of about 80\% and finally decreasing to about 10\%. Interestingly, the thermal flux of GRB 160625B initially increases from about 40\% and it peaks at about 50\% at around 0.3 s, and finally decreasing to about 10\%. The case is similar to GRB 110721A \citep{2013MNRAS.433.2739I}.

\subsubsection{Parameter, $\Re$}
Under the spherical symmetry \citep{2007RSPTA.365.1171P}, the ratio between the observed quantity $F_{\rm BB}$ and $T$, which is denoted as $\Re$, can be measured by (for $r_{\rm ph}>r_{s}$)
\begin{equation}
\Re=\left(\frac{F_{\rm BB}}{\sigma_{\rm SB} T^{4}}\right)^{1/2}=\xi \frac{(1+z)^{2}}{d_{\rm L}} \frac{r_{\rm ph}}{\Gamma},
\end{equation}
where $\sigma_{\rm SB}$ is the Stefan-Boltzmann constant, and $\xi$ is a numerical factor of the order of unity that can be obtained from angular integration, $z$ is the redshift, and $D_{L}$ is the luminosity distance.
The $\xi$ value is adopted as 1.06 in \cite{2007ApJ...664L...1P}.
The observed BB normalization, $A(t)$, is related to $\Re$, $\Re=2\pi c\hbar^{3/2}A(t)^{1/2}$, where $\hbar$ is the reduced Planck constant. 

\begin{figure*}
\includegraphics[angle=0, scale=0.45]{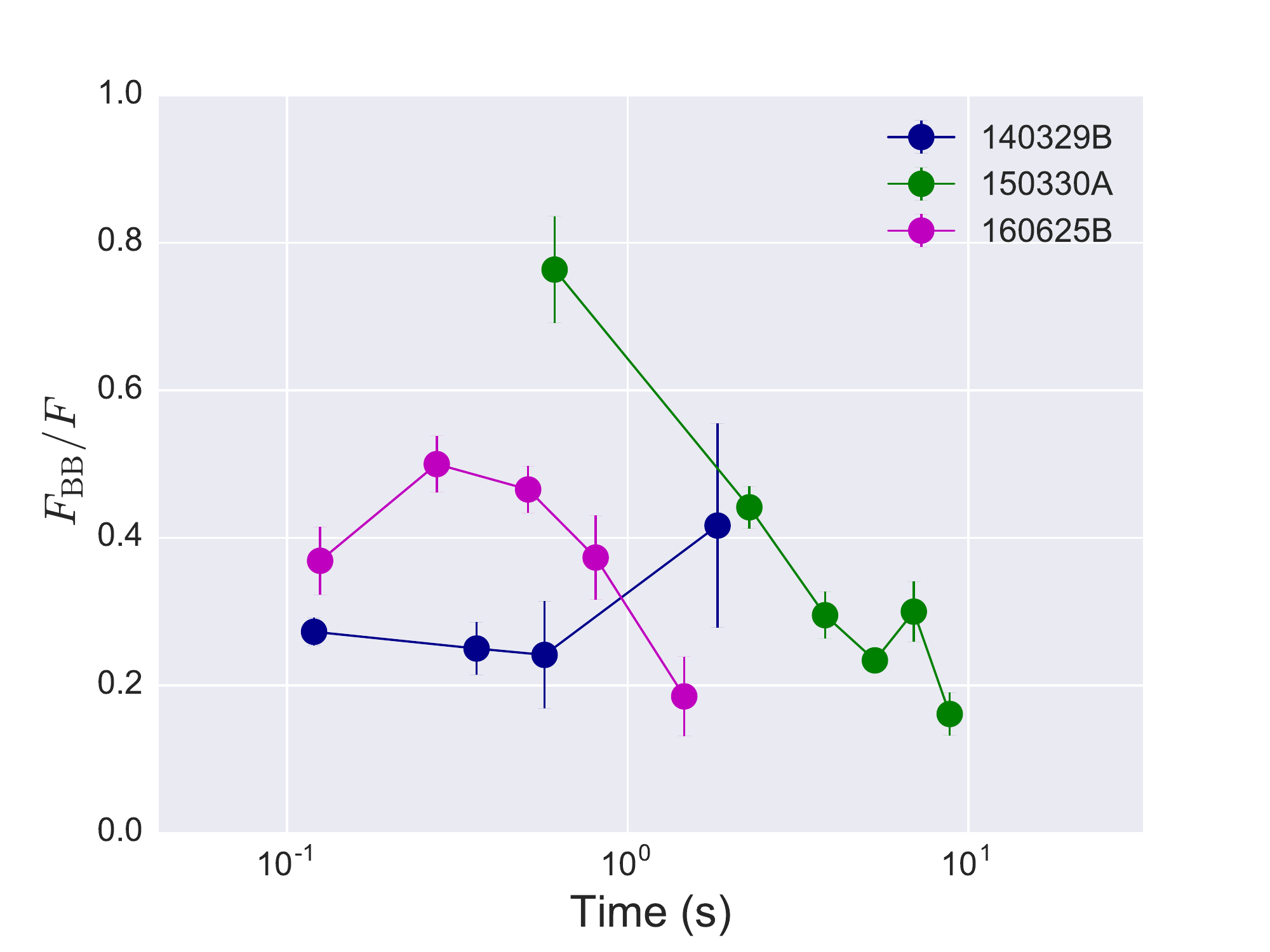}
\includegraphics[angle=0, scale=0.45]{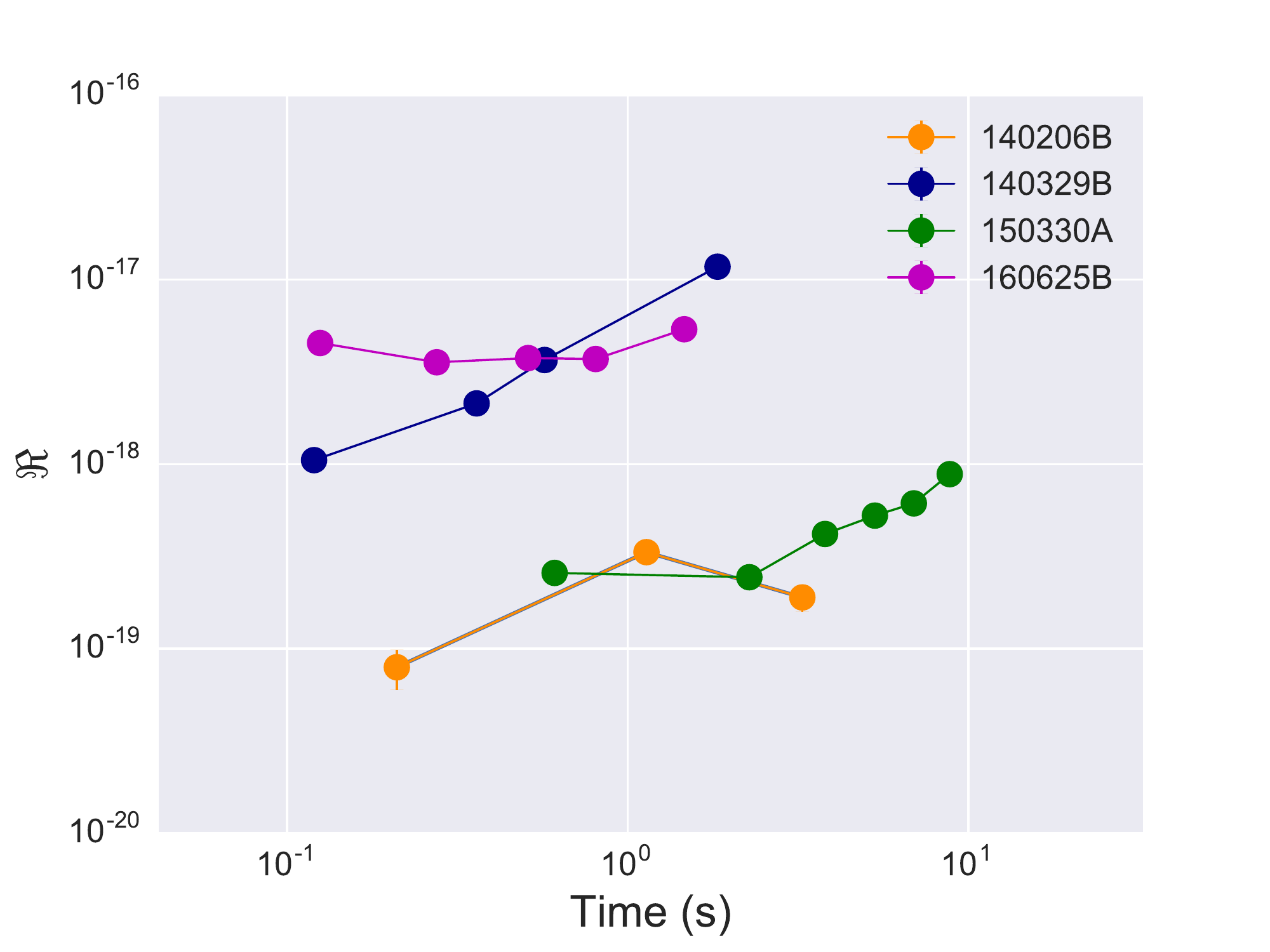}
\caption{Left panel: evolution of the fraction of thermal flux to total flux, $F_{\rm BB}/F$. Right panel: evolution of the parameter $\Re=(F_{\rm BB}/\sigma T^{4})^{1/2}$. Different colors represent in each individual burst: GRB 140206B (orange), GRB 140329B (blue), GRB 150330A (green), GRB 160625B (pink).}
\label{RatioReTime}
\end{figure*}

For bursts with known redshifts, the parameters $\Re$ can be interpreted as an effective transverse size of the emitting region \citep{2009ApJ...702.1211R}. Therefore, a constant $\Re$ means that the effective emitting area of the photosphere is time independent. This is the case for GRB 160625B as presented in the right panel of Figure \ref{RatioReTime}. 
Moreover, $\Re$ typically is observed to increase over a pulse \citep{2009ApJ...702.1211R}, this is the case for most bursts in our sample, with two prominent cases of GRB 140329B and GRB 150330A.

To derive the parameters in the rest frame, the selected bursts should have known redshift. 
A redshift measurement of 2.73 is found for GRB 140206B \citep{2014GCN.15800....1M}, and 1.406 is for GRB 160625B \citep{2018NatAs...2...69Z}.
However, GRB 140329B and GRB 150330A are without known redshift. On the other hand, \cite{2013MNRAS.433.2739I} investigated how the different values of redshift affect the results and pointed out that, for different values of $z$, the estimated outflow parameters change within a factor, but the time evolution of the behavior of the parameters remains the same. We therefore assume an average value of GRBs \citep{2006A&A...453..797B}, {\it z}=2, for the calculation (see also in \citealt{2013MNRAS.433.2739I}).

\subsubsection{Lorentz Factor, $\Gamma$}\label{Sec:Gamma}

The coasting values of the Lorentz factor ($r_{\rm ph}>r_{\rm s}$) are derived by (see also Equation(4) in \citealt{2007ApJ...664L...1P})
\begin{equation}
\Gamma \propto (F/\Re)^{1/4} Y^{1/4},
\label{Eq:Gamma}
\end{equation}
where $Y$ is the ratio of total fireball energy to the energy emitted in gamma-rays.

\begin{figure*}
\centering
\includegraphics[angle=0, scale=0.50]{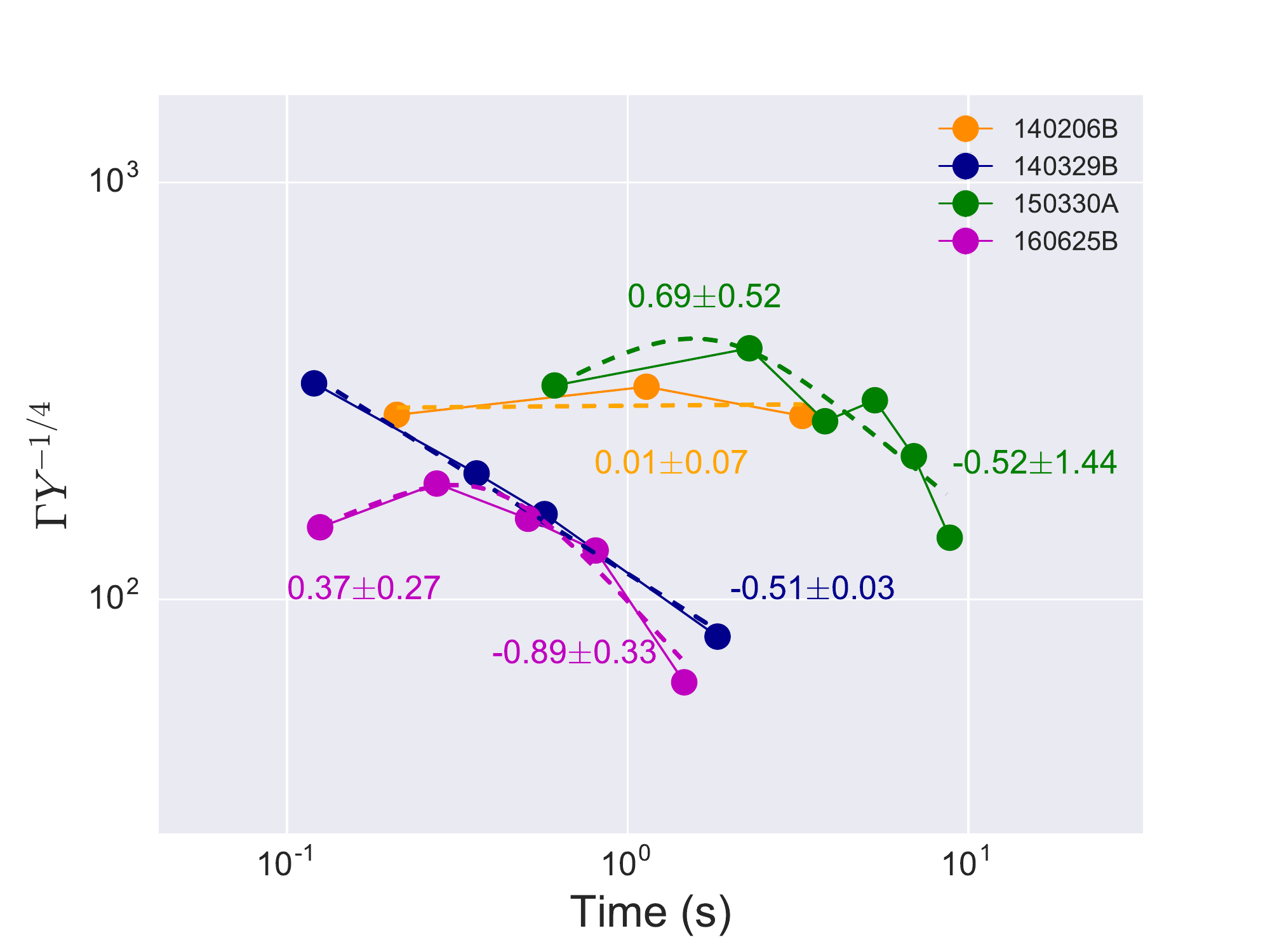}
\caption{Temporal evolution of the Lorentz factor, $\Gamma$. Color conventions are the same as in Figure \ref{RatioReTime}. GRB 140206B presents a constant behavior with best-fitting power-law index 0.01$\pm$0.07, while GRB 140329B shows monotonic decay with the index -0.51$\pm$0.03. GRB 150330A and GRB 160625B exhibit a smoothly broken power-law decay feature, with the power-law indices 0.69$\pm$0.52 and -0.52$\pm$1.44 before and after break for GRB 150330A; and 0.37$\pm$0.27 and -0.89$\pm$0.33 before and after break for GRB 160625B.}\label{GammaTime}
\end{figure*}

As depicted in Figure \ref{GammaTime}, temporal evolution of the Lorentz factor either an approximately constant (e.g. GRB 140206B) or a monotonic decay behavior (e.g. GRB 140329B exhibits a monotonic decline throughout the pulse), or both (initially approximately constant and then decaying faster, e.g. GRB 150330A, GRB 160625B). 
The best fitting gives the power-law indices of 0.01$\pm$0.07 for GRB 140206B and -0.51$\pm$0.03 for GRB 140329B.
We fit GRB 150330A and GRB 160625B with a smoothly broken power-law function, which gives the best fitting power-law indices of 0.69$\pm$0.52 (before the break) and -0.52$\pm$1.44 (after the break) for GRB150330A, and 0.37$\pm$0.27 (before the break) and -0.89$\pm$0.33 (after the break) for GRB 160625B.

The global view on the evolution of the Lorentz factor decays with time from a few hundred $Y^{1/4}$ down to below 100 $Y^{1/4}$ \citep[e.g.,][]{2011ApJ...739...47F, 2014ApJ...795..155P}. The decreasing Lorentz factor may have several implications \citep{2013MNRAS.433.2739I}.
Since $\Re$ typically is observed to increase over a pulse, the decrease of the values of $\Gamma=\Gamma(t)$ is not surprising by realizing the fact that $\Re\sim L_{0}/\Gamma^{4}$; such a decrease must, therefore, be a common characteristic form over individual pulse structures in GRBs \citep{2009ApJ...702.1211R}. 
This is because, during the rising phase of the pulse, both the total flux and $\Re$ increase with nearly an approximate rate; the Lorentz factor thus is close to a constant or shows only a moderate decrease with time. However, $\Gamma$ decays much faster during the decay phase since $\Re$ continues to increase whereas the flux, instead, decreases \citep{2016MNRAS.456.2157I}.

\subsubsection{Nozzle Radius, $r_{0}$}

After obtaining $\Re$, the calculation of $r_{0}$ for the case of $r_{\rm ph}>r_{\rm s}$ is given by (see also in Equation (5) in \citealt{2007ApJ...664L...1P}),
\begin{equation}
r_{0} \propto (F_{\rm BB}/F Y)^{3/2} \Re.
\label{Eq:r0}
\end{equation}
Figure \ref{RTime} presents the evolution of the radii ($r_{\rm 0}$, $r_{\rm s}$, $r_{\rm ph}$). 
The nozzle radius, $r_{0}$, presents a moderate decrease with time for GRB 140206B, GRB 150330A, and GRB 160625B, with a size of the order of from 10$^{8}$ to 10$^{9}$ cm. Interestingly, the $r_{0}$ of GRB 140329B increases by nearly two orders of magnitude throughout the first pulse.  

\subsubsection{Saturation Radius, $r_{\rm s}$}

Once we know $r_{0}$, we can get an estimate of the saturation radius, $r_{\rm s}$, which is given by 
\begin{equation}
r_{s} =\Gamma r_{0}.
\end{equation}
We find that $r_{\rm ph}$ is above $r_{\rm s}$ in the bursts of our sample, with a typical size of $\sim$10$^{11}$ cm (see Figure \ref{RTime}).

\subsubsection{Photospheric Radius, $r_{\rm ph}$}

Considering a relativistic bulk motion, for a photon propagating a distance $ds$, the optical depth, $\tau$, is given by $\tau= \int^{\infty}_{r_{\rm ph}} \frac{n \sigma_{\rm T}}{2\Gamma^{2}}dr$, where $\sigma_{\rm T}$ is the Thomson cross section, $n$ is the electron number density, and $ds=(1-\beta \rm cos \theta) dr/cos \theta$, with $\theta$ the angle from the line of sight, and considering the case of electron motion in the direction of photon, $\theta=0$.
Assuming that the Lorentz factor is constant (i.e. in the coast phase: $r>r_{\rm s}$, $\Gamma\equiv\eta$), the optical depth can be integrated by covering the distance $\tau= \frac{L_{0}\sigma_{T}}{8 \pi m_{\rm p} c^{3} \Gamma^{2} \eta}\frac{1}{r}$; the photosphere radius ($\tau=1$) therefore can be calculated by
\begin{equation}
r_{\rm ph}=\frac{L_{0}\sigma_{T}}{8 \pi m_{\rm p} c^{3} \Gamma^{3}_{\rm ph} },
\end{equation}
where $L_{0}$ is the burst luminosity, which is given by
$L_{0}=4\pi d^{2}_{L} Y F_{\rm tot}$, where $d_{L}$ is the luminosity distance and $F_{\rm tot}$ is the total observed $\gamma-$ray flux (thermal+nonthermal).

\begin{figure*}
\includegraphics[angle=0, scale=0.45]{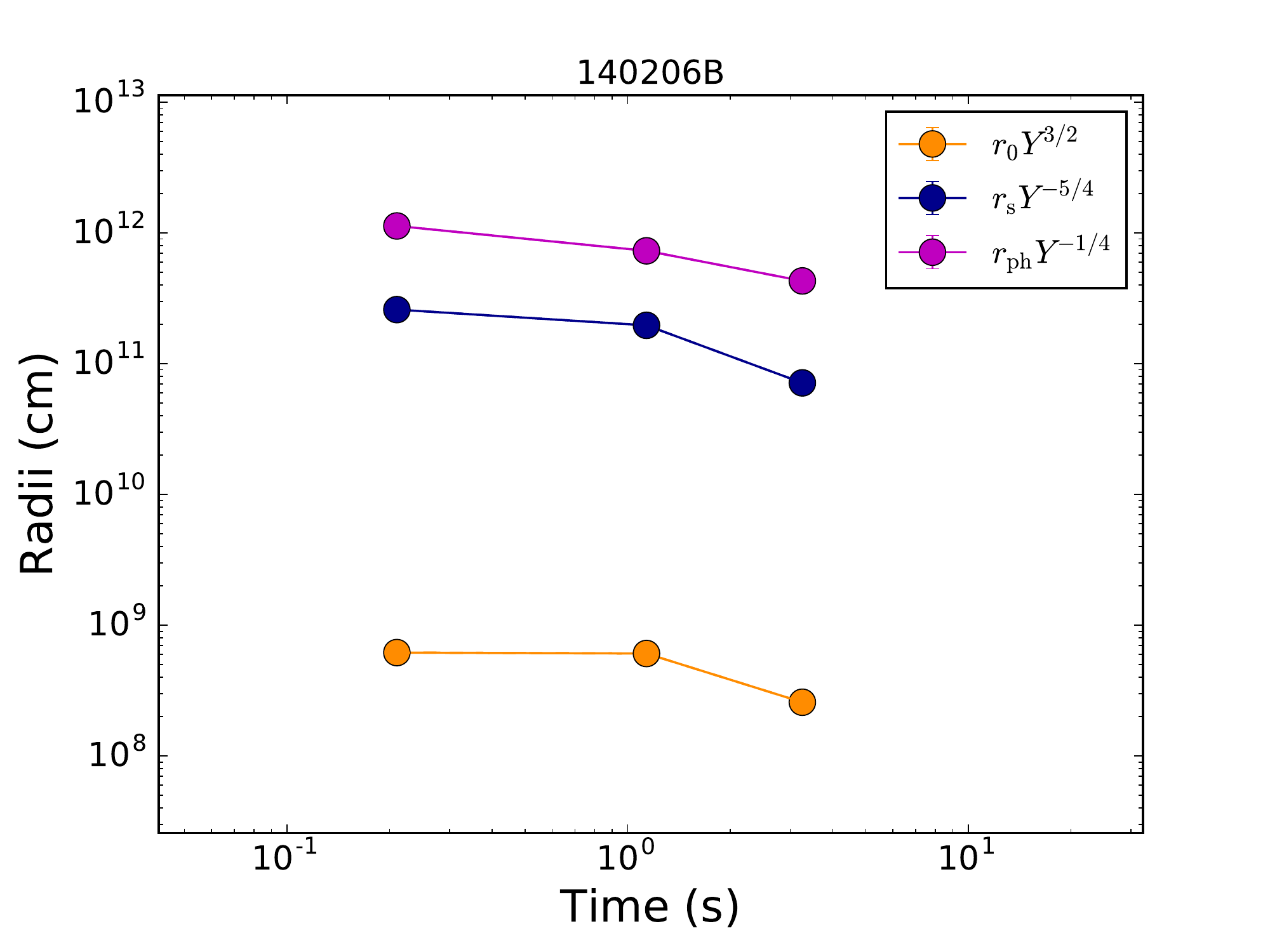}
\includegraphics[angle=0, scale=0.45]{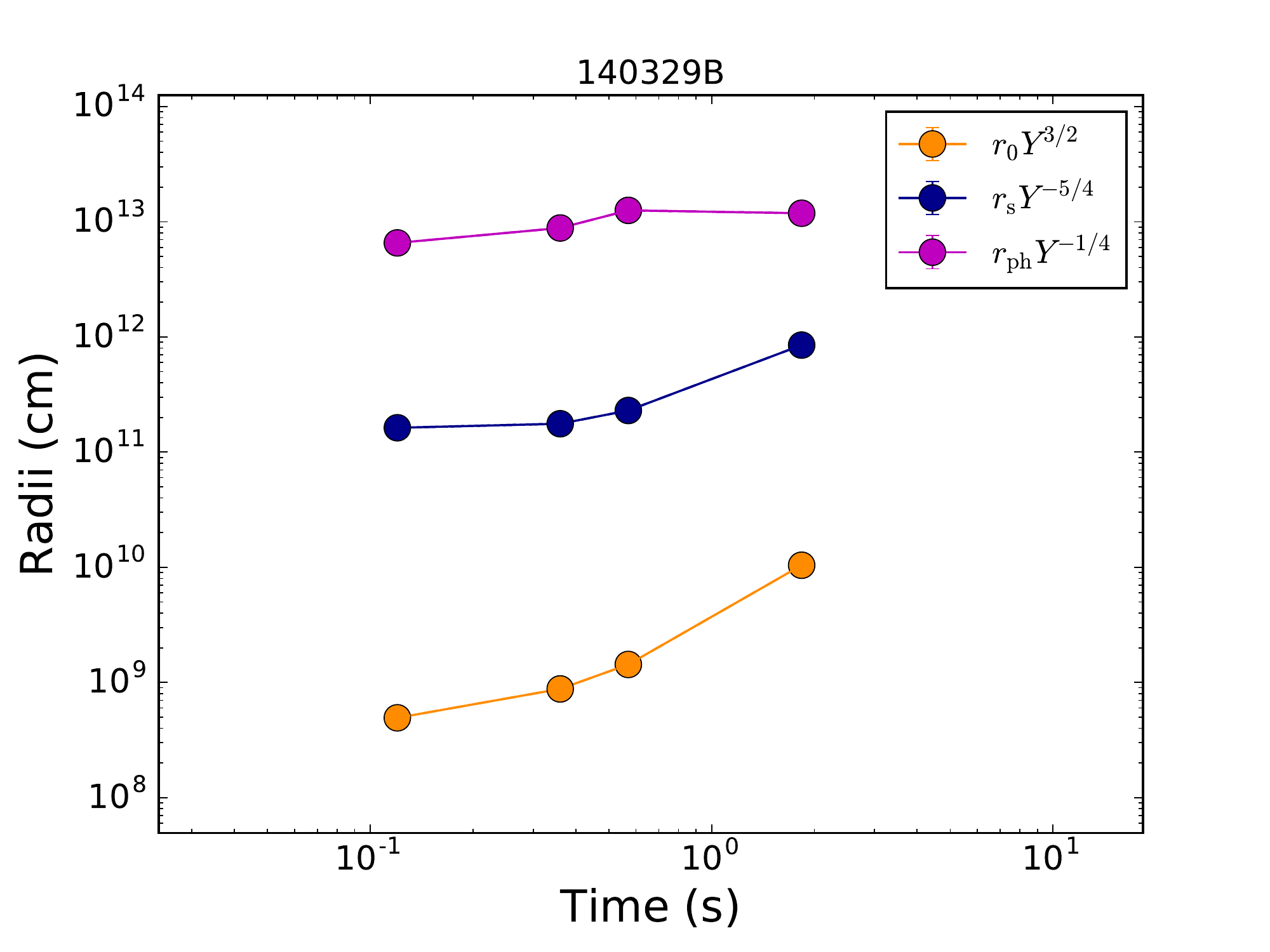}
\includegraphics[angle=0, scale=0.45]{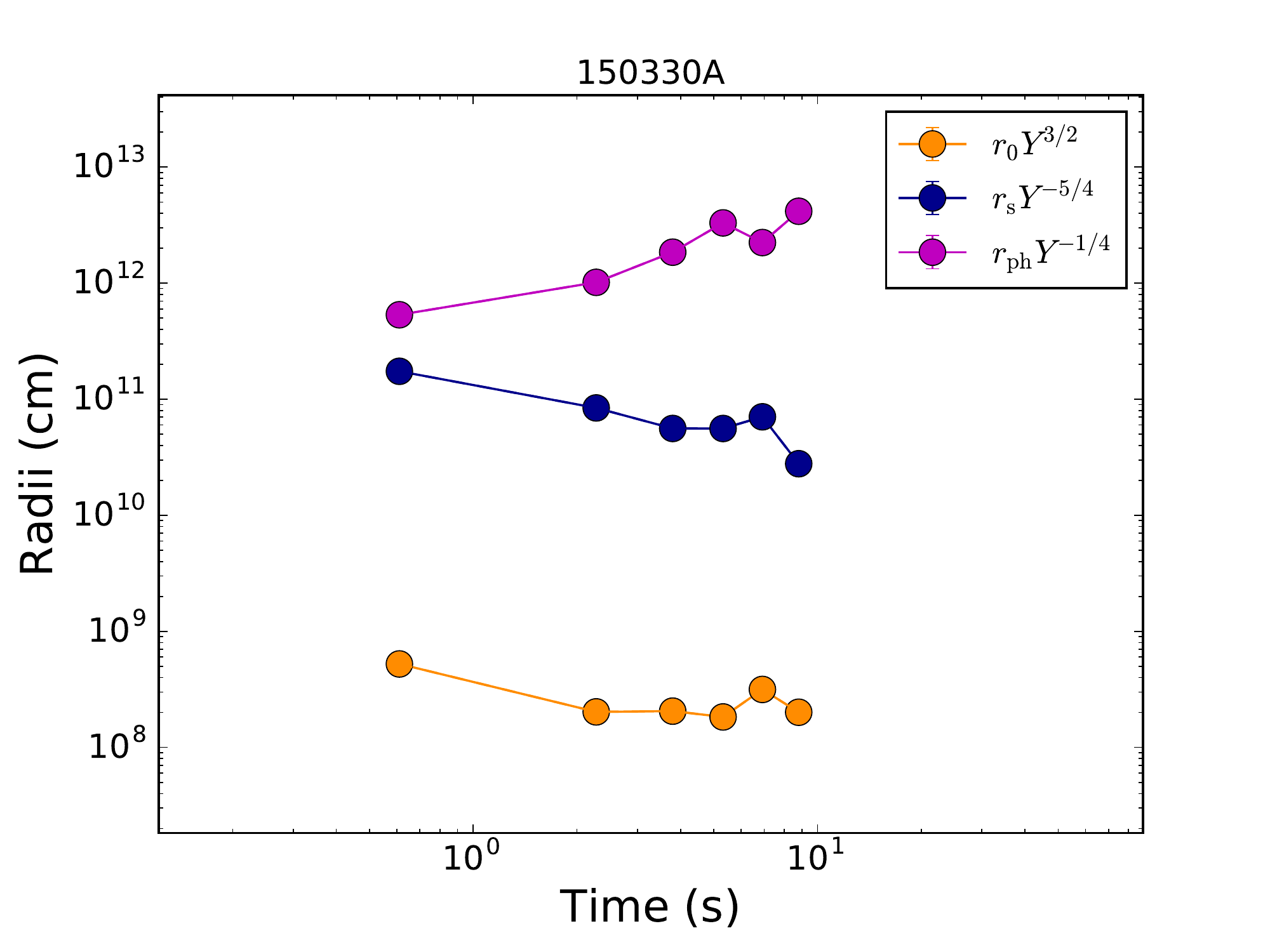}
\includegraphics[angle=0, scale=0.45]{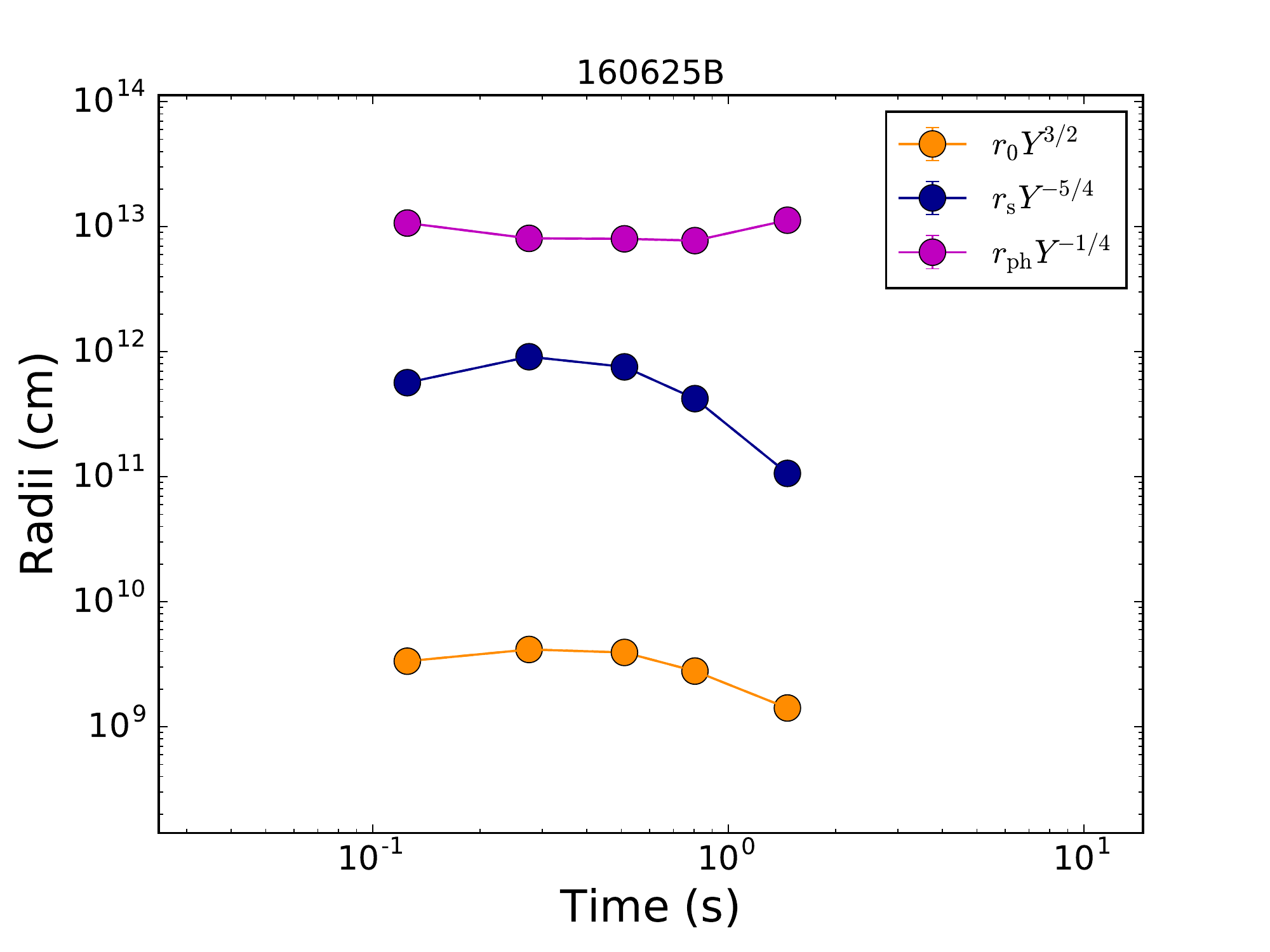}
\caption{Temporal evolution of the photospheric radius $r_{\rm ph}$, saturation radius $r_{\rm s}$, and nozzle radius $r_{\rm 0}$. Different colors represent in different characteristic radii: $r_{\rm 0}$ (orange), $r_{\rm s}$ (blue), $r_{\rm ph}$ (pink).}\label{RTime}
\end{figure*}

As shown in Figure \ref{RTime}, the photospheric radius, $r_{\rm ph}$, shows an increase with time for GRB 150330A while a moderate decrease with time for GRB 140206B, with the size ranging from $\sim$10$^{12}$ $Y^{1/4}$ cm to $\sim$10$^{13}$ $Y^{1/4}$ cm. GRB 140329B and GRB 160625B are without a significant variation throughout the pulses, exhibiting constant values of $r_{\rm ph}$, with a size of the order of 10$^{13}$ $Y^{1/4}$ cm. The size scale and the moderate variation for the bursts are similar to the results found in some previous studies \citep{2010ApJ...709L.172R, 2013ApJ...770...32G, 2013MNRAS.433.2739I}.

\subsubsection{Parameter $Y$}

Parameter $Y$ relates to the radiative efficiency ($Y^{-1}$) of the bursts, which is defined as
\begin{equation}
Y =\frac{L_{0}}{L_{obs,\gamma}},
\end{equation}
where $L_{\rm obs,\gamma}$ is the observed $\gamma$-ray luminosity. As discussed in \S \ref{Sec:Gamma}, since the estimations of $\Gamma$ depend on the parameter $Y$ (Equation \ref{Eq:Gamma}), the determined evolution in $\Gamma(t)$ therefore could be attributed to corresponding variations in $Y(t)$. 

Parameter $Y$ can be inferred from afterglow measurement \citep{2011ApJ...738..138R, 2016ApJ...824..127W}.
\cite{2016ApJ...824..127W} found that there is relatively small scatter in the estimates of the radiative efficiency of bursts with an average value of $Y\sim$ 2, which suggests that large variations in $Y$ within individual bursts are unlikely. 
Moreover, to investigate this further, \cite{2013MNRAS.433.2739I} assumed a constant $\Gamma$ throughout the burst and found a large variation in $Y$ within individual bursts in all analyzed bursts in their sample; the corresponding estimated value of $Y$ increases with time from nearly 1 to 1000. Since the evolution in $Y(t)$ also affects the value of $r_{0}$, they then calculated the $r_{0}$ based on such evolution of $Y$, and found that in all bursts $r_{0}$ decreases from $\sim$10$^{8}$ to 10$^{3}$ cm, which is inapposite with the Schwarzschild radius of the central black hole, since the Schwarzschild radius is of the order of about 10$^{6.5}$ cm \citep{1998AIPC..428..783P}, and the inferred $r_{0}$  values are smaller than the Schwarzschild radius.
Therefore, \cite{2013MNRAS.433.2739I} concluded that a large variation of $Y$ cannot account for the observed decrease in $\Gamma$. The result confirms the finding in \cite{2016ApJ...824..127W}.

None of the bursts were observed with afterglow \citep[e.g.,][]{2012ApJ...758...27L, 2015ApJ...805...13L,2018ApJS..234...26L} in \cite{2013MNRAS.433.2739I}; this is why they made such an assumption. 
Fortunately, we find that one burst in our sample, GRB 140206B, has a good afterglow measurement. This could provide us with a clue to determine the parameter $Y$ directly. 
GRB 140206B was found to have a black hole central engine \citep{2018ApJS..236...26L}, with prompt $\gamma$-ray energy $E_{\gamma,\rm iso}= 30.68\pm 1.12$, and isotropic kinetic energy $E_{\rm K, iso}=46.28 \pm 8.34$, in units of 10$^{52}$ erg. We therefore obtain an estimation of the parameter $Y$ with afterglow emission for GRB 140206B as $\sim$ 2.51. This result also confirms well the findings in \cite{2013MNRAS.433.2739I} and \cite{2016ApJ...824..127W}. It should also be noted that estimating $Y$ from afterglow modeling may critically depend on the estimation of kinetic energy $E_{\rm K}$.  Here we consider only the X-ray band \citep{2016ApJ...824..127W, 2018ApJS..236...26L} to estimate $E_{\rm K}$, which could also result in a significant underestimate of the true energy of the blast wave \cite[e.g.,][]{2015MNRAS.454.1073B, 2016MNRAS.461...51B}. 

\subsubsection{Correlation Analysis}

\begin{figure*}
\includegraphics[angle=0, scale=0.45]{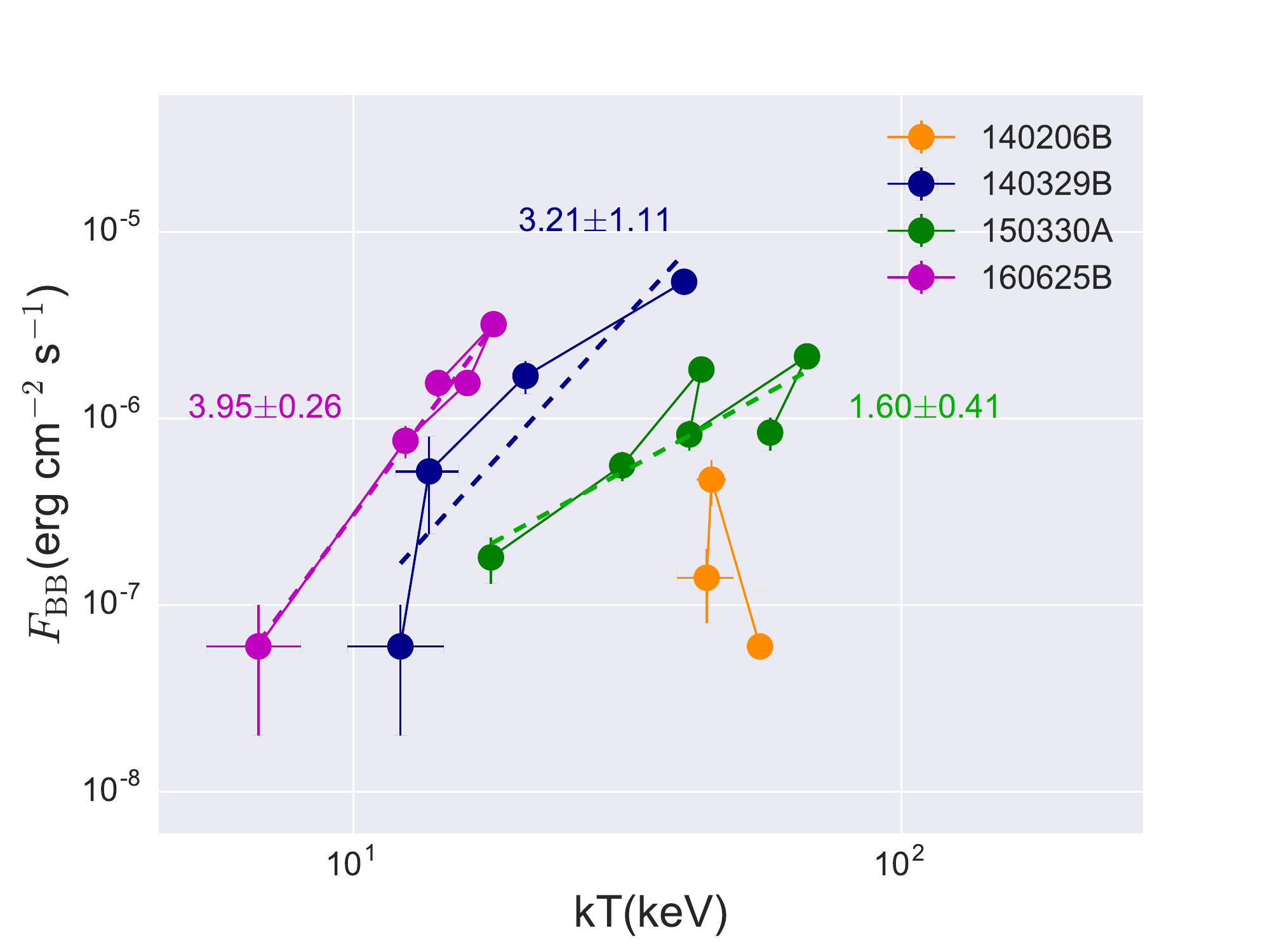}
\includegraphics[angle=0, scale=0.45]{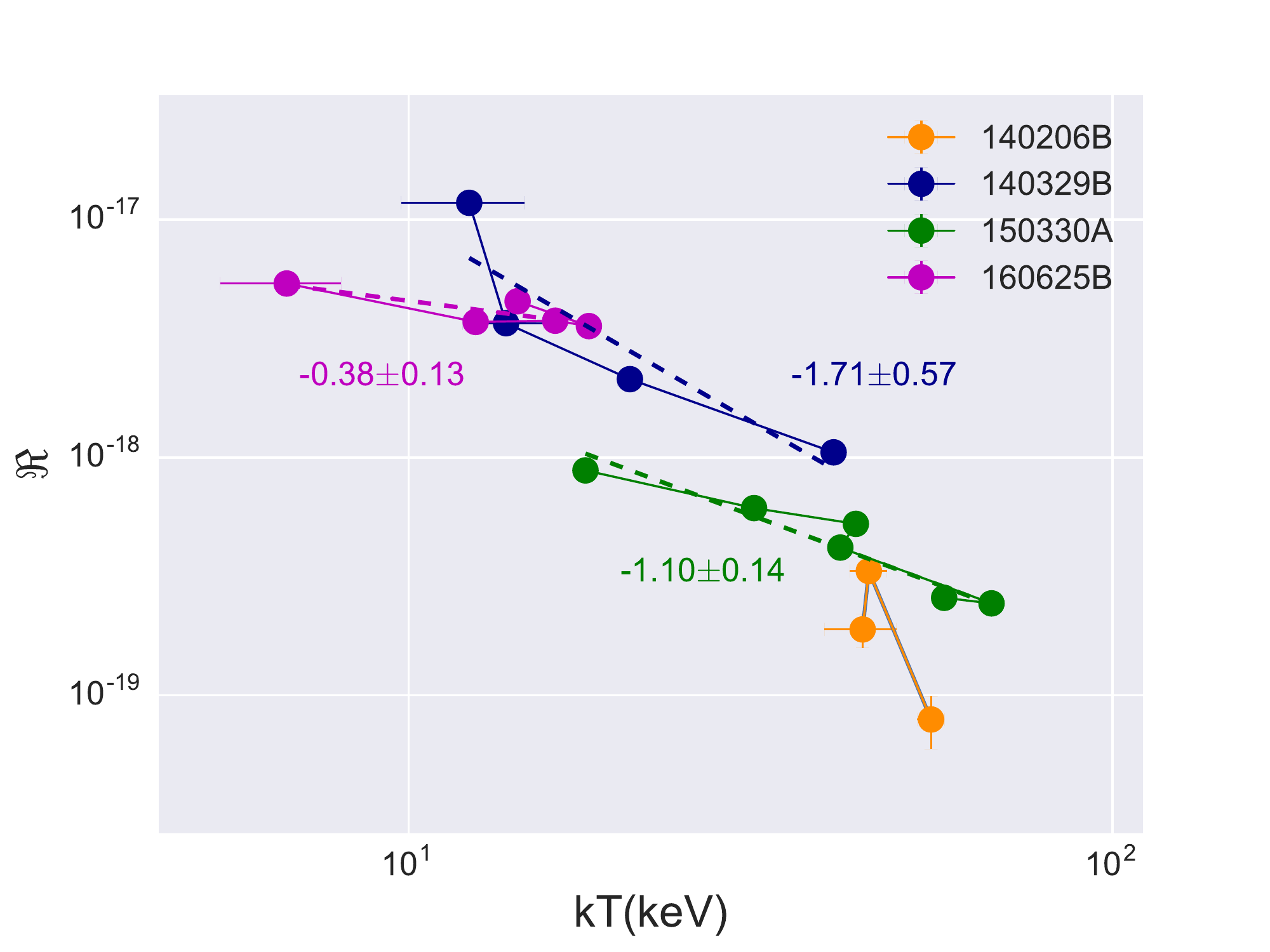}
\includegraphics[angle=0, scale=0.45]{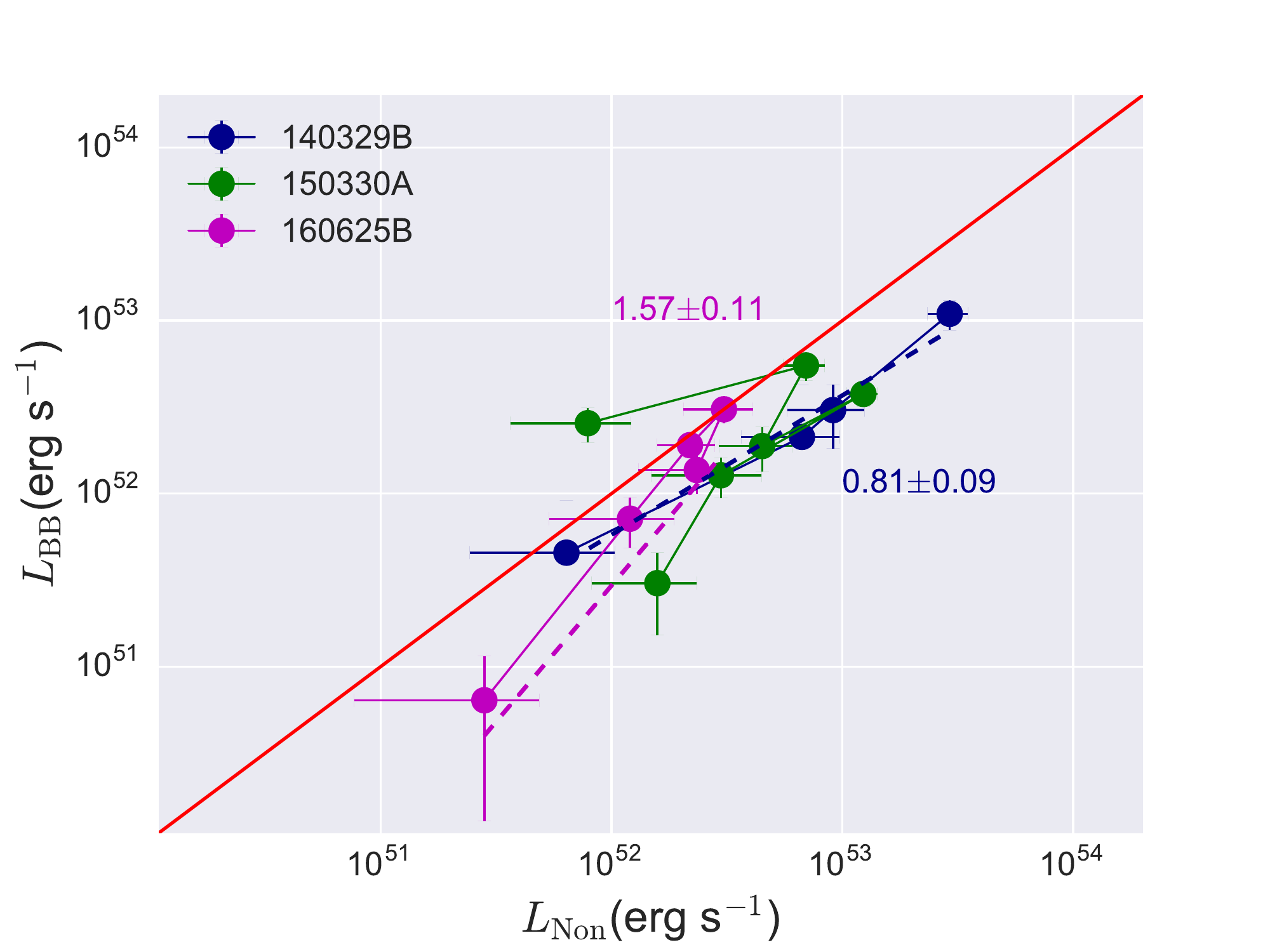}
\includegraphics[angle=0, scale=0.45]{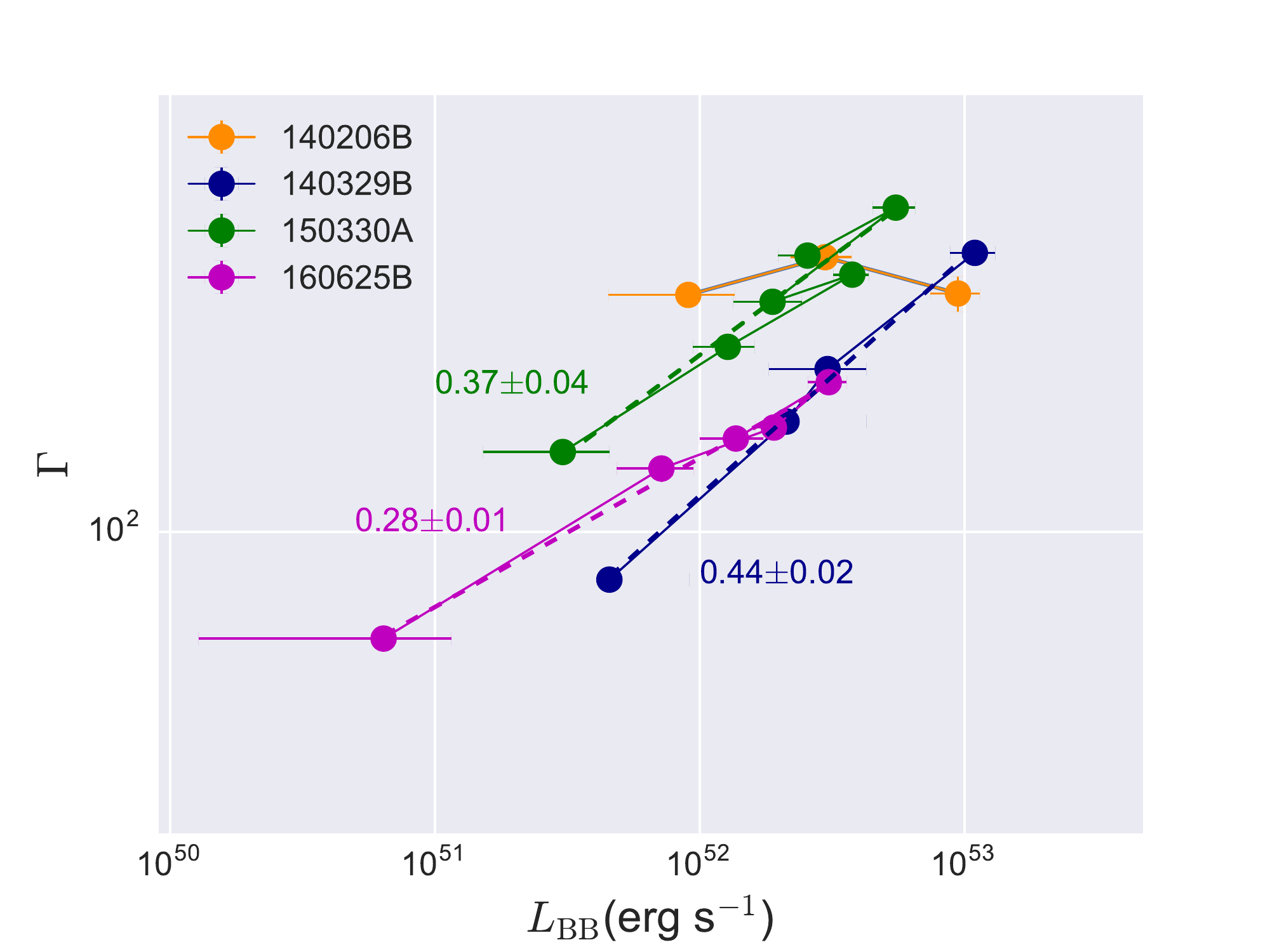}
\caption{Scatter plots of several characteristic parameters of photosphere emission:
$F_{\rm BB}$ vs. kT (top left panel), $\Re$ vs. $kT$ (top right panel), $L_{\rm BB}$ vs. $L_{\rm Non}$ (bottom right panel), and $\Gamma$ vs. $L_{\rm BB}$ (bottom left panel). Color conventions are the same as in Figure \ref{RatioReTime}. $F_{\rm BB}$-$kT$, $\Re$-$kT$, $L_{\rm BB}$-$L_{\rm Non}$ and $\Gamma$-$L_{\rm BB}$ are all found to have the hardness-intensity correlations.
For $F_{\rm BB}$-$kT$ correlation, the best-fitting results give a power-law index of 3.21$\pm$0.25 for GRB 140329B, 1.60$\pm$0.41 for GRB 150330A, and 3.95$\pm$0.26 for GRB 160625B;
For $\Re$-$kT$ correlation, we have -1.71$\pm$0.57 for GRB 140329B, -1.10$\pm$0.14 for GRB 150330A, and -0.38$\pm$0.13 for GRB 160625B;
For $L_{\rm BB}$-$L_{\rm Non}$ correlation, we obtain 0.81$\pm$0.09 for GRB 140329B, and 1.57$\pm$0.11 for GRB 160625B, the solid line in the panel represents the equal line.
For $\Gamma$-$L_{\rm BB}$ correlation, we have 0.44$\pm$0.02 for GRB 140329B, 0.37$\pm$0.04 for GRB 150330A, and 0.28$\pm$0.01 for GRB 160625B.}\label{Relations}
\end{figure*}

Correlation analysis plays an important role in the understanding of GRB physics, as it provides clues to help for us to reveal their nature \citep[e.g. the Amati relation;][]{2002A&A...390...81A}. Figure \ref{Relations} displays several correlations in the flow parameters. 

In the top left panel of Figure \ref{Relations}, we show the observed BB flux, $F_{\rm BB}$, as a function of the temperature, $kT$. This correlation is also known as the hardness-intensity correlation \citep{2001ApJ...548..770B, 2009ApJ...702.1211R}.
The energy flux is found to increase with the temperature in most cases in our sample (except for GRB 140206B, with only three time bins). \cite{2009ApJ...702.1211R} analyze the correlation with a BATSE sample and point out that such a hardness--intensity correlation can be described as $F_{\rm BB}\propto T^{\delta}$ with a power-law index of $\delta\simeq$ 4, and this is the fundamental property of a BB emitter. In our cases, we find similar results; the best-fit results (see the dashed lines in Figure \ref{Relations}) show a power-law index of 3.95$\pm$0.25 for GRB 160625B, 3.21$\pm$1.11 for GRB 140329B, and 1.60$\pm$0.41 for GRB 150330A.  

In the top right panel of Figure \ref{Relations}, we display the parameter $\Re$ plotted versus its temperature $kT$ for these bursts. We find a tight negative correlation for all the bursts. The best fitting gives the power-law indices as -1.71$\pm$0.57 for GRB 140329B, -1.10$\pm$0.14 for GRB 150330A, and -0.38$\pm$0.13 for GRB 160625B. 

In the left bottom panel of Figure \ref{Relations}, we compare the BB luminosity $L_{\rm BB}$ with the nonthermal luminosity $L_{\rm Non-thermal}$ within the bursts. We also find a hardness--intensity correlation in this; the BB luminosity is found to increase with the nonthermal luminosity, and the BB luminosity for all our bursts are found to be slightly below to the nonthermal luminosity. The best fitting gives the power-law indices as 0.81$\pm$0.09 for GRB 140329B, and 1.57$\pm$0.11 for GRB 160625B.

In the bottom right panel of Figure \ref{Relations}, we present the Lorentz factor $\Gamma_{\rm ph}$ against the the BB luminosity $L_{\rm BB}$. A tight positive correlation is also found in our sample; the power-law indices typically range within [0.25, 0.50]. We obtain 0.44$\pm$0.02 for GRB 140329B, 0.37$\pm$0.04 for GRB 150330A, and 0.28$\pm$0.02 for GRB 160625B.

\subsection{Synchrotron Emission Component}

Phenomenologically, most GRBs can be explained by an observed nonthermal spectrum, that is, the synchrotron component.
In this section, we study another case in which a cold Poynting flux fully dominated outflow for the following pulses (Part II) of our sample.

We revisited the method that was applied in \cite{2016MNRAS.456.2157I} for our sample. We first simply review the basic points of synchrotron emission\footnote{Here note that the model is based on implicitly assuming isotropic synchrotron emission in the comoving frame. However, there have been various works in recent years suggesting that there should be bulk relativistic motions in the comoving frame, either due to ``mini-jets" \cite[e.g.,][]{2003astro.ph.12347L, 2009MNRAS.395L..29G, 2009MNRAS.395..472K, 2009ApJ...695L..10L, 2016MNRAS.455L...6B, 2018MNRAS.476.1785B} or due to a striped wind magnetic field geometry \cite[e.g.,][]{2016MNRAS.459.3635B}. Specifically, this effect can have important implications on the pulse width.}.
The dissipation of the kinetic energy of the outflow at a certain radius accelerates the electrons to some characteristic Lorentz factor, and the observed peak energy of synchrotron emission from these electrons can be written as
\begin{equation}
E_{\rm syn} =\frac{3}{2}\hbar \frac{qB}{m_{e}c}\gamma_{e}^{2} \frac{\Gamma}{(1+z)},
\label{Eq:Syn}
\end{equation}
where $\hbar$ is the reduced Plank constant, $q$ and $m_{e}$ are the charge and the mass of an electron, respectively, and $B$ is the magnetic field intensity in the comoving frame. The observed flux from synchrotron emission is given by 
\begin{equation}
F_{\rm syn} =\frac{\sigma_{T}c\Gamma^{2}\gamma_{e}^{2}B^{2}N_{e}}{24\pi^{2}d^{2}_{L}},
\end{equation}
where $N_{e}$ is the number of radiating electrons.
The electrons will cool when they emit radiation, and the radiative cooling time is given by 

\begin{equation}
t_{\rm cool} =\frac{6 \pi m_{e}c}{\sigma_{T}B^{2}\pi^{2}\Gamma \gamma_{e}^{2}(1+\mathcal{Y})},
\end{equation}
where $\mathcal{Y}$ is the Compton $\mathcal{Y}$-parameter.
The dynamical time can be defined as\footnote{Equation (\ref{Eq:tdyn}) only holds if the width of the emitting shell is determined by the causal length scale \citep{2013ApJ...769...69B}.}
\begin{equation}
t_{\rm dyn} \simeq\frac{R}{2\Gamma^{2}c}.
\label{Eq:tdyn}
\end{equation}
There are two cases: 
(i) fast-cooling, $t_{\rm dyn}>t_{\rm cool}$, where the electrons lose all their energy by synchrotron radiation within the dynamical time, and all the electrons cool fast down to $\nu_{c}$ (characteristic frequency); and (ii) slow-cooling, $t_{\rm dyn}<t_{\rm cool}$, where the electrons do not efficiently radiate and therefore do not lose their energy within the dynamical time.

The synchrotron emission model is troubled by many uncertain parameters. 
In general, it is difficult to constrain all relevant parameters precisely, since there are only a few observed qualities. One thing we usually need to think is how to effectively constrain the synchrotron emission model parameters through a finite number of observations.

\begin{figure*}
\includegraphics[angle=0, scale=0.45]{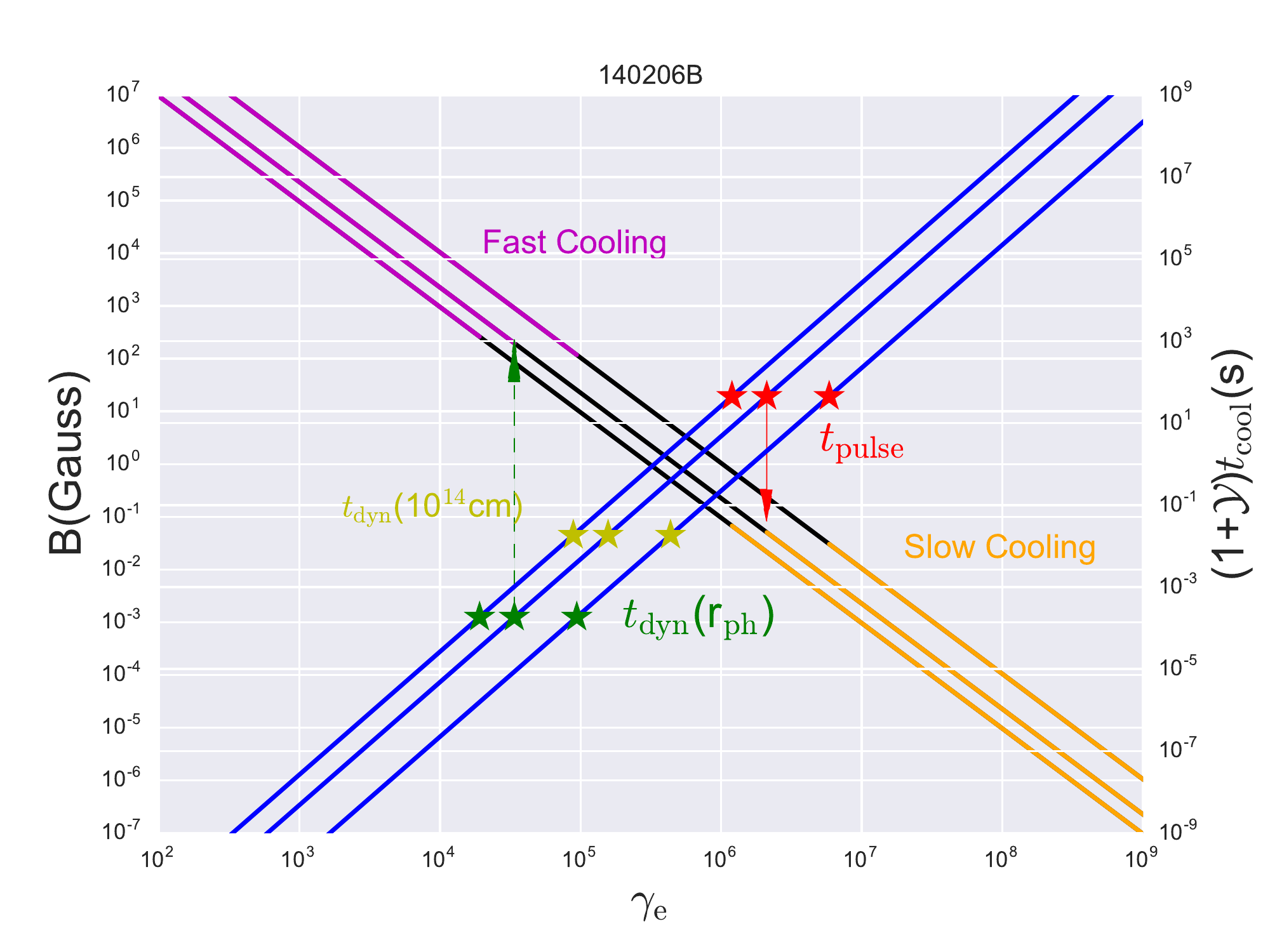}
\includegraphics[angle=0, scale=0.45]{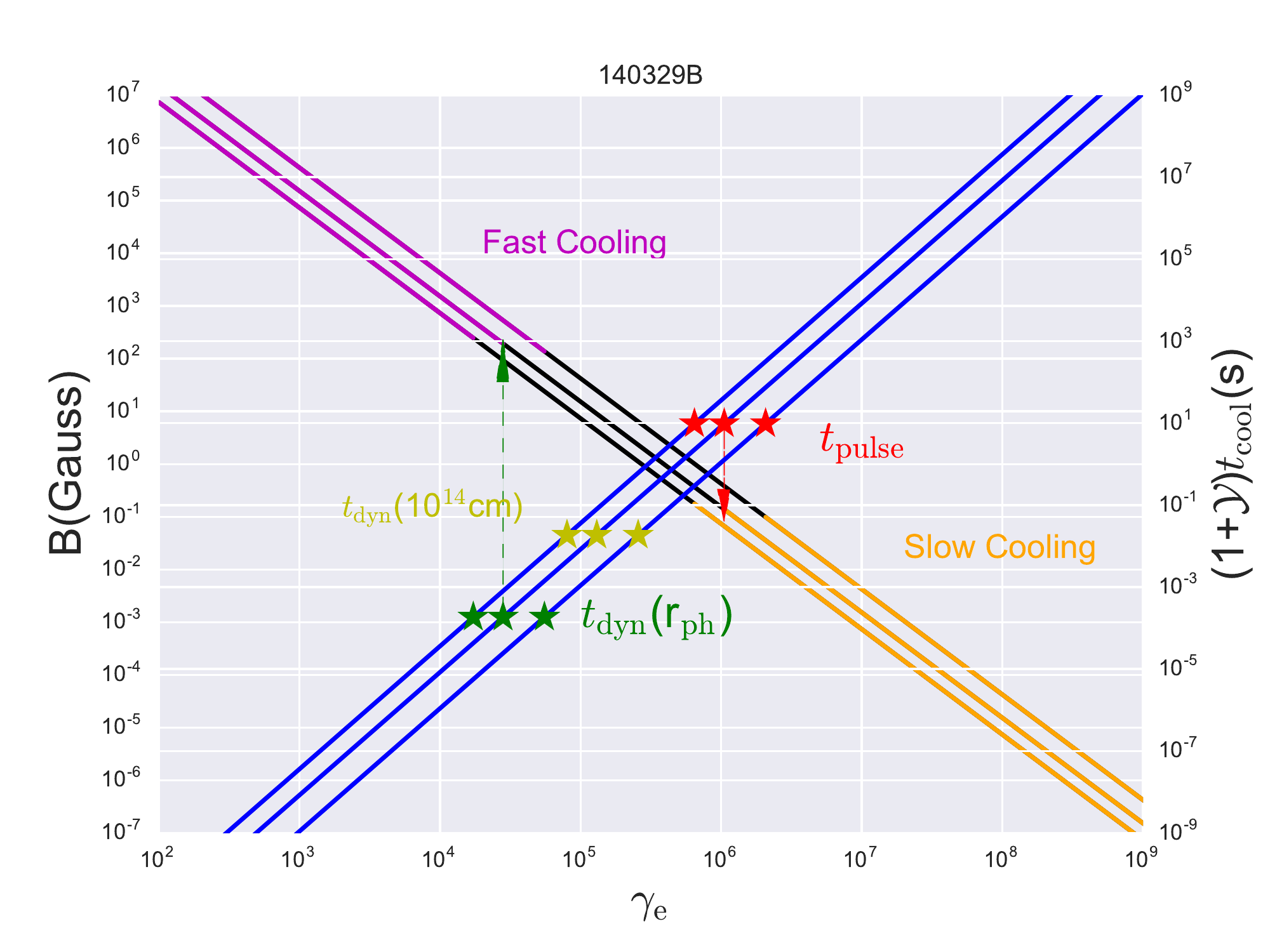}
\includegraphics[angle=0, scale=0.45]{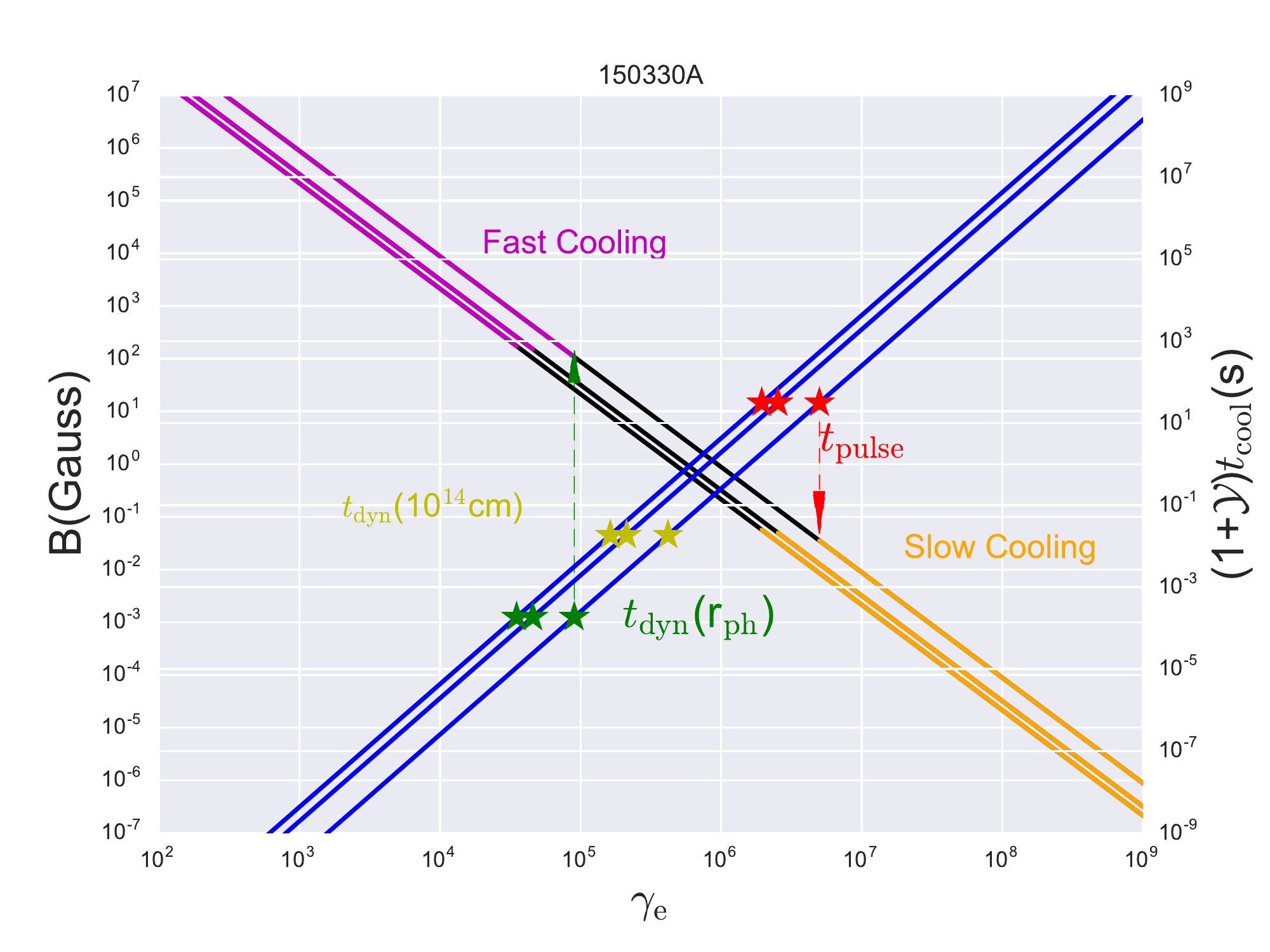}
\includegraphics[angle=0, scale=0.45]{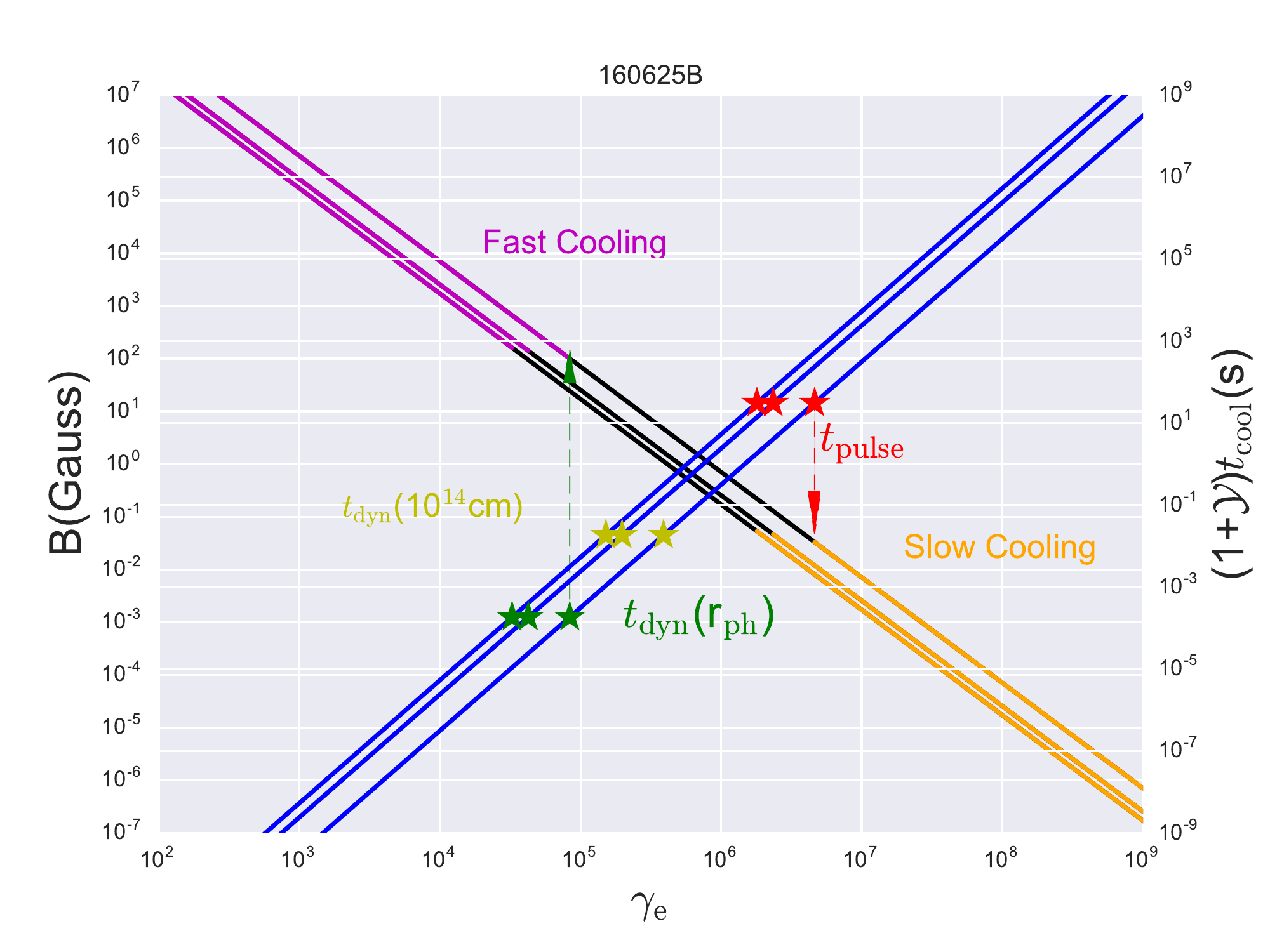}
\caption{Allowed relations between $B$ and $\gamma_{e}$ based on synchrotron emission. The left y-axis ($B$ as a function of $\gamma_{e}$: purple, black, and orange lines) shows the constraints obtained for $B\gamma^{2}_{e}$ from Equation (\ref{Eq:Bgamma}) for three time bins: one before (earliest), one at, and one after (latest) the peak of the prompt emission light curve. The right y-axis displays $t_{\rm cool}$ as a function of $\gamma_{e}$ (blue lines).
The dynamical time for different characteristic radii ($r_{\rm ph}$: grey star; 10$^{14}$ cm: blue star) and $t_{\rm pulse}$ (red star) are marked with different colors.
The red area lines show the values of $\gamma_{e}$ and $B$ that result in $t_{\rm cool} < t_{\rm dyn}(r_{\rm ph})$ for all allowed values of $r_{\rm d} > r_{\rm ph}$, which indicates that the electrons are always in the fast-cooling regime. 
The orange area lines represent that the values of $\gamma_{e}$ and $B$ will always result in $t_{\rm cool}<t_{\rm pulse}$, which is in the slow-cooling regime for the allowed values of $r_{\rm d}$. 
The black area lines represent the values of $\gamma_{e}$ and $B$ for the case $t_{\rm dyn}(r_{\rm ph}) < t_{\rm cool} < t_{\rm pulse}$, which can result in synchrotron emission for electrons cooling either fast or slow depending on what the corresponding dynamical time is and where the dissipation occurs.}\label{Btcoolgammae}
\end{figure*}

Assuming that the properties of the outflow are the same at the photosphere and at the dissipation site, we can give a constraint for the $B\gamma_{e}^{2}$ in each time bin from Equation (\ref{Eq:Syn}),
\begin{equation}
B\gamma_{e}^{2} =\frac{E_{\rm syn}(1+z)4\pi m_{e}c}{\Gamma 3h q}.
\label{Eq:Bgamma}
\end{equation}
We show them as the black lines in Figure \ref{Btcoolgammae}, where constraints obtained for three time bins are plotted: one before (earliest), one at, and one after (latest) the peak photon flux in order to capture the time evolution.

Furthermore, Equation (\ref{Eq:Syn}) is substituted into Equation (\ref{Eq:Bgamma}), and therefore we can obtain the expression of the cooling time, $t_{\rm cool}$, as a function of $\gamma_{e}$ (see the right-hand y-axis in Figure \ref{Btcoolgammae}),
\begin{equation}
\frac{\gamma_{e}^{3}}{(1+\mathcal{Y})t_{\rm cool}} =\frac{8\pi m_{\rm e}c \sigma_{\rm T}E^{2}_{\rm syn}(1+z)^{2}}{27 \Gamma h^{2} q^{2}}.
\label{Eq:tcoolgamma}
\end{equation}
The cooling timescale, $t_{\rm cool}$, is very sensitive to the change in  $\gamma_{e}$ as described in Equation (\ref{Eq:tcoolgamma}), resulting in longer cooling times for large values of $\gamma_{e}$. 

Over the blue lines, we mark the dynamical time given by Equation (\ref{Eq:tdyn}) for $r_{\rm ph}$ and $t_{\rm pulse}$ with green and red stars, as well as the dynamical time for 10$^{14}$ cm (gray star).
There are three regimes: (i) The red area lines show the values of $\gamma_{e}$ and $B$ that result in $t_{\rm cool}<t_{\rm dyn}(r_{\rm ph})$ for all allowed values of $r_{\rm d}>r_{\rm ph}$, which indicates that the electrons are always in the fast-cooling regime.
(ii) The width of the pulse of the bursts, $t_{\rm pulse}$, can give an upper limit of the dynamical time, which corresponds to an upper limit of the allowed dissipation radius, $r_{\rm d,max}=2\Gamma^{2}ct_{\rm pulse}$.
Therefore, the orange area lines represent that the values of $\gamma_{e}$ and $B$ will always result in $t_{\rm cool}<t_{\rm pulse}$, which is in the slow-cooling regime for the allowed values of $r_{\rm d}$.
(iii) Finally, the black area lines represent the values of $\gamma_{e}$ and $B$ for the case $t_{\rm dyn}(r_{\rm ph})<t_{\rm cool}<t_{\rm pulse}$, which can result in synchrotron emission for electrons cooling either fast or slow depending on what the corresponding dynamical time is and where the dissipation occurs.

On the other hand, since the observed peak energy of synchrotron emission (here assuming that the peak energies of synchrotron emission are the same as derived from the CPL model) can be derived from the spectral fits. According to Equation (\ref{Eq:Bgamma}), one has $B\gamma^{2}_{e}\propto \Gamma^{-1}$.
Analytically in prior articles \citep{2013ApJ...769...69B, 2014NatPh..10..351U} it is suggested that both $B$, and $\gamma_{e}$, evolve during the burst.
One question is, if we assume a constant $\Gamma$ throughout a burst, how is the evolution characteristic of $B\gamma^{2}_{e}$? Since $\Gamma$ has a typical value of a few hundred, one can therefore reliably estimate the typical values of $B\gamma^{2}_{e}$. We thus can study the evolution of $B\gamma^{2}_{e}$ in all time bins within bursts.

To investigate this further, we illustrate the evolution of $B\gamma^{2}_{e}$ within bursts for different $\Gamma$ values ($\Gamma$=100, 300, 600, 1000) for all the analyzed bursts in Figure \ref{Bgammae}.
We find that $B\gamma^{2}_{e}$ indeed decreases with time and has a narrow distribution, typically ranging from 10$^{11}$ to 10$^{12}$ for all the cases. 
This, in turn, suggests that $\Gamma$ decreases during the bursts.
Also, if considering a lower limit on $B$ being between 10$^{-3}$ and 10$^{-4}$ G, and therefore an upper limit on $\gamma_{e}$ lying at $\sim$few time 10$^{7}$, this result is consistent with the finding in \cite{2016MNRAS.456.2157I}, or if considering a typical value of $B$ $\sim$ 10$^{3}$ G, thereby $\gamma_{e}$ lying at $\sim$10$^{4}$.
Furthermore, we also show the distributions of $B\gamma^{2}_{e}$ for these typical values of $\Gamma$ in Figure \ref{Bgammae}. All the distributions can well fitted by the Gaussian function. The typical fits give log $B\gamma^{2}_{e}$=11.81$\pm$0.19 for $\Gamma$=100, log $B\gamma^{2}_{e}$=11.26$\pm$0.18 for $\Gamma$=300, log $B\gamma^{2}_{e}$=10.95$\pm$0.19 for $\Gamma$=600, and log $B\gamma^{2}_{e}$=10.83$\pm$0.18 for $\Gamma$=1000.

\section{Discussion and Conclusion}

\subsection{Magnetization parameter, $\sigma_{0}$, and the Hybrid jet system}
 
The physical scenarios discussed above indeed always focus on the extreme cases: either a pure hot fireball dominated outflow or a Poynting flux fully dominated outflow.  However, another more natural possibility is that the central engine of GRB jets is more likely to be a hybrid system, which is involved in both a hot fireball component and a cold Poynting flux component simultaneously \cite[e.g.,][]{2005ApJ...625L..95R, 2006A&A...457..763G, 2007AIPC..921..478B, 2009ApJ...702.1211R, 2011ApJ...727L..33G, 2012ApJ...757L..31A, 2013MNRAS.433.2739I, 2014ApJ...784...17B, 2015ApJ...801..103G, 2015ApJ...802..134B, 2017A&A...598A..23N, 2017MNRAS.468.3202B}. 
Based on this fact, a theory of photosphere emission for a hybrid outflow with a hot fireball component and a cold Poynting flux component has been developed in \cite{2015ApJ...801..103G}.
Two parameters are used to describe the hybrid outflow: the dimensionless entropy for the hot component $\eta=L_{\rm hot}/\dot{M}c^{2}$ and the magnetization parameter for the Poynting flux component $\sigma_{0} \equiv L_{\rm Poynting}/L_{\rm hot}$.
The different combinations of the parameters ($\eta$, $\sigma_{0}$)\footnote{Note that in the ``hybrid" models, changing ($\eta$, $\sigma_{0}$) may result in completely different dynamics for the jet. For example, for magnetically dominated jets, the Lorentz factor no longer evolves linearly with radius \citep{2002A&A...387..714D, 2019MNRAS.484.1378G}. This can have significant effects on both the resulting radiation of both the thermal component \citep{2017MNRAS.468.3202B} and the application to data in \cite{2018ApJ...867...52X}.} mainly correspond to three types of observation:
(i) if $\eta \gg 1$ and $\sigma_{0} \ll 1$, a pure hot fireball is obtained (e.g., the most typical case, GRB 090902B with dominated BB component);
(ii) if $\eta$ is smaller and $\sigma_{0}$ is larger, the photosphere emission component becomes subdominant with a nonthermal component (e.g., GRB 110721A, which can be best fitted by the Band function plus a BB.); 
(iii) if $\eta$ is close to unity and $\sigma_{0} \gg 1$, the outflow is fully dominated by the Poynting flux (e.g., GRB 080916C).
In \cite{2015ApJ...801..103G}, the photosphere parameters for these different regimes of ($\eta$, $\sigma_{0}$) are derived.
However, for the hybrid model, further constraints by observation are required.

\subsection{Polarization, Future Prospects}

Theoretically, photon polarization is an extra important tool to diagnose jet composition and radiation mechanism \citep{2009ApJ...698.1042T, 2013MNRAS.428.2430L, 2014IJMPD..2330002Z}. 
Thus, the transition of the jet composition in quite a number of GRBs found in this work can be tested by the observation of polarization. This test cannot be achieved by the current polarization data (e.g., the POLAR detect; \citealt{2019NatAs.tmp..188Z}), which is eager for future observations.
Interestingly, two bursts (GRB 160625B, and GRB 160820A) in our sample are reported to have the observation of polarization \citep{2017Natur.547..425T, 2018ApJ...862..154C}.
Besides, the corresponding X-ray, optical, and other energy-band afterglows for these bursts may also be different, which requires further observations to confirm in the future.
 
\subsection{Conclusion}

In this paper, we systematically searched for evidence of the transition from fireball to Poynting-flux-dominated outflow of GRBs with {\it Fermi} data. 
We first fit the time-integrated spectra for the bursts with the Band and the CPL models, respectively, and found that $\alpha$ is much harder (close to 0) for Part I (the first pulse/sub-burst).
We then further fit the time-resolved spectra in each slice with the CPL model and also found that the results are consistent with the former analysis. Such a photon index is beyond the so-called synchrotron line of death ($\alpha$=-2/3, \citealt{1998ApJ...506L..23P}), and is much harder than the typical $\alpha$ value ($\alpha$=-1.0) of long GRBs, thus suggesting a significant contribution of thermal emission from the fireball photosphere \citep{2010ApJ...709L.172R, 2000ApJ...530..292M}.
Furthermore, it is interesting to find that three bursts exhibit a thermal precursor before the main bursts \citep{1991Natur.350..592M}. 
Part II (the following pulses/sub-bursts) cannot be fitted by a Plank function but can be well fitted by the empirical CPL function, suggesting a synchrotron origin. 
The average $\alpha$ of the time-integrated spectra for Part II is consistent with the typical $\alpha$ distribution.

In summary, after performing a detailed spectral analysis, except for GRB 160625B, we find three more bursts whose spectral properties of the pulses/sub-bursts are quite different, showing the transition from thermal emission to nonthermal emission between well-separated pulses/sub-bursts within a single GRB.
Such a transition is likely to be the evidence of the change of jet composition from a hot fireball to a cold Poynting-flux-dominated outflow. The main results can be drawn as follows:
\begin{itemize}
\item 43 bright {\it Fermi} bursts exhibit at least two clear pulses in their prompt emission light curves in our sample. In total, 9 out of 43 bursts ($\sim$ 21\%) are found to have a transition from fireball (thermal) to Poynting-flux-dominated (non-thermal) outflow based on the time-integrated spectra. This indicates that such a transition is commonly detected in the bright multipulse {\it Fermi} bursts. 

\item Among these bursts, we select four cases (including GRB 160625B) with adequate data to perform the detailed time-resolved spectral analysis. Three out of four bursts present an early precursor before the main burst, which shows that a thermal characteristic exists in both their time-integrated spectrum and time-resolved spectra.
Moreover, we also find that the duration of the early thermal component is much shorter than that of the nonthermal component, which implies that the thermal process proceeds more rapidly than the nonthermal one.

\item Part I, indeed, can be well fitted by the model that contains a BB component.
Based on the model of thermal+ nonthermal components, we derive outflow properties, which are consistent with the typical observation from the photosphere emission. 
Part II can be well explained by the optically thin synchrotron emission component.

\item Other interesting results:

(1) The spectral evolution could be more prominent to occur in later pulses.

(2) The low-energy power-law index $\alpha$, obtained from the CPL model, is slightly softer than that from the Band model.

(3) We have introduced a new, more reasonable time-bin method of the time-resolved spectra, which combines the advantages of both the traditional S/N and the BBlocks techniques.

(4) We find that, in comparing with the distributions of the different empirical models (the Band and the CPL) based on the same type of spectrum (time-integrated), the distributions of different type of spectra (time-integrated and time-resolved) based on the same model (the CPL) are more consistent with each other, whether by the $E_{\rm pk(c)}$-distribution or by the $\alpha$-distribution.

\end{itemize}

We conclude that {\it a good fraction of the multipulse {\it Fermi} bursts present implication of a transition from fireball to Poynting-flux-dominated outflow}.

\acknowledgments

I am grateful to Felix Ryde, Shabnam Iyyani, Damien B\'egu\'e, Bing Zhang, Yu Wang, Bin-Bin Zhang, Asaf Pe{\textquoteright}er, Pawan Kumar, Jin-Jun Geng, Yan-Zhi Meng, Yong-Feng Huang, Li-Ping Xin, Shuang-Nan Zhang, and Gregory Vereshchagin for useful discussions, and I appreciate to the referee for the constructive report. I also thank the following people who provided indirect help on the subject: Xue-Feng Wu, Xiang-Yu Wang, Zi-Gao Dai, Thomas Pak-Hin, Yi-Zhong Fan, Da-Ming Wei, Wei-Min Gu, Jian-Yan Wei, and Remo Ruffini. This research made use of the High Energy Astrophysics Science Archive Research Center (HEASARC) Online Service at the NASA/Goddard Space Flight Center (GSFC). Part of this work made use of our personal Python library. 

%\clearpage
\vspace{5mm}
\facilities{{\it Fermi}/GBM}
\software{3ML\citep{2015arXiv150708343V}}
\bibliography{../../MyBibFiles/MyBibFile.bib}

\clearpage
\startlongtable
\begin{deluxetable*}{ccccccccc}
%\rotate
\tablewidth{0pt}
\tabletypesize{\footnotesize}
\tablecaption{Results of the Time-integrated Spectral Fits of Our Total Sample}
%\tablenum{1}
\tablehead{
\colhead{Burst ID}
&\colhead{$t_{1}$$\sim$$t_{2}$}
&\colhead{Sequence}
&\colhead{$\alpha$}
&\colhead{$E_{\rm c}$}
&\colhead{Model}
&\colhead{Detector}
&\colhead{Selected by}
&\colhead{-ln(likelihood)/dof}\\
&
(s)
&&&
(keV)
}
\colnumbers
\startdata
081009140&0.0$\sim$8.0&1st&-1.12$^{+0.03}_{-0.03}$&55$^{+18}_{-17}$&CPL&n3,b1&PeakFlux&1134/212\\
081009140&36.0$\sim$50.0&2nd&-1.50$^{+0.70}_{-0.70}$&19$^{+14}_{-8}$&CPL&n3,b1&PeakFlux&1248/212\\
081215784&0.0$\sim$2.9&1st&-0.72$^{+0.18}_{-0.18}$&688$^{+40}_{-35}$&CPL&n9,na,nb,b1&PeakFlux&1838/438\\
081215784&2.9$\sim$4.4&2nd&-0.58$^{+0.04}_{-0.04}$&210$^{+15}_{-14}$&CPL&n9,na,nb,b1&PeakFlux&1295/438\\
081215784&4.4$\sim$8.0&3rd&-0.75$^{+0.03}_{-0.03}$&250$^{+16}_{-15}$&CPL&n9,na,nb,b1&PeakFlux&1844/438\\
090618353&0.0$\sim$45.0&1st&-0.87$^{+0.09}_{-0.09}$&154$^{+26}_{-22}$&CPL&n4,b0&PeakFlux&1659/212\\
090618353&45.0$\sim$76.0&2nd&-1.03$^{+0.02}_{-0.02}$&241$^{+11}_{-10}$&CPL&n4,b0&PeakFlux&1606/212\\
090618353&76.0$\sim$103.0&3rd&-1.11$^{+0.02}_{-0.02}$&141$^{+6}_{-6}$&CPL&n4,b0&PeakFlux&1494/212\\
090618353&103.0$\sim$116.0&4th&-1.33$^{+0.06}_{-0.06}$&100$^{+12}_{-11}$&CPL&n4,b0&PeakFlux&1246/212\\
090926181&0.0$\sim$8.0&1st&-0.60$^{+0.01}_{-0.01}$&267$^{+6}_{-6}$&CPL&n3,n6,n7,b1&Fluence&2202/343\\
090926181&8.0$\sim$20.0&2nd&-0.89$^{+0.01}_{-0.01}$&225$^{+6}_{-6}$&CPL&n3,n6,n7,b1&Fluence&2266/343\\
091127976&0.0$\sim$1.0&1st&-1.15$^{+0.03}_{-0.03}$&108$^{+7}_{-6}$&CPL&n6,n7,n9,b1&PeakFlux&905/341\\
091127976&1.0$\sim$2.0&2nd&-1.37$^{+0.03}_{-0.03}$&276$^{+32}_{-28}$&CPL&n6,n7,n9,b1&PeakFlux&881/341\\
091127976&6.5$\sim$8.5&3rd&-1.89$^{+0.08}_{-0.08}$&60$^{+9}_{-8}$&CPL&n6,n7,n9,b1&PeakFlux&1112/341\\
100719989&0.0$\sim$3.4&1st&-0.34$^{+0.04}_{-0.04}$&235$^{+17}_{-16}$&CPL&n4,n5,b0&PeakFlux&1015/228\\
100719989&3.4$\sim$10.0&2nd&-0.51$^{+0.04}_{-0.04}$&148$^{+10}_{-9}$&CPL&n4,n5,b0&PeakFlux&1197/228\\
100719989&20.0$\sim$24.0&3rd&-0.85$^{+0.11}_{-0.11}$&230$^{+60}_{-50}$&CPL&n4,n5,b0&PeakFlux&945/228\\
100826957&-1.0$\sim$50.0&1st&-0.79$^{+0.02}_{-0.02}$&323$^{+22}_{-20}$&CPL&n7,n8,b1&Fluence&2146/230\\
100826957&58.0$\sim$110.0&2nd&-1.12$^{+0.03}_{-0.03}$&278$^{+32}_{-29}$&CPL&n7,n8,b1&Fluence&2034/230\\
100829876&0.0$\sim$1.4&1st&-0.42$^{+0.07}_{-0.07}$&113$^{+11}_{-10}$&CPL&n2,b0&PeakFlux&805/221\\
100829876&1.4$\sim$1.9&2nd&-0.69$^{+0.11}_{-0.11}$&200$^{+50}_{-40}$&CPL&n2,b0&PeakFlux&577/199\\
101014175&0.0$\sim$16.0&1st&-0.98$^{+0.01}_{-0.01}$&200$^{+8}_{-7}$&CPL&n6,n7,b1&Fluence&2108/325\\
101014175&16.0$\sim$40.0&2nd&-1.38$^{+0.02}_{-0.02}$&280$^{+25}_{-23}$&CPL&n6,n7,b1&Fluence&2242/325\\
101014175&98.0$\sim$120.0&3rd&-1.15$^{+0.05}_{-0.05}$&162$^{+20}_{-18}$&CPL&n6,n7,b1&Fluence&2234/325\\
101014175&155.0$\sim$170.0&4th&-1.12$^{+0.08}_{-0.08}$&370$^{+159}_{-110}$&CPL&n6,n7,b1&Fluence&2056/324\\
101014175&195.0$\sim$230.0&4th&-0.88$^{+0.29}_{-0.29}$&390$^{+140}_{-40}$&CPL&n6,n7,b1&Fluence&2600/324\\
110301214&0.0$\sim$3.5&1st&-0.81$^{+0.02}_{-0.02}$&108$^{+3}_{-3}$&CPL&n7,n8,nb,b1&PeakFlux&1686/341\\
110301214&3.7$\sim$8.0&2nd&-1.11$^{+0.03}_{-0.03}$&103$^{+5}_{-5}$&CPL&n7,n8,nb,b1&PeakFlux&1755/341\\
110625881&-1.0$\sim$9.7&1st&-0.66$^{+0.12}_{-0.12}$&193$^{+40}_{-34}$&CPL&n7,n8,nb,b1&PeakFlux&2556/439\\
110625881&9.7$\sim$18.0&2nd&-0.87$^{+0.03}_{-0.03}$&204$^{+15}_{-14}$&CPL&n7,n8,nb,b1&PeakFlux&2492/439\\
110625881&20.0$\sim$27.0&3rd&-0.79$^{+0.02}_{-0.02}$&142$^{+5}_{-5}$&CPL&n7,n8,nb,b1&PeakFlux&2455/439\\
110625881&27.0$\sim$32.0&4th&-1.04$^{+0.04}_{-0.04}$&179$^{+15}_{-14}$&CPL&n7,n8,b1&PeakFlux&2128/439\\
110825102&0.0$\sim$14.0&1st&-1.21$^{+0.05}_{-0.05}$&182$^{+22}_{-20}$&CPL&n3,n6,n7,b1&PeakFlux&2341/342\\
110825102&14.0$\sim$21.0&2nd&-0.91$^{+0.02}_{-0.02}$&260$^{+14}_{-13}$&CPL&n3,n6,n7,b1&PeakFlux&2122/342\\
110903009&0.0$\sim$2.7&1st&-1.30$^{+0.23}_{-0.23}$&54$^{+19}_{-14}$&CPL&n2,n5,b0&PeakFlux&1408/326\\
110903009&2.7$\sim$7.0&2nd&-1.36$^{+0.11}_{-0.11}$&33$^{+4}_{-3}$&CPL&n2,n5,b0&PeakFlux&1551/326\\
110903009&19.9$\sim$28.0&3rd&-1.42$^{+0.08}_{-0.08}$&270$^{+10}_{-70}$&CPL&n2,n5,b0&PeakFlux&1860/326\\
120204054&1.0$\sim$40.0&1st&-0.95$^{+0.03}_{-0.03}$&172$^{+11}_{-10}$&CPL&n0,n1,n3,b0&PeakFlux&2616/441\\
120204054&40.0$\sim$52.0&2nd&-1.19$^{+0.02}_{-0.02}$&240$^{+17}_{-15}$&CPL&n0,n1,n3,b0&PeakFlux&3765/441\\
120328268&0.0$\sim$16.8&1st&-0.78$^{+0.02}_{-0.02}$&222$^{+12}_{-11}$&CPL&n7,n9,nb,b1&PeakFlux&3111/437\\
120328268&16.8$\sim$40.0&2nd&-1.02$^{+0.03}_{-0.03}$&198$^{+14}_{-13}$&CPL&n7,n9,nb,b1&PeakFlux&3214/437\\
120728434&-1.0$\sim$4.0&1st&-0.69$^{+0.27}_{-0.27}$&120$^{+60}_{-40}$&CPL&n1,n2,n5,b0&Fluence&2203/438\\
120728434&4.0$\sim$55.0&2nd&-0.12$^{+0.05}_{-0.05}$&53$^{+2}_{-2}$&CPL&n1,n2,n5,b0&Fluence&6463/438\\
120728434&65.0$\sim$95.0&3rd&-0.26$^{+0.04}_{-0.04}$&447$^{+14}_{-14}$&CPL&n1,n2,n5,b0&Fluence&5956/438\\
130504978&0.0$\sim$22.0&1st&-0.94$^{+0.02}_{-0.02}$&810$^{+80}_{-80}$&CPL&n2,n9,na,b1&Fluence&3020/440\\
130504978&22.0$\sim$42.0&2nd&-1.15$^{+0.02}_{-0.02}$&700$^{+80}_{-70}$&CPL&n2,n9,na,b1&Fluence&2919/440\\
130504978&45.0$\sim$59.0&3rd&-1.48$^{+0.05}_{-0.05}$&830$^{+400}_{-280}$&CPL&n2,n9,na,b1&Fluence&2596/440\\
130504978&59.0$\sim$74.0&4th&-1.22$^{+0.02}_{-0.02}$&640$^{+80}_{-70}$&CPL&n2,n9,na,b1&Fluence&2739/440\\
130504978&74.0$\sim$84.0&5th&-1.39$^{+0.07}_{-0.07}$&260$^{+70}_{-60}$&CPL&n2,n9,na,b1&Fluence&2361/440\\
130606497&0.0$\sim$12.0&1st&-0.87$^{+0.02}_{-0.02}$&580$^{+40}_{-40}$&CPL&n7,n8,b1&Fluence&1509/228\\
130606497&12.0$\sim$22.0&2nd&-1.33$^{+0.01}_{-0.01}$&3100$^{+700}_{-600}$&CPL&n7,n8,b1&Fluence&1454/228\\
130606497&35.0$\sim$48.0&3rd&-1.04$^{+0.02}_{-0.02}$&290$^{+20}_{-18}$&CPL&n7,n8,b1&Fluence&1438/228\\
130606497&48.0$\sim$63.0&4th&-0.92$^{+0.03}_{-0.03}$&207$^{+12}_{-11}$&CPL&n7,n8,b1&Fluence&1484/228\\
131127592&0.0$\sim$5.4&1st&-1.03$^{+0.03}_{-0.03}$&309$^{+28}_{-26}$&CPL&n1,n2,n5,b0&PeakFlux&2132/442\\
131127592&5.4$\sim$15.5&2nd&-1.15$^{+0.03}_{-0.03}$&241$^{+18}_{-16}$&CPL&n1,n2,n5,b0&PeakFlux&2523/442\\
131127592&15.5$\sim$20.0&3rd&-1.01$^{+0.04}_{-0.04}$&135$^{+11}_{-10}$&CPL&n1,n2,n5,b0&PeakFlux&2040/442\\
140206275&0.0$\sim$4.7&1st&0.23$^{+0.10}_{-0.10}$&109$^{+11}_{-10}$&CPL&n0,n1,n3,b0&PeakFlux&1902/441\\
140206275&4.7$\sim$25.0&2nd&-0.86$^{+0.01}_{-0.01}$&349$^{+14}_{-13}$&CPL&n0,n1,n3,b0&PeakFlux&3136/441\\
140206275&25.0$\sim$50.0&3rd&-1.21$^{+0.03}_{-0.03}$&219$^{+17}_{-15}$&CPL&n0,n1,n3,b0&PeakFlux&3065/441\\
140213807&-1.0$\sim$5.0&1st&-1.19$^{+0.03}_{-0.03}$&255$^{+24}_{-22}$&CPL&n0,n1,b0&PeakFlux&2356/442\\
140213807&5.0$\sim$12.0&2nd&-1.15$^{+0.04}_{-0.04}$&83$^{+5}_{-5}$&CPL&n0,n1,b0&PeakFlux&2454/442\\
140329295&-1.0$\sim$1.6&1st&-0.59$^{+0.14}_{-0.14}$&114$^{+24}_{-20}$&CPL&n8,nb,b1&PeakFlux&1101/326\\
140329295&19.0$\sim$22.8&2nd&-0.77$^{+0.03}_{-0.03}$&203$^{+11}_{-10}$&CPL&n8,nb,b1&PeakFlux&1428/326\\
140329295&22.8$\sim$28.7&3rd&-0.86$^{+0.02}_{-0.02}$&227$^{+9}_{-9}$&CPL&n8,nb,b1&PeakFlux&1687/326\\
140416060&-1.0$\sim$5.4&1st&-1.14$^{+0.07}_{-0.07}$&123$^{+16}_{-14}$&CPL&n2,b0&PeakFlux&989/225\\
140416060&5.4$\sim$18.0&2nd&-1.20$^{+0.04}_{-0.04}$&144$^{+11}_{-10}$&CPL&n2,b0&PeakFlux&1221/225\\
140416060&18.0$\sim$50.0&3rd&-1.26$^{+0.06}_{-0.06}$&148$^{+19}_{-17}$&CPL&n2,b0&PeakFlux&1513/225\\
140508128&0.0$\sim$10.0&1st&-0.91$^{+0.03}_{-0.03}$&325$^{+28}_{-26}$&CPL&na,b1&PeakFlux&1200/211\\
140508128&23.0$\sim$29.0&2nd&-0.72$^{+0.05}_{-0.05}$&152$^{+15}_{-14}$&CPL&na,b1&PeakFlux&1011/211\\
140508128&38.0$\sim$42.0&3rd&-1.23$^{+0.07}_{-0.07}$&200$^{+40}_{-40}$&CPL&na,b1&PeakFlux&878/211\\
140508128&46.5$\sim$48.2&4th&-0.51$^{+0.20}_{-0.20}$&51$^{+11}_{-9}$&CPL&na,b1&PeakFlux&577/211\\
140523129&0.0$\sim$2.6&1st&-0.64$^{+0.05}_{-0.05}$&292$^{+35}_{-31}$&CPL&n3,n4,n5,b0&PeakFlux&1809/442\\
140523129&2.6$\sim$6.5&2nd&-0.53$^{+0.04}_{-0.04}$&237$^{+17}_{-16}$&CPL&n3,n4,n5,b0&PeakFlux&2088/442\\
140523129&6.5$\sim$11.7&3rd&-0.82$^{+0.03}_{-0.03}$&185$^{+11}_{-10}$&CPL&n3,n4,n5,b0&PeakFlux&2240/442\\
140523129&11.7$\sim$23.0&4th&-1.20$^{+0.04}_{-0.04}$&120$^{+11}_{-10}$&CPL&n3,n4,n5,b0&PeakFlux&2681/442\\
140810782&0.0$\sim$20.0&1st&-0.67$^{+0.07}_{-0.07}$&207$^{+30}_{-27}$&CPL&n2,n5,b0&Fluence&2335/327\\
140810782&20.0$\sim$41.0&2nd&-0.66$^{+0.05}_{-0.05}$&278$^{+29}_{-26}$&CPL&n2,n5,b0&Fluence&2428/327\\
140810782&42.0$\sim$55.0&3rd&-0.76$^{+0.04}_{-0.04}$&171$^{+14}_{-13}$&CPL&n2,n5,b0&Fluence&2165/327\\
150118409&0.0$\sim$40.0&1st&-0.94$^{+0.01}_{-0.01}$&631$^{+35}_{-33}$&CPL&n1,n2,n5,b0&PeakFlux&3555/441\\
150118409&40.0$\sim$52.0&2nd&-0.88$^{+0.03}_{-0.03}$&580$^{+50}_{-50}$&CPL&n1,n2,n5,b0&PeakFlux&2714/441\\
150201574&0.0$\sim$4.2&1st&-0.81$^{+0.02}_{-0.02}$&161$^{+6}_{-6}$&CPL&n3,n4,b0&PeakFlux&1591/316\\
150201574&4.2$\sim$8.5&2nd&-1.03$^{+0.02}_{-0.02}$&119$^{+4}_{-4}$&CPL&n3,n4,b0&PeakFlux&2180/316\\
150330828&-1.0$\sim$10.0&1st&-0.30$^{+0.08}_{-0.08}$&190$^{+22}_{-20}$&CPL&n1,n2,n5,b0&PeakFlux&2600/444\\
150330828&123.5$\sim$142.2&2nd&-0.95$^{+0.01}_{-0.01}$&330$^{+11}_{-10}$&CPL&n1,n2,n5,b0&PeakFlux&3178/444\\
150330828&142.2$\sim$155.0&3rd&-1.10$^{+0.02}_{-0.02}$&214$^{+13}_{-12}$&CPL&n1,n2,n5,b0&PeakFlux&2737/444\\
151227218&0.0$\sim$4.2&1st&-1.02$^{+0.08}_{-0.08}$&250$^{+60}_{-50}$&CPL&n1,n2,n5,b0&PeakFlux&1883/442\\
151227218&20.0$\sim$39.0&2nd&-1.26$^{+0.02}_{-0.02}$&530$^{+50}_{-50}$&CPL&n1,n2,n5,b0&PeakFlux&2978/442\\
151227218&42.0$\sim$46.0&3rd&-1.25$^{+0.08}_{-0.08}$&190$^{+50}_{-40}$&CPL&n1,n2,n5,b0&PeakFlux&1856/442\\
160422499&0.0$\sim$3.1&1st&-1.17$^{+0.02}_{-0.02}$&378$^{+26}_{-24}$&CPL&n0,n1,n5,b0&PeakFlux&1878/440\\
160422499&3.2$\sim$15.0&2nd&-0.93$^{+0.01}_{-0.01}$&225$^{+6}_{-6}$&CPL&n0,n1,n5,b0&PeakFlux&2887/440\\
160509374&-1.0$\sim$5.0&1st&-1.12$^{+0.08}_{-0.08}$&500$^{+200}_{-140}$&CPL&n0,n1,n3,b0&Fluence&2015/443\\
160509374&5.0$\sim$30.0&2nd&-0.89$^{+0.01}_{-0.01}$&382$^{+10}_{-10}$&CPL&n0,n1,n3,b0&Fluence&3509/443\\
160625945&0.0$\sim$2.0&1st&-0.29$^{+0.13}_{-0.13}$&40$^{+4}_{-4}$&CPL&n6,n7,n9,b1&PeakFlux&1400/439\\
160625945&180.0$\sim$192.0&2nd&-0.68$^{+0.01}_{-0.01}$&565$^{+12}_{-12}$&CPL&n6,n7,n9,b1&PeakFlux&4308/439\\
160625945&192.0$\sim$198.0&3rd&-0.63$^{+0.01}_{-0.01}$&324$^{+5}_{-5}$&CPL&n6,n7,n9,b1&PeakFlux&3318/439\\
160625945&198.0$\sim$215.0&4th&-0.71$^{+0.01}_{-0.01}$&352$^{+6}_{-6}$&CPL&n6,n7,n9,b1&PeakFlux&3947/439\\
160802259&0.0$\sim$7.0&1st&-0.53$^{+0.03}_{-0.03}$&210$^{+10}_{-10}$&CPL&n2,b0&PeakFlux&1220/225\\
160802259&14.0$\sim$20.0&2nd&-0.88$^{+0.06}_{-0.06}$&126$^{+15}_{-14}$&CPL&n2,b0&PeakFlux&1099/225\\
160816730&0.0$\sim$7.0&1st&-0.80$^{+0.04}_{-0.04}$&163$^{+12}_{-11}$&CPL&n6,n7,n9,b1&PeakFlux&2169/441\\
160816730&7.0$\sim$17.0&2nd&-0.64$^{+0.03}_{-0.03}$&182$^{+9}_{-8}$&CPL&n6,n7,n9,b1&PeakFlux&2430/441\\
170808936&0.0$\sim$13.5&1st&-0.88$^{+0.01}_{-0.01}$&2350$^{+9}_{-8}$&CPL&n1,n2,n5,b0&Fluence+Peak&2729/442\\
170808936&13.5$\sim$20.0&2nd&-0.98$^{+0.01}_{-0.01}$&280$^{+9}_{-9}$&CPL&n1,n2,n5,b0&Fluence+Peak&2328/442\\
171102107&-2.0$\sim$3.0&1st&-0.48$^{+0.19}_{-0.19}$&37$^{+7}_{-6}$&CPL&n0,n1,n2,b0&Fluence&2061/442\\
171102107&3.0$\sim$10.0&2nd&-1.05$^{+0.29}_{-0.29}$&160$^{+17}_{-80}$&CPL&n0,n1,n2,b0&Fluence&2264/442\\
171102107&25.0$\sim$42.0&3rd&-1.22$^{+0.09}_{-0.09}$&530$^{+300}_{190}$&CPL&n0,n1,n2,b0&Fluence&2869/442\\
171102107&42.0$\sim$62.0&4th&-0.83$^{+0.03}_{-0.03}$&141$^{+8}_{-7}$&CPL&n0,n1,n2,b0&Fluence&3083/442\\
171119992&0.0$\sim$20.0&1st&-0.93$^{+0.20}_{-0.20}$&290$^{+150}_{-100}$&CPL&n2,b0&PeakFlux&1399/225\\
171119992&20.0$\sim$35.0&2nd&-1.21$^{+0.24}_{-0.24}$&210$^{+160}_{-90}$&CPL&n2,b0&PeakFlux&1301/225\\
171120556&0.0$\sim$7.0&1st&-0.86$^{+0.07}_{-0.07}$&209$^{+35}_{-30}$&CPL&n0,n1,n3,b0&PeakFlux&2371/442\\
171120556&15.4$\sim$20.0&2nd&-1.50$^{+0.00}_{-0.00}$&125$^{+1}_{-1}$&CPL&n0,n1,n3,b0&PeakFlux&2001/442\\
171227000&0.0$\sim$24.0&1st&-0.71$^{+0.01}_{-0.01}$&884$^{+30}_{-29}$&CPL&n5,b0&PeakFlux&1687/223\\
171227000&24.0$\sim$60.0&2nd&-1.05$^{+0.03}_{-0.03}$&346$^{+34}_{-31}$&CPL&n5,b0&PeakFlux&1580/223\\
180113011&0.0$\sim$2.8&1st&-0.83$^{+0.03}_{-0.03}$&630$^{+70}_{-60}$&CPL&n3,n4,b0&PeakFlux&1342/326\\
180113011&2.8$\sim$5.4&2nd&-0.75$^{+0.04}_{-0.04}$&460$^{+50}_{-50}$&CPL&n3,n4,b0&PeakFlux&1328/326\\
180113418&0.0$\sim$19.6&1st&-0.73$^{+0.01}_{-0.01}$&388$^{+12}_{11}$&CPL&n1,n2,n9,b0&PeakFlux&3402/443\\
180113418&19.6$\sim$33.2&2nd&-0.74$^{+0.02}_{-0.02}$&223$^{+9}_{-8}$&CPL&n1,n2,n9,b0&PeakFlux&2930/443\\
180120207&0.0$\sim$13.0&1st&-0.96$^{+0.02}_{-0.02}$&147$^{+6}_{-6}$&CPL&n9,na,nb,b1&PeakFlux&2830/438\\
180120207&13.0$\sim$30.0&2nd&-1.19$^{+0.02}_{-0.02}$&160$^{+7}_{-7}$&CPL&n9,na,nb,b1&PeakFlux&2998/438\\
\enddata 
\end{deluxetable*}\label{Table:Total}

\clearpage 
\begin{deluxetable*}{cc|cc|ccc|cccc}
\rotate
\tablewidth{0pt}
\setlength{\tabcolsep}{0.35em}
\tabletypesize{\scriptsize}
\tablecaption{Results of the Time-resolved Spectral Fits of GRB 140206B}
\tablehead{
\multirow{3}{0.8cm}{$t_{1}$$\sim$$t_{2}$} 
&\multirow{3}{1cm}{S/N} 
&\multicolumn{2}{|c}{Blackbody Fitting}
&\multicolumn{3}{|c}{Cutoff Power-Law Fitting}
&\multicolumn{4}{|c}{Band Function Fitting}\\
\cline{3-11}
&
&\colhead{kT}
&\colhead{-ln(posterior)/AIC/BIC/DIC}
&\colhead{$\alpha$}
&\colhead{$E_{\rm c}$}
&\colhead{-ln(posterior)/AIC/BIC/DIC}
&\colhead{$\alpha$}
&\colhead{$E_{\rm pk}$}
&\colhead{$\beta$}
&\colhead{-ln(posterior)/AIC/BIC/DIC}\\
(s)
&&
(keV)
&&&
(keV)
&&&
(keV)
}
\startdata
\hline
1st\\
\hline
-1.00$\sim$-0.04&2.73&...&...&...&...&...&...&...&...&...\\
-0.04$\sim$0.46&13.87&55.23$^{+2.52}_{-2.48}$&-599/1202/1210/1231&0.53$^{+0.12}_{-0.11}$&93.5$^{+5.0}_{-5.4}$&-609/1225/1237/1212&0.34$^{+0.21}_{-0.20}$&256.2$^{+19.2}_{-18.9}$&-6.77$^{+2.20}_{-2.24}$&-614/1237/1253/1227\\
0.46$\sim$1.81&9.94&45.05$^{+2.71}_{-2.73}$&-1179/2361/2369/2391&0.35$^{+0.15}_{-0.14}$&88.5$^{+8.4}_{-8.3}$&-1191/2388/2401/2370&0.47$^{+0.38}_{-0.39}$&206.2$^{+30.6}_{-34.0}$&5.47$^{+2.94}_{-3.07}$&-1196/2400/2416/2370\\
1.81$\sim$4.70&5.50&44.15$^{+5.26}_{-5.18}$&-1676/3356/3364/3387&0.23$^{+0.20}_{-0.19}$&87.3$^{+9.4}_{-9.8}$&-1692/3390/3403/3376&0.03$^{+0.60}_{-0.56}$&444.4$^{+123.6}_{-269.4}$&-5.78$^{+2.69}_{-2.81}$&-1696/3399/3416/2851\\
\hline
2nd\\
\hline
4.70$\sim$5.81&2.58&...&...&...&...&...&...&...&...&...\\
5.81$\sim$7.63&17.32&...&...&...&...&...&...&...&...&...\\
7.63$\sim$8.79&24.76&...&...&-0.84$^{+0.05}_{-0.05}$&1458.8$^{+284.4}_{-298.5}$&-1254/2514/2526/2492&-0.66$^{+0.07}_{-0.07}$&885.5$^{+153.3}_{-157.9}$&-2.09$^{+0.15}_{-0.13}$&-1237/2482/2498/2473\\
8.79$\sim$11.84&59.56&...&...&-0.85$^{+0.03}_{-0.03}$&623.9$^{+52.9}_{-52.9}$&-2058/4122/4135/4103&-0.71$^{+0.04}_{-0.04}$&458.1$^{+39.1}_{-39.1}$&-1.94$^{+0.05}_{-0.05}$&-2013/4035/4051/4026\\
11.84$\sim$12.57&38.80&...&...&-0.92$^{+0.03}_{-0.03}$&729.5$^{+85.4}_{-87.9}$&-991/1988/2000/1966&-0.92$^{+0.03}_{-0.03}$&772.8$^{+73.9}_{-73.2}$&-6.61$^{+2.29}_{-2.28}$&-984/1976/1992/1963\\
12.57$\sim$13.24&54.54&...&...&-0.61$^{+0.04}_{-0.04}$&324.5$^{+21.8}_{-21.8}$&-995/1997/2009/1979&-0.59$^{+0.04}_{-0.04}$&440.2$^{+24.1}_{-23.7}$&-4.93$^{+2.03}_{-2.82}$&-989/1987/2004/1975\\
13.24$\sim$14.60&94.85&...&...&-0.68$^{+0.02}_{-0.02}$&314.6$^{+13.8}_{-13.4}$&-1608/3221/3234/3203&-0.58$^{+0.03}_{-0.03}$&336.1$^{+16.2}_{-16.0}$&-2.32$^{+0.08}_{-0.08}$&-1581/3170/3187/3159\\
14.60$\sim$15.82&76.58&...&...&-0.81$^{+0.03}_{-0.02}$&306.6$^{+17.3}_{-17.4}$&-1434/2875/2887/2855&-0.68$^{+0.04}_{-0.04}$&278.5$^{+20.7}_{-20.6}$&-2.25$^{+0.10}_{-0.10}$&-1414/2836/2853/2826\\
15.82$\sim$17.34&68.64&...&...&-0.85$^{+0.03}_{-0.03}$&252.7$^{+15.9}_{-15.9}$&-1529/3064/3077/3045&-0.60$^{+0.06}_{-0.06}$&189.1$^{+13.5}_{-13.3}$&-2.11$^{+0.06}_{-0.06}$&-1501/3009/3026/3000\\
17.34$\sim$18.84&56.33&...&...&-0.97$^{+0.04}_{-0.03}$&247.0$^{+20.0}_{-20.6}$&-1487/2980/2993/2960&-0.66$^{+0.07}_{-0.07}$&149.3$^{+11.1}_{-10.9}$&-2.05$^{+0.05}_{-0.05}$&1456/2920/2937/2912\\
18.84$\sim$19.62&31.30&...&...&-1.05$^{+0.06}_{-0.06}$&285.8$^{+50.3}_{-49.8}$&-939/1884/1897/1863&-0.90$^{+0.13}_{-0.13}$&203.2$^{+50.2}_{-45.3}$&-3.18$^{+1.14}_{-1.81}$&-927/1861/1878/1845\\
19.62$\sim$25.00&60.64&...&...&-1.23$^{+0.03}_{-0.03}$&256.9$^{+23.6}_{-23.8}$&-2307/4620/4632/4598&-1.04$^{+0.05}_{-0.05}$&129.1$^{+10.0}_{-10.0}$&-2.10$^{+0.06}_{-0.05}$&-2272/4552/4569/4546\\
\hline
3rd\\
\hline
25.00$\sim$27.66&36.38&...&...&-1.30$^{+0.05}_{-0.05}$&270.2$^{+40.8}_{-40.8}$&-1764/3534/3546/3512&-1.09$^{+0.09}_{-0.08}$&120.2$^{+14.5}_{-15.3}$&-2.12$^{+0.09}_{-0.08}$&-1745/3498/3515/3492\\
27.66$\sim$29.21&41.52&...&...&-0.89$^{+0.05}_{-0.05}$&177.5$^{+15.9}_{-16.2}$&-1436/2878/2891/2860&-0.87$^{+0.05}_{-0.05}$&192.8$^{+12.1}_{-11.9}$&-5.87$^{+2.62}_{-2.77}$&-1429/2866/2882/2857\\
29.21$\sim$31.53&62.81&...&...&-0.92$^{+0.03}_{-0.03}$&160.5$^{+10.4}_{-10.2}$&-1751/3509/3521/3490&-0.90$^{+0.04}_{-0.04}$&166.9$^{+8.7}_{-9.1}$&-4.90$^{+2.07}_{-3.00}$&-1743/3493/3510/3484\\
31.53$\sim$33.11&39.98&...&...&-1.27$^{+0.05}_{-0.05}$&240.5$^{+31.9}_{-32.3}$&-1398/2801/2814/2779&-1.24$^{+0.06}_{-0.06}$&162.8$^{+18.4}_{-20.5}$&-5.09$^{+2.66}_{-3.12}$&-1385/2778/2795/2772\\
33.11$\sim$34.99&33.28&...&...&-0.88$^{+0.05}_{-0.05}$&175.7$^{+16.4}_{-16.1}$&-1436/2878/2891/2860&-1.23$^{+0.07}_{-0.07}$&159.7$^{+20.4}_{-21.1}$&-4.62$^{+2.27}_{-3.11}$&-1510/3027/3044/3022\\
34.99$\sim$38.15&33.63&...&...&-1.16$^{+0.07}_{-0.07}$&142.3$^{+19.0}_{-18.8}$&-1859/3724/3737/3704&-1.13$^{+0.07}_{-0.07}$&114.1$^{+8.2}_{-8.4}$&-5.50$^{+2.61}_{-2.95}$&-1848/3705/3722/3700\\
38.15$\sim$43.48&30.88&...&...&-1.42$^{+0.08}_{-0.08}$&133.7$^{+22.5}_{-23.2}$&-2174/4354/4367/4332&-1.33$^{+0.10}_{-0.10}$&69.5$^{+6.5}_{-6.7}$&-3.96$^{+1.46}_{-2.56}$&-2159/4327/4343/4321\\
43.48$\sim$50.00&13.14&...&...&...&...&...&...&...&...&...\\
\enddata
\end{deluxetable*}\label{Table:140206B}

\clearpage
\begin{deluxetable*}{cc|cc|cccc}
\tablewidth{0pt}
\setlength{\tabcolsep}{0.35em}
\tabletypesize{\footnotesize}
\tablecaption{Results of the Time-resolved Spectral Fits of GRB 140329B}
%\tablenum{4}
\tablehead{
\multirow{3}{0.8cm}{$t_{1}$$\sim$$t_{2}$} 
&\multirow{3}{1cm}{S/N} 
&\multicolumn{2}{|c}{Blackbody Fitting}
&\multicolumn{3}{|c}{Cutoff Power-Law Fitting}\\
\cline{3-7}
&
&\colhead{kT}
&\colhead{-ln(posterior)/AIC/BIC/DIC}
&\colhead{$\alpha$}
&\colhead{$E_{\rm c}$}
&\colhead{-ln(posterior)/AIC/BIC/DIC}\\
(s)
&&
(keV)
&&&
(keV)
}
\startdata
\hline
The 1st Sub-Burst\\
\hline
-1.00$\sim$-0.12&0.69&...&...&...&...&...\\
-0.12$\sim$-0.01&8.09&...&...&...&...&...\\
-0.01$\sim$0.25&39.54&40.15$^{+1.00}_{-1.00}$&-509/1022/1030/1037&-0.47$^{+0.07}_{-0.07}$&160.1$^{+14.9}_{-14.8}$&-377/760/771/747&\\
0.25$\sim$0.47&20.46&20.62$^{+0.80}_{-0.81}$&-221/446/454/458&-0.56$^{+0.17}_{-0.16}$&76.7$^{+13.5}_{-13.7}$&-202/409/421/392&\\
0.47$\sim$0.67&9.51&13.75$^{+1.79}_{-1.81}$&-145/294/302/303&Unconstrained&Unconstrained&Unconstrained&\\
0.67$\sim$3.00&4.17&12.19$^{+2.42}_{-2.43}$&-1114/2231/2239/2237&Unconstrained&Unconstrained&Unconstrained&\\
\hline
The 2nd Sub-Burst\\
\hline
19.00$\sim$19.53&2.97&...&...&...&...&...\\
19.53$\sim$20.06&31.94&...&...&-0.97$^{+0.05}_{-0.05}$&734.8$^{+138.2}_{-138.4}$&-561/1128/1140/1112\\
20.06$\sim$20.93&56.75&...&...&-0.86$^{+0.04}_{-0.04}$&282.7$^{+24.3}_{-24.3}$&-848/1702/1714/1688\\
20.93$\sim$21.17&41.04&...&...&-0.96$^{+0.06}_{-0.06}$&293.0$^{+38.3}_{-39.3}$&-321/647/659/632\\
21.17$\sim$22.33&110.39&...&...&-0.91$^{+0.02}_{-0.02}$&266.0$^{+13.0}_{-13.2}$&-1171/2348/2360/2334\\
22.33$\sim$22.77&52.77&...&...&-1.00$^{+0.04}_{-0.04}$&256.8$^{+25.9}_{-25.4}$&-557/1120/1132/1105\\
22.77$\sim$22.80&19.71&...&...&...&...&...\\
\hline
The 3rd Sub-Burst\\
\hline
22.80$\sim$23.64&94.55&...&...&-0.93$^{+0.02}_{-0.02}$&325.6$^{+18.6}_{-18.8}$&-982/1970/1982/1955\\
23.64$\sim$23.95&92.21&...&...&-0.75$^{+0.03}_{-0.03}$&440.5$^{+25.5}_{-25.7}$&-613/1232/1244/1217\\
23.95$\sim$24.10&59.71&...&...&-0.86$^{+0.04}_{-0.04}$&405.0$^{+36.7}_{-37.3}$&-308/622/634/606\\
24.10$\sim$24.14&38.36&...&...&-0.67$^{+0.07}_{-0.07}$&312.0$^{+40.3}_{-41.0}$&-22/49/61/33\\
24.14$\sim$24.34&76.44&...&...&-0.90$^{+0.03}_{-0.03}$&389.9$^{+32.9}_{-32.7}$&-408/822/833/805\\
24.34$\sim$24.61&73.29&...&...&-0.86$^{+0.04}_{-0.04}$&268.7$^{+20.0}_{-20.0}$&-457/919/931/904\\
24.61$\sim$25.00&71.01&...&...&-0.94$^{+0.04}_{-0.04}$&266.2$^{+22.2}_{-22.2}$&-579/1164/1176/1149\\
25.00$\sim$25.28&47.44&...&...&-1.01$^{+0.06}_{-0.05}$&242.6$^{+30.7}_{-30.4}$&-371/749/760/733\\
25.28$\sim$25.86&52.16&...&...&-1.12$^{+0.05}_{-0.05}$&217.5$^{+25.2}_{-26.0}$&-591/1188/1200/1172\\
25.86$\sim$26.59&43.36&...&...&-1.02$^{+0.07}_{-0.07}$&161.5$^{+21.3}_{-21.1}$&-684/1374/1386/1359\\
26.59$\sim$27.63&34.66&...&...&-1.16$^{+0.08}_{-0.08}$&168.0$^{+28.4}_{-28.4}$&-768/1542/1553/1526\\
27.63$\sim$28.70&20.92&...&...&-1.16$^{+0.15}_{-0.15}$&127.6$^{+33.9}_{-37.0}$&-785/1576/1588/1550\\
\enddata
\end{deluxetable*}\label{Table:140329B}

\clearpage
\begin{deluxetable*}{cc|cc|ccc}
\tablewidth{0pt}
\tabletypesize{\footnotesize}
\tablecaption{Results of the Time-resolved Spectral Fits of GRB 150330A}
%\tablenum{5}
\tablehead{
\multirow{3}{0.8cm}{$t_{1}$$\sim$$t_{2}$} 
&\multirow{3}{1cm}{S/N} 
&\multicolumn{2}{|c}{Blackbody Fitting}
&\multicolumn{3}{|c}{Cutoff Power-Law Fitting}\\
\cline{3-7}
&
&\colhead{kT}
&\colhead{-ln(posterior)/AIC/BIC/DIC}
&\colhead{$\alpha$}
&\colhead{$E_{\rm c}$}
&\colhead{-ln(posterior)/AIC/BIC/DIC}\\
(s)
&&
(keV)
&&&
(keV)
}
\startdata
\hline
The 1st Sub-Burst\\
\hline
-1.00$\sim$-0.17&1.95&...&...&...&...&...\\
-0.17$\sim$1.39&16.58&57.64$^{+2.42}_{-2.44}$&-1371/2745/2754/2776&0.25$^{+0.19}_{-0.20}$&130.9$^{+22.4}_{-20.5}$&-1382/2771/2783/2679\\
1.39$\sim$3.16&36.78&67.30$^{+1.76}_{-1.77}$&-1664/3331/3340/3361&-0.25$^{+0.07}_{-0.07}$&246.9$^{+23.0}_{-23.3}$&-1567/3140/3152/3124\\
3.16$\sim$4.43&20.37&41.04$^{+1.56}_{-1.56}$&-1287/2578/2587/2606&-0.29$^{+0.12}_{-0.11}$&145.8$^{+21.6}_{-20.9}$&-1272/2551/2563/2533\\
4.43$\sim$6.20&50.40&43.18$^{+0.83}_{-0.82}$&-1787/3578/3586/3603&-0.40$^{+0.05}_{-0.05}$&175.7$^{+13.5}_{-13.6}$&-1581/3168/3180/3153\\
6.20$\sim$7.63&22.11&30.95$^{+1.08}_{-1.09}$&-1326/2656/2664/2681&-0.20$^{+0.14}_{-0.14}$&88.3$^{+12.2}_{-11.9}$&-1319/2644/2656/2622\\
7.63$\sim$10.00&12.71&17.83$^{+1.08}_{-1.08}$&-1653/3309/3318/3332&-0.81$^{+0.26}_{-0.24}$&107.3$^{+21.7}_{-43.6}$&-1654/3314/3326/3224\\
\hline
The 2nd Sub-Burst\\
\hline
123.50$\sim$124.51&8.70&...&...&...&...&...\\
124.51$\sim$125.37&18.72&...&...&...&...&...\\
125.37$\sim$127.19&42.29&...&...&-0.85$^{+0.05}_{-0.05}$&433.7$^{+59.3}_{-59.0}$&-1589/3185/3197/3162\\
127.19$\sim$128.90&73.12&...&...&-0.91$^{+0.03}_{-0.03}$&654.4$^{+63.9}_{-63.2}$&-1663/3333/3345/3312\\
128.90$\sim$129.34&45.93&...&...&-0.82$^{+0.05}_{-0.05}$&455.2$^{+63.5}_{-64.1}$&-788/1583/1595/1562\\
129.34$\sim$130.08&69.90&...&...&-0.84$^{+0.03}_{-0.03}$&379.7$^{+32.9}_{-33.6}$&-1073/2153/2165/2132\\
130.08$\sim$131.01&96.65&...&...&-0.85$^{+0.02}_{-0.02}$&475.4$^{+32.9}_{-32.9}$&-1371/2749/2761/2127\\
131.01$\sim$131.52&102.21&...&...&-1.05$^{+0.02}_{-0.02}$&1529.5$^{+197.2}_{-197.5}$&-1087/2180/2193/2153\\
131.52$\sim$131.78&58.68&...&...&-1.00$^{+0.04}_{-0.04}$&478.3$^{+74.0}_{-75.4}$&-588/1182/1194/1158\\
131.78$\sim$132.53&81.67&...&...&-0.98$^{+0.03}_{-0.03}$&350.4$^{+29.6}_{-29.8}$&-1108/2223/2235/2200\\
132.53$\sim$133.41&97.59&...&...&-0.95$^{+0.03}_{-0.03}$&322.7$^{+22.4}_{-21.9}$&-1259/2523/2536/2501\\
133.41$\sim$133.89&97.39&...&...&-0.92$^{+0.03}_{-0.03}$&337.8$^{+25.2}_{-25.6}$&-960/1926/1939/1904\\
133.89$\sim$134.25&68.98&...&...&-1.01$^{+0.04}_{-0.04}$&347.5$^{+37.9}_{-37.5}$&-695/1397/1409/1374\\
134.25$\sim$134.88&66.71&...&...&-1.05$^{+0.04}_{-0.04}$&264.2$^{+28.7}_{-30.0}$&-964/1935/1947/1912\\
134.88$\sim$135.70&57.96&...&...&-0.90$^{+0.05}_{-0.05}$&585.5$^{+100.5}_{-101.7}$&-789/1584/1597/1562\\
135.70$\sim$136.55&74.48&...&...&-0.99$^{+0.04}_{-0.04}$&217.8$^{+18.6}_{-18.6}$&-1179/2364/2376/2343\\
136.55$\sim$136.95&58.99&...&...&-0.98$^{+0.05}_{-0.05}$&225.3$^{+25.0}_{-25.6}$&-681/1368/1380/1346\\
136.95$\sim$137.56&87.39&...&...&-0.96$^{+0.03}_{-0.03}$&265.1$^{+21.2}_{-20.8}$&-1025/2056/2069/2034\\
137.56$\sim$138.00&57.35&...&...&-1.10$^{+0.05}_{-0.05}$&278.9$^{+36.1}_{-36.3}$&-702/1409/1422/1386\\
138.00$\sim$139.06&66.38&...&...&-1.03$^{+0.04}_{-0.04}$&198.8$^{+19.4}_{-19.2}$&-1236/2477/2490/2456\\
139.06$\sim$140.95&64.34&...&...&-1.05$^{+0.04}_{-0.04}$&172.4$^{+16.0}_{-15.7}$&-1607/3221/3233/3201\\
140.95$\sim$142.20&39.32&...&...&-1.20$^{+0.06}_{-0.06}$&246.8$^{+45.1}_{-45.8}$&-1252/2510/2523/2488\\
\hline
The 3rd Sub-Burst\\
\hline
142.20$\sim$144.93&50.59&...&...&-1.24$^{+0.05}_{-0.05}$&225.0$^{+32.1}_{-32.8}$&-1825/3656/3669/3635\\
144.93$\sim$145.56&34.18&...&...&-1.07$^{+0.08}_{-0.08}$&210.1$^{+37.9}_{-39.5}$&-858/1723/1735/1701\\
145.56$\sim$145.70&23.86&...&...&-1.30$^{+0.13}_{-0.12}$&927.6$^{+197.5}_{-612.5}$&-192/391/403/338\\
145.70$\sim$146.93&86.01&...&...&-0.98$^{+0.03}_{-0.03}$&267.2$^{+20.6}_{-20.7}$&-1427/2859/2872/2838\\
146.93$\sim$148.17&70.48&...&...&-1.09$^{+0.04}_{-0.04}$&226.9$^{+23.6}_{-23.2}$&-1373/2753/2765/2731\\
148.17$\sim$150.52&78.14&...&...&-1.14$^{+0.04}_{-0.04}$&191.2$^{+17.0}_{-17.2}$&-1797/3599/3612/3578\\
150.52$\sim$151.31&34.03&...&...&-1.28$^{+0.09}_{-0.09}$&245.6$^{+60.9}_{-63.9}$&-960/1927/1939/1902\\
151.31$\sim$153.26&36.32&...&...&-1.21$^{+0.07}_{-0.07}$&185.2$^{+33.0}_{-33.4}$&-1568/3141/3154/3120\\
153.26$\sim$155.00&23.30&...&...&-1.39$^{+0.12}_{-0.12}$&370.7$^{+132.8}_{-174.8}$&-1446/2897/2909/2862\\
\enddata
\end{deluxetable*}\label{Table:150330A}

\clearpage
%\startlongtable
\begin{deluxetable*}{cc|cc|ccc}
\tablewidth{0pt}
\tabletypesize{\footnotesize}
\tablecaption{Results of the Time-resolved Spectral Fits of GRB 160625B}
%\tablenum{6}
\tablehead{
\multirow{3}{0.8cm}{$t_{1}$$\sim$$t_{2}$} 
&\multirow{3}{1cm}{S/N} 
&\multicolumn{2}{|c}{Blackbody Fitting}
&\multicolumn{3}{|c}{Cutoff Power-Law Fitting}\\
\cline{3-7}
&
&\colhead{kT}
&\colhead{-ln(posterior)/AIC/BIC/DIC}
&\colhead{$\alpha$}
&\colhead{$E_{\rm c}$}
&\colhead{-ln(posterior)/AIC/BIC/DIC}\\
(s)
&&
(keV)
&&&
(keV)
}
\startdata
\hline
The 1st Sub-Burst\\
\hline
-1.00$\sim$-0.12&2.72&...&...&...&...&...\\
-0.12$\sim$0.04&11.51&15.67$^{+0.80}_{-0.80}$&-106/217/225/234&-0.21$^{+0.27}_{-0.28}$&38.9$^{+7.9}_{-7.4}$&-115/236/249/181\\
0.04$\sim$0.21&21.50&14.28$^{+0.44}_{-0.43}$&-211/426/434/440&-0.22$^{+0.19}_{-0.18}$&36.0$^{+4.7}_{-4.7}$&-203/412/424/386\\
0.21$\sim$0.34&29.48&18.03$^{+0.43}_{-0.42}$&-170/343/352/359&-0.18$^{+0.14}_{-0.13}$&44.2$^{+4.2}_{-4.4}$&-139/283/296/265\\
0.34$\sim$0.68&36.63&16.15$^{+0.29}_{-0.29}$&-535/1073/1082/1089&-0.09$^{+0.12}_{-0.12}$&37.0$^{+3.1}_{-3.0}$&-489/983/996/966\\
0.68$\sim$0.93&15.92&12.45$^{+0.49}_{-0.49}$&-255/514/522/529&-0.29$^{+0.24}_{-0.26}$&31.4$^{+5.3}_{-5.2}$&-256/519/531/481\\
0.93$\sim$2.00&2.83&6.71$^{+1.26}_{-1.31}$&-990/1985/1993/1995&Unconstrained&Unconstrained&Unconstrained\\
\hline
The 2nd Sub-Burst\\
\hline
187.00$\sim$187.36&27.16&...&...&-0.91$^{+0.03}_{-0.03}$&2056.0$^{+353.7}_{-351.8}$&-724/1453/1466/1430\\
187.36$\sim$187.88&50.06&...&...&-0.85$^{+0.05}_{-0.05}$&1459.3$^{+356.0}_{-367.7}$&-1109/2223/2236/2196\\
187.88$\sim$188.09&52.80&...&...&-0.84$^{+0.02}_{-0.02}$&1581.2$^{+163.9}_{-163.0}$&-638/1283/1295/1258\\
188.09$\sim$188.23&53.09&...&...&-0.83$^{+0.03}_{-0.03}$&1728.5$^{+191.1}_{-190.0}$&-496/998/1010/972\\
188.23$\sim$188.35&62.07&...&...&-0.79$^{+0.02}_{-0.02}$&1472.5$^{+120.3}_{-122.2}$&-459/924/937/899\\
188.35$\sim$188.71&132.93&...&...&-0.76$^{+0.01}_{-0.01}$&1255.0$^{+53.0}_{-52.6}$&-1491/2989/3001/2963\\
188.71$\sim$189.48&222.36&...&...&-0.71$^{+0.01}_{-0.01}$&798.3$^{+22.6}_{-23.0}$&-2630/5266/5279/5242\\
189.48$\sim$189.88&156.48&...&...&-0.68$^{+0.01}_{-0.01}$&595.1$^{+22.1}_{-22.2}$&-1570/3145/3157/3121\\
189.88$\sim$190.27&129.50&...&...&-0.71$^{+0.02}_{-0.02}$&561.6$^{+24.2}_{-24.2}$&-1366/2739/2751/2716\\
190.27$\sim$190.64&105.03&...&...&-0.72$^{+0.02}_{-0.02}$&485.0$^{+23.1}_{-23.3}$&-1152/2310/2322/2287\\
190.64$\sim$191.46&131.97&...&...&-0.70$^{+0.02}_{-0.02}$&377.7$^{+13.4}_{-13.1}$&-1684/3374/3387/3353\\
191.46$\sim$192.98&149.30&...&...&-0.74$^{+0.01}_{-0.01}$&325.7$^{+9.8}_{-9.7}$&-2204/4414/4427/4393\\
192.98$\sim$193.69&124.78&...&...&-0.64$^{+0.02}_{-0.02}$&321.4$^{+10.6}_{-10.5}$&-1504/3015/3027/2994\\
193.69$\sim$194.24&124.26&...&...&-0.64$^{+0.02}_{-0.02}$&394.8$^{+13.5}_{-13.5}$&-1412/2830/2843/2809\\
194.24$\sim$195.41&195.01&...&...&-0.64$^{+0.01}_{-0.01}$&390.3$^{+8.6}_{-8.5}$&-2342/4691/4703/4669\\
195.41$\sim$195.91&121.36&...&...&-0.65$^{+0.02}_{-0.02}$&378.3$^{+13.4}_{-13.5}$&-1360/2727/2739/2706\\
195.91$\sim$196.69&138.84&...&...&-0.65$^{+0.01}_{-0.01}$&359.2$^{+10.8}_{-10.8}$&-1698/3402/3414/3381\\
196.69$\sim$197.26&90.53&...&...&-0.73$^{+0.02}_{-0.02}$&334.9$^{+15.6}_{-15.4}$&-1252/2509/2522/2488\\
197.26$\sim$197.92&117.08&...&...&-0.68$^{+0.02}_{-0.02}$&418.7$^{+14.8}_{-14.9}$&-1442/2890/2902/2868\\
197.92$\sim$198.30&98.33&...&...&-0.68$^{+0.02}_{-0.02}$&438.9$^{+19.2}_{-19.3}$&-1050/2105/2118/2084\\
198.30$\sim$198.45&50.87&...&...&-0.70$^{+0.04}_{-0.04}$&337.7$^{+26.7}_{-26.8}$&-420/846/858/825\\
198.45$\sim$198.71&53.71&...&...&-0.74$^{+0.03}_{-0.03}$&318.6$^{+23.4}_{-23.8}$&-691/1387/1400/1366\\
198.71$\sim$198.97&67.75&...&...&-0.64$^{+0.03}_{-0.03}$&390.0$^{+23.1}_{-23.4}$&-732/1471/1483/1450\\
198.97$\sim$199.83&139.11&...&...&-0.65$^{+0.01}_{-0.01}$&412.6$^{+12.0}_{-12.3}$&-1740/3487/3499/3466\\
199.83$\sim$200.11&89.67&...&...&-0.58$^{+0.02}_{-0.02}$&438.8$^{+19.1}_{-19.1}$&-884/1773/1786/1753\\
200.11$\sim$201.20&195.06&...&...&-0.70$^{+0.01}_{-0.01}$&615.2$^{+16.3}_{-16.0}$&-2687/5379/5392/5356\\
201.20$\sim$202.49&178.57&...&...&-0.67$^{+0.01}_{-0.01}$&421.8$^{+9.8}_{-9.6}$&-2303/4613/4625/4591\\
202.49$\sim$203.38&122.69&...&...&-0.74$^{+0.01}_{-0.01}$&398.2$^{+13.8}_{-13.8}$&-1718/3442/3454/3420\\
203.38$\sim$204.65&108.96&...&...&-0.79$^{+0.02}_{-0.02}$&369.8$^{+14.1}_{-14.2}$&-1880/3767/3779/3745\\
204.65$\sim$204.91&63.36&...&...&-0.70$^{+0.03}_{-0.03}$&449.0$^{+29.1}_{-29.2}$&-704/1413/1426/1392\\
204.91$\sim$206.30&113.25&...&...&-0.70$^{+0.02}_{-0.02}$&326.4$^{+10.9}_{-10.9}$&-1924/3855/3867/3835\\
206.30$\sim$207.80&84.56&...&...&-0.76$^{+0.02}_{-0.02}$&253.9$^{+11.3}_{-11.3}$&-1846/3698/3710/3679\\
207.80$\sim$208.65&53.47&...&...&-0.83$^{+0.03}_{-0.04}$&238.1$^{+18.4}_{-18.3}$&-1340/2685/2698/2665\\
208.65$\sim$211.64&80.40&...&...&-0.89$^{+0.02}_{-0.02}$&282.8$^{+13.3}_{-13.6}$&-2277/4560/4572/4540\\
211.64$\sim$212.58&31.33&...&...&-0.99$^{+0.06}_{-0.06}$&224.7$^{+30.5}_{-30.5}$&-1231/2467/2480/2447\\
212.58$\sim$213.88&22.09&...&...&-1.13$^{+0.06}_{-0.06}$&260.5$^{+46.4}_{-47.5}$&-1392/2791/2803/2770\\
213.88$\sim$218.00&22.46&...&...&-1.16$^{+0.06}_{-0.06}$&364.2$^{+77.1}_{-76.7}$&-2185/4377/4389/4356\\
\enddata
\end{deluxetable*}\label{Table:160625B}

\clearpage 
\begin{deluxetable*}{cc|cc|ccc|cc}
\setlength{\tabcolsep}{0.35em}
\tablewidth{0pt}
\tabletypesize{\scriptsize}
\tablecaption{Results of the Time-resolved Spectral Fits of Our Sample (Part I): BB vs. PL+BB}
%\tablenum{8}
\tablehead{
\multirow{3}{0.8cm}{$t_{1}$$\sim$$t_{2}$} 
&\multirow{3}{0.8cm}{S/N} 
&\multicolumn{2}{|c}{Blackbody Fitting}
&\multicolumn{3}{|c}{Power-law Plus Blackbody Fitting}
&\multirow{3}{2cm}{Null Hypothesis Probability}
&\multirow{3}{*}{Support Model}\\
\cline{3-7}
&
&\colhead{kT}
&\colhead{-ln(posterior)/AIC/BIC/DIC}
&\colhead{Index}
&\colhead{kT}
&\colhead{-ln(posterior)/AIC/BIC/DIC}\\
(s)
&&
(keV)
&&&
(keV)
}
\startdata
\hline
140206B\\
\hline
-0.04$\sim$0.46&13.87&55.23$^{+2.52}_{-2.48}$&-599/1202/1210/1231&-2.99$^{+0.53}_{-0.43}$&56.8$^{+2.6}_{-2.6}$&-602/1211/1228/1225&$<10^{-4}$&BB\\
0.46$\sim$1.81&9.94&45.05$^{+2.71}_{-2.73}$&-1179/2361/2369/2391&-3.90$^{+0.91}_{-0.84}$&45.2$^{+2.6}_{-2.6}$&-1182/2371/2388/2396&$<10^{-4}$&BB\\
1.81$\sim$4.70&5.50&44.15$^{+5.26}_{-5.18}$&-1676/3356/3364/3387&-3.87$^{+0.65}_{-0.72}$&45.0$^{+5.3}_{-5.8}$&-1681/3369/3386/3392&$<10^{-4}$&BB\\
\hline
140329B\\
\hline
-0.01$\sim$0.25&39.54&40.15$^{+1.00}_{-1.00}$&-509/1022/1030/1037&-1.54$^{+0.03}_{-0.03}$&42.2$^{+1.8}_{-1.8}$&-394/796/811/798&1&PL+BB\\
0.25$\sim$0.47&20.46&20.62$^{+0.80}_{-0.81}$&-221/446/454/458&-1.73$^{+0.08}_{-0.08}$&21.5$^{+1.6}_{-1.6}$&-197/402/417/402&1&PL+BB\\
0.47$\sim$0.67&9.51&13.75$^{+1.79}_{-1.81}$&-145/294/302/303&-1.90$^{+0.13}_{-0.07}$&15.0$^{+26.2}_{-13.7}$&-109/226/241/-55259&1&PL+BB\\
0.67$\sim$3.00&4.17&12.19$^{+2.42}_{-2.43}$&-1114/2231/2239/2237&-2.11$^{+0.29}_{-0.28}$&5.7$^{+8.7}_{-4.5}$&-1106/2220/2235/-5331&1&PL+BB\\
\hline
150330A\\
\hline
-0.17$\sim$1.39&16.58&57.64$^{+2.42}_{-2.44}$&-1371/2745/2754/2776&-1.69$^{+0.20}_{-0.22}$&59.3$^{+2.9}_{-2.9}$&-1374/2755/2772/2774&$<10^{-4}$&BB\\
1.39$\sim$3.16&36.78&67.30$^{+1.76}_{-1.77}$&-1664/3331/3340/3361&-1.46$^{+0.04}_{-0.04}$&73.8$^{+2.5}_{-2.5}$&-1595/3198/3214/3213&1&PL+BB\\
3.16$\sim$4.43&20.37&41.04$^{+1.56}_{-1.56}$&-1287/2578/2587/2606&-1.49$^{+0.07}_{-0.07}$&43.1$^{+2.2}_{-2.3}$&-1269/2547/2563/2561&1&PL+BB\\
4.43$\sim$6.20&50.40&43.18$^{+0.83}_{-0.82}$&-1787/3578/3586/3603&-1.45$^{+0.02}_{-0.02}$&45.7$^{+1.4}_{-1.4}$&-1612/3232/3249/3244&1&PL+BB\\
6.20$\sim$7.63&22.11&30.95$^{+1.08}_{-1.09}$&-1326/2656/2664/2681&-1.62$^{+0.10}_{-0.10}$&32.3$^{+1.6}_{-1.6}$&-1311/2629/2646/2642&1&PL+BB\\
7.63$\sim$10.00&12.71&17.83$^{+1.08}_{-1.08}$&-1653/3309/3318/3332&-1.70$^{+0.11}_{-0.11}$&18.8$^{+2.1}_{-2.2}$&-1639/3287/3303/3295&1&PL+BB\\
\hline
160625B\\
\hline
-0.12$\sim$0.04&11.51&15.67$^{+0.80}_{-0.80}$&-106/217/225/234&-2.59$^{+0.40}_{-0.38}$&17.5$^{+1.2}_{-1.2}$&-111/230/246/238&$<10^{-4}$&BB\\
0.04$\sim$0.21&21.50&14.28$^{+0.44}_{-0.43}$&-211/426/434/440&-1.94$^{+0.13}_{-0.13}$&15.0$^{+0.9}_{-0.8}$&-200/409/425/405&0.99&PL+BB\\
0.21$\sim$0.34&29.48&18.03$^{+0.43}_{-0.42}$&-170/343/352/359&-2.24$^{+0.15}_{-0.16}$&20.7$^{+0.7}_{-0.7}$&-142/291/308/284&1&PL+BB\\
0.34$\sim$0.68&36.63&16.15$^{+0.29}_{-0.29}$&-535/1073/1082/1089&-2.09$^{+0.10}_{-0.10}$&17.9$^{+0.5}_{-0.5}$&-489/986/1003/982&1&PL+BB\\
0.68$\sim$0.93&15.92&12.45$^{+0.49}_{-0.49}$&-255/514/522/529&-2.42$^{+0.28}_{-0.31}$&14.1$^{+0.9}_{-0.9}$&-254/517/533/509&0.63&PL+BB\\
0.93$\sim$2.00&2.83&6.71$^{+1.26}_{-1.31}$&-990/1985/1993/1995&-3.75$^{+0.99}_{-0.85}$&6.4$^{+1.7}_{-1.5}$&-986/1981/1997/1981&0.98&PL+BB\\
\enddata
\end{deluxetable*}\label{Table:Hypothesis}

\clearpage
\begin{deluxetable*}{c|ccccc|c}
\tablewidth{0pt}
\tabletypesize{\footnotesize}
\tablecaption{Comparison of Different Models of Our Sample (Part I) with the Brightest Time Bin}
%\tablenum{7}
\tablehead{
\colhead{GRB Name} 
&\colhead{Single BB\tablenotemark{a}}
&\colhead{CPL\tablenotemark{b}}
&\colhead{PL+BB\tablenotemark{c}}
&\colhead{CPL+BB\tablenotemark{d}}
&\colhead{Band+BB\tablenotemark{e}}
&\colhead{Best Model Suggested}
}
\colnumbers
\startdata
\hline
140206B\\
\hline
-ln(posterior)&-599&-609&-601&Unconstrained\tablenotemark{f}&Unconstrained&\\
AIC&1202&1225&1211&Unconstrained&Unconstrained&\\
BIC&1210&1237&1227&Unconstrained&Unconstrained&Single BB\\
DIC&1231&1212&1225&Unconstrained&Unconstrained&\\
dof&439&438&437&436&435&\\
\hline
\hline
140329B\\
\hline
-ln(posterior)&-509&-377&-394&-371&Unconstrained&\\
AIC&1022&760&796&753&Unconstrained&\\
BIC&1030&771&811&772&Unconstrained&PL+BB\\
DIC&1037&747&798&750&Unconstrained&\\
dof&324&323&322&321&320&\\
\hline
150330A\\
\hline
-ln(posterior)&-1787&-1582&-1608&-1574&Unconstrained&\\
AIC&3578&3169&3223&3158&Unconstrained&\\
BIC&3586&3182&3240&3178&Unconstrained&CPL+BB\\
DIC&3603&3153&3244&2866&Unconstrained&\\
dof&442&441&440&439&438&\\
\hline
160625B\\
\hline
-ln(posterior)&-535&-489&-489&-479&Unconstrained&\\
AIC&1073&983&986&968&Unconstrained&\\
BIC&1082&996&1003&988&Unconstrained&PL+BB\\
DIC&1089&966&982&970&Unconstrained&\\
dof&437&436&435&434&433&\\
\enddata
\tablenotetext{a}{Single BB model fitting.}
\tablenotetext{b}{Cut power-law model fitting.}
\tablenotetext{c}{Simple power-law plus BB model fitting.}
\tablenotetext{d}{Cut power-law plus BB model fitting.}
\tablenotetext{e}{Band plus BB model fitting.}
\tablenotetext{f}{Parameters cannot be constrained with the model.}
\end{deluxetable*}\label{Table:Comparison}

\clearpage

\clearpage

\clearpage

\clearpage

\clearpage

\clearpage

\clearpage

\clearpage

\clearpage

\clearpage

\clearpage

\clearpage

\clearpage
\appendix
\setcounter{figure}{0}    
\setcounter{section}{0}
\setcounter{table}{0}
\renewcommand{\thesection}{S\arabic{section}}
\renewcommand{\thefigure}{S\arabic{figure}}
\renewcommand{\thetable}{S\arabic{table}}
\renewcommand{\theequation}{S\arabic{equation}}

\begin{figure*}
\includegraphics[angle=0, scale=0.50]{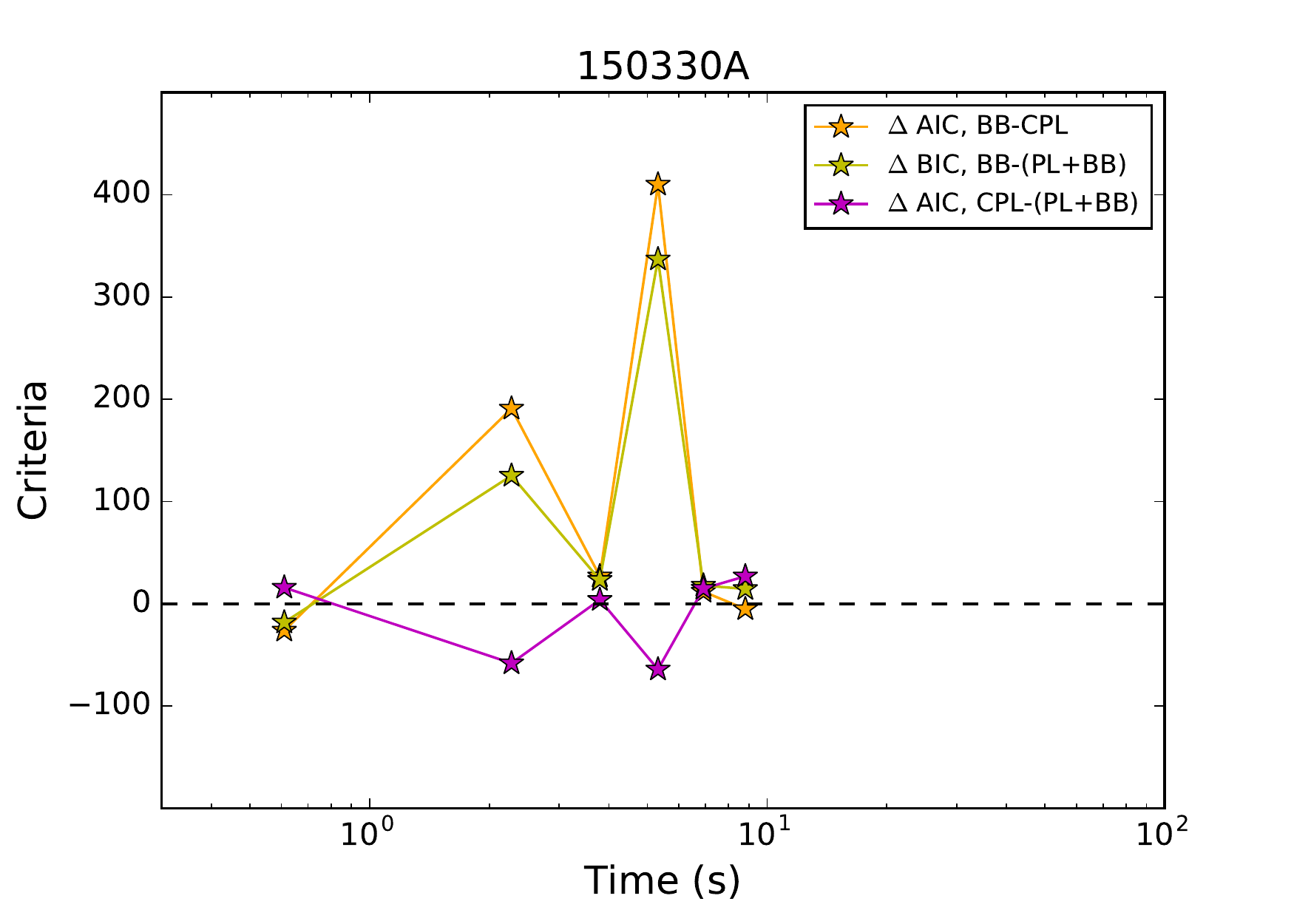}
\includegraphics[angle=0, scale=0.50]{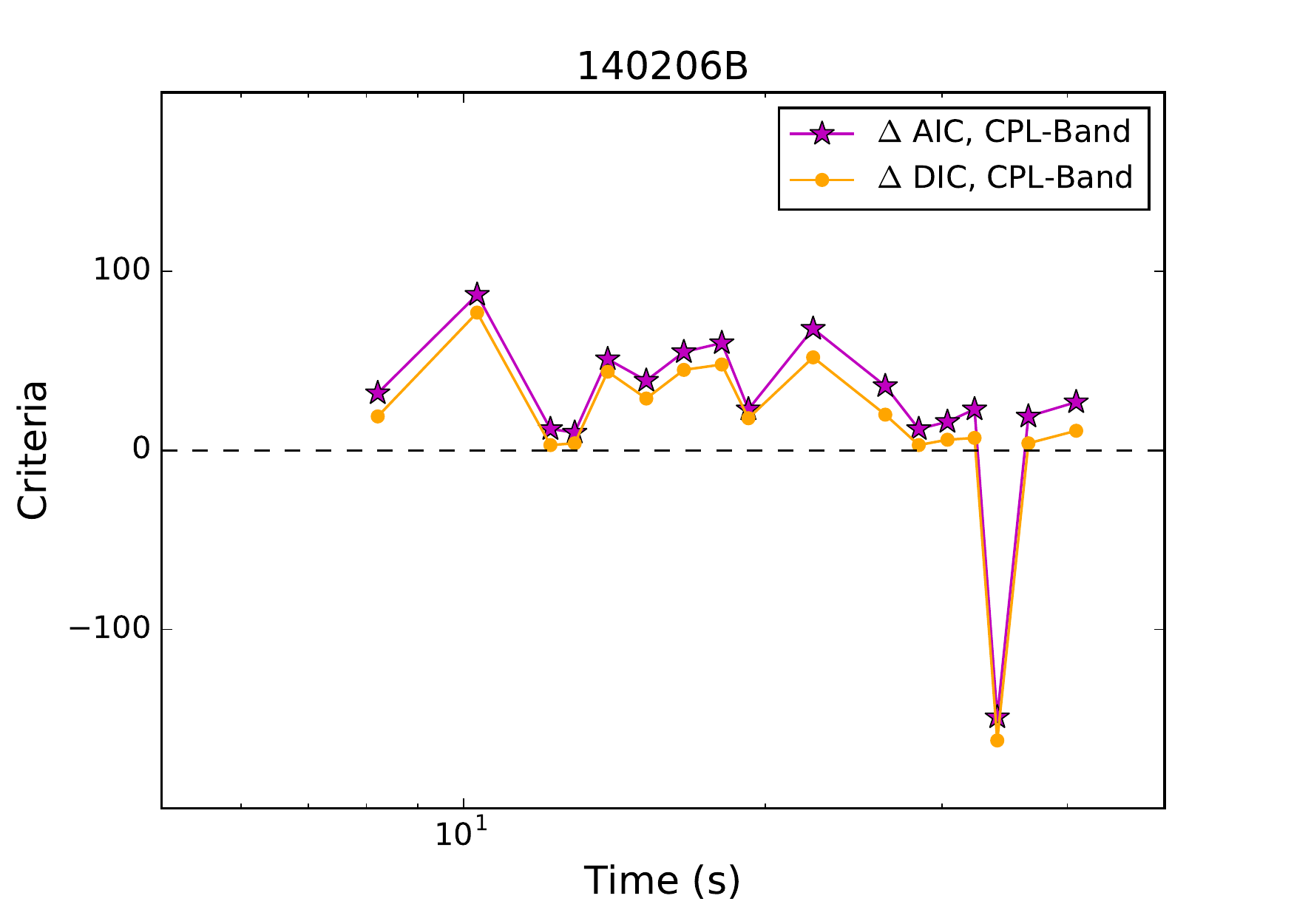}
\caption{Temporal evolution of the $\Delta$AIC/BIC/DIC, which is derived from comparing two different empirical models and based on the time-resolved spectral fitting results. The left-hand panel: data is derived from GRB 150330A (Part I); the right-hand panel: data is derived from   GRB 140206B (Part II). Different color represents different empirical models, and the horizontal dashed lines either represent the value of $\Delta$AIC/BIC/DIC=0.}\label{BIC}
\end{figure*}

\begin{figure*}
\includegraphics[angle=0, scale=0.45]{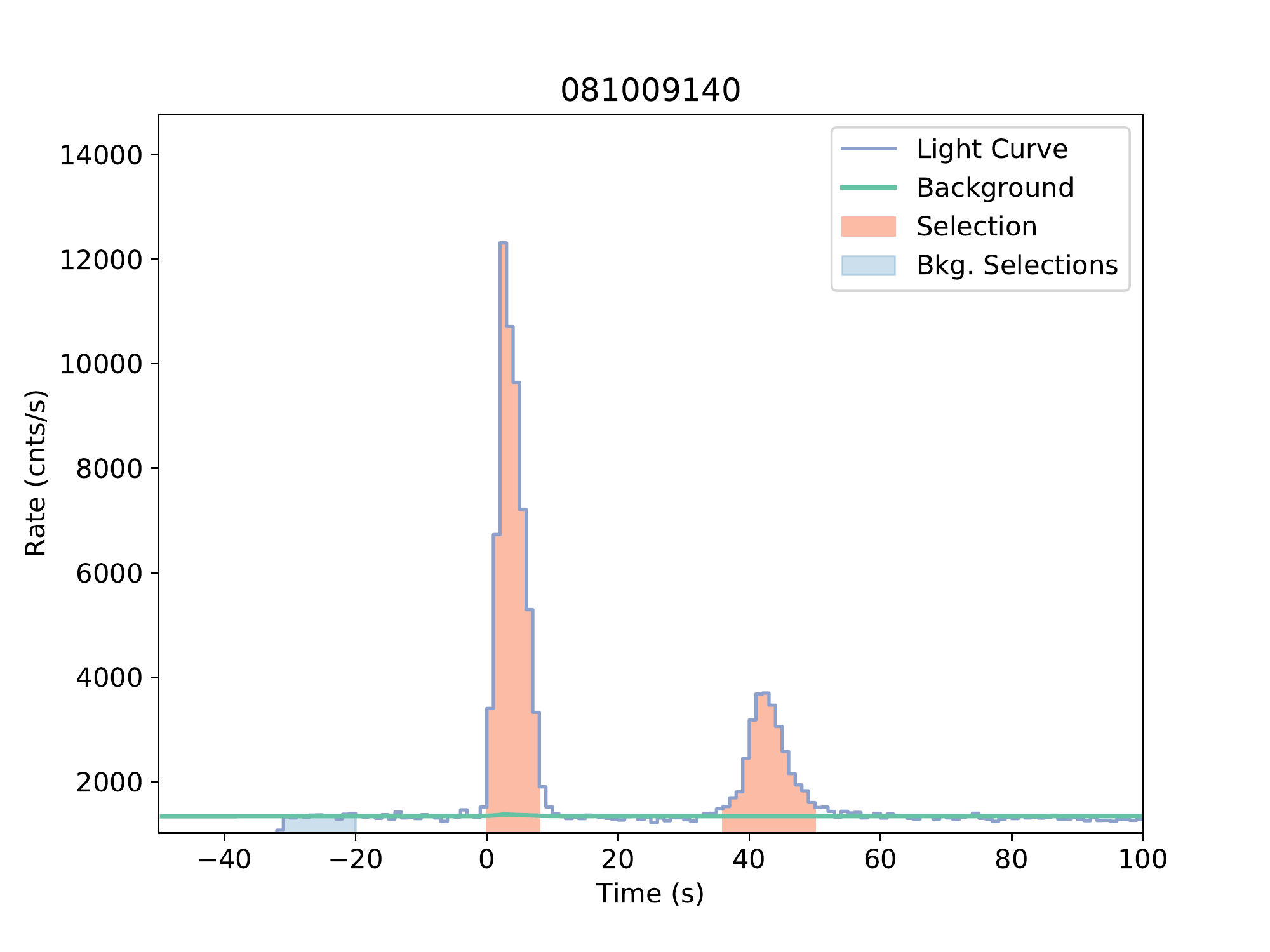}
\includegraphics[angle=0, scale=0.45]{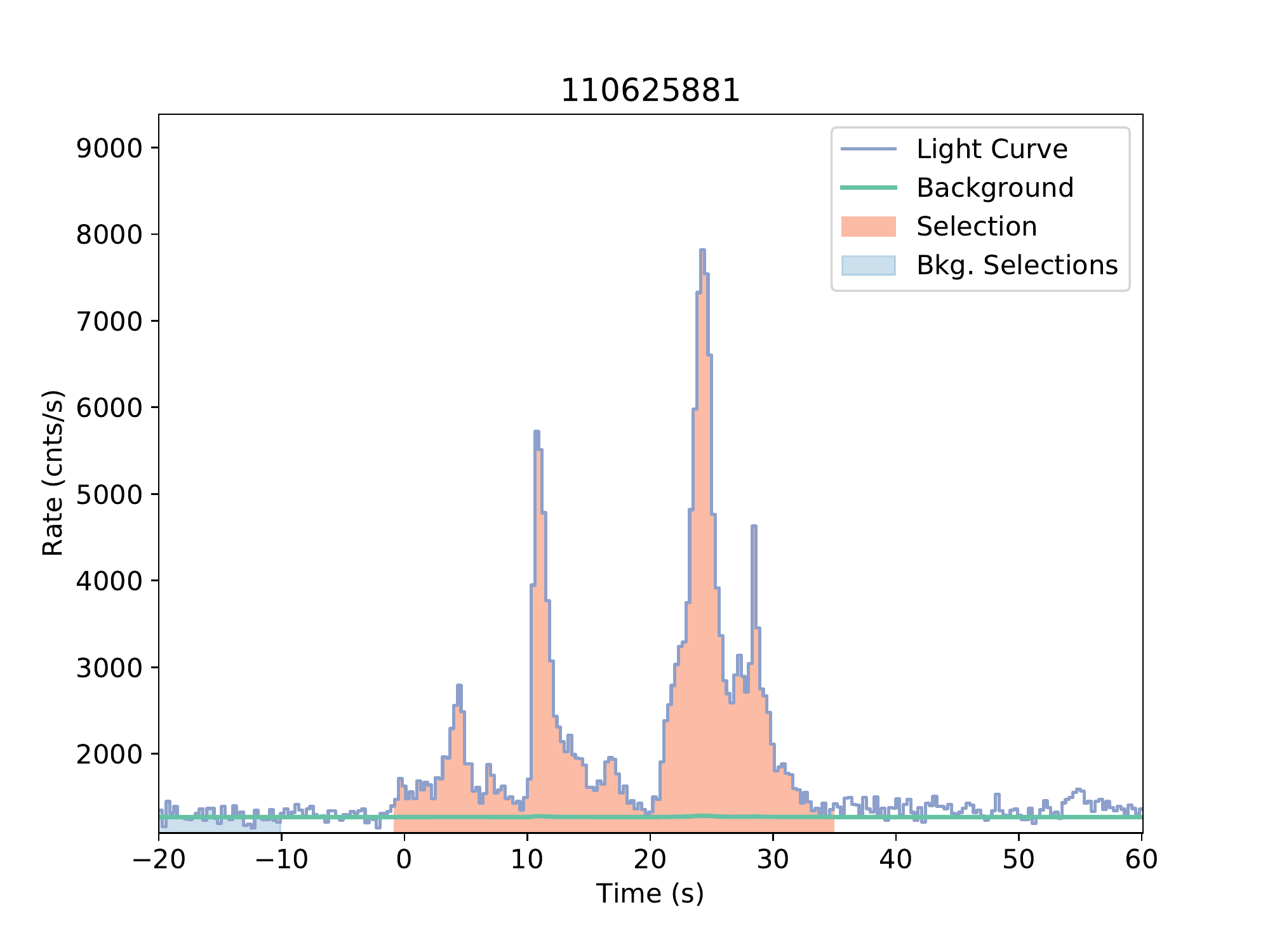}
\includegraphics[angle=0, scale=0.45]{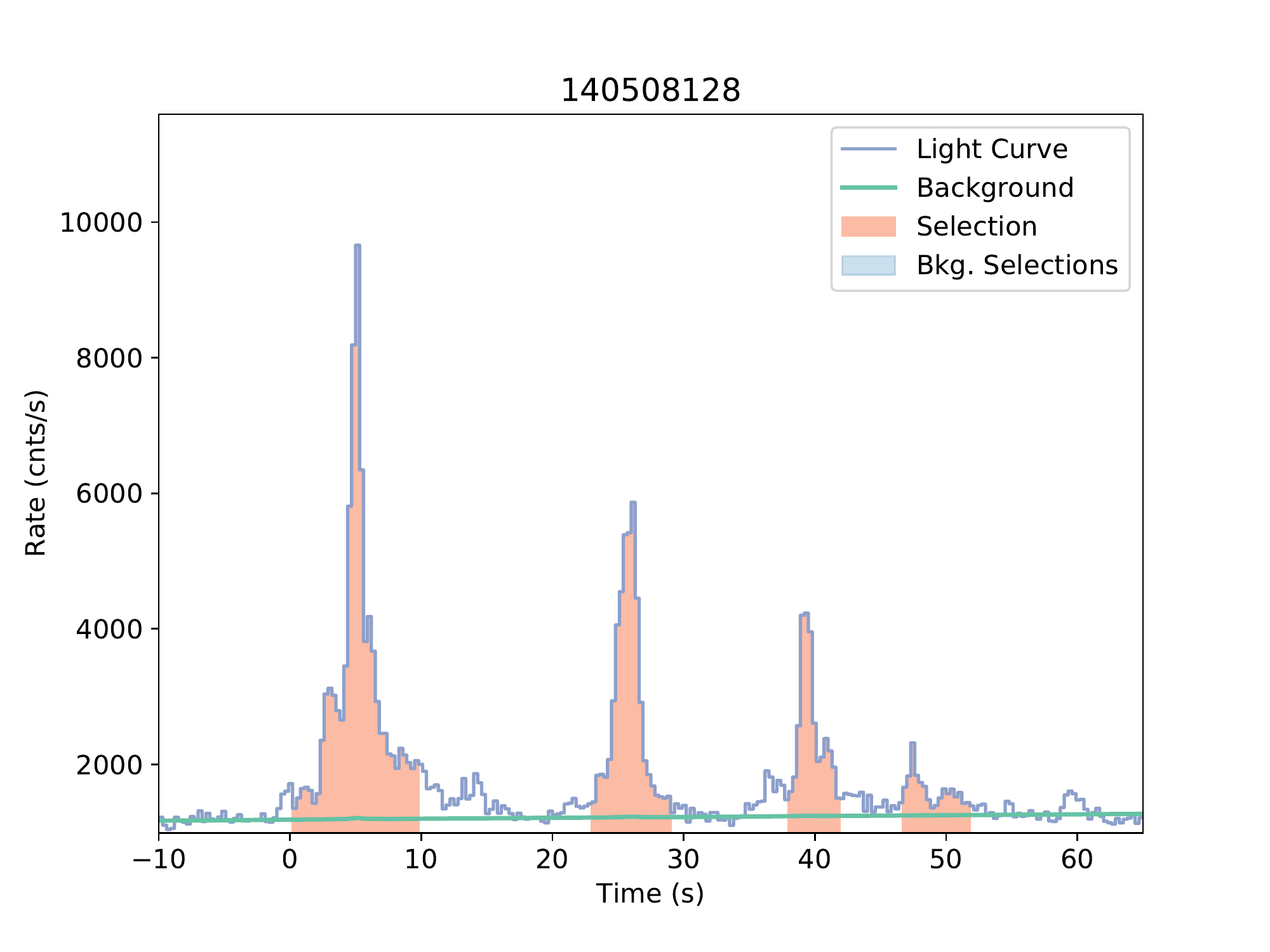}
\includegraphics[angle=0, scale=0.45]{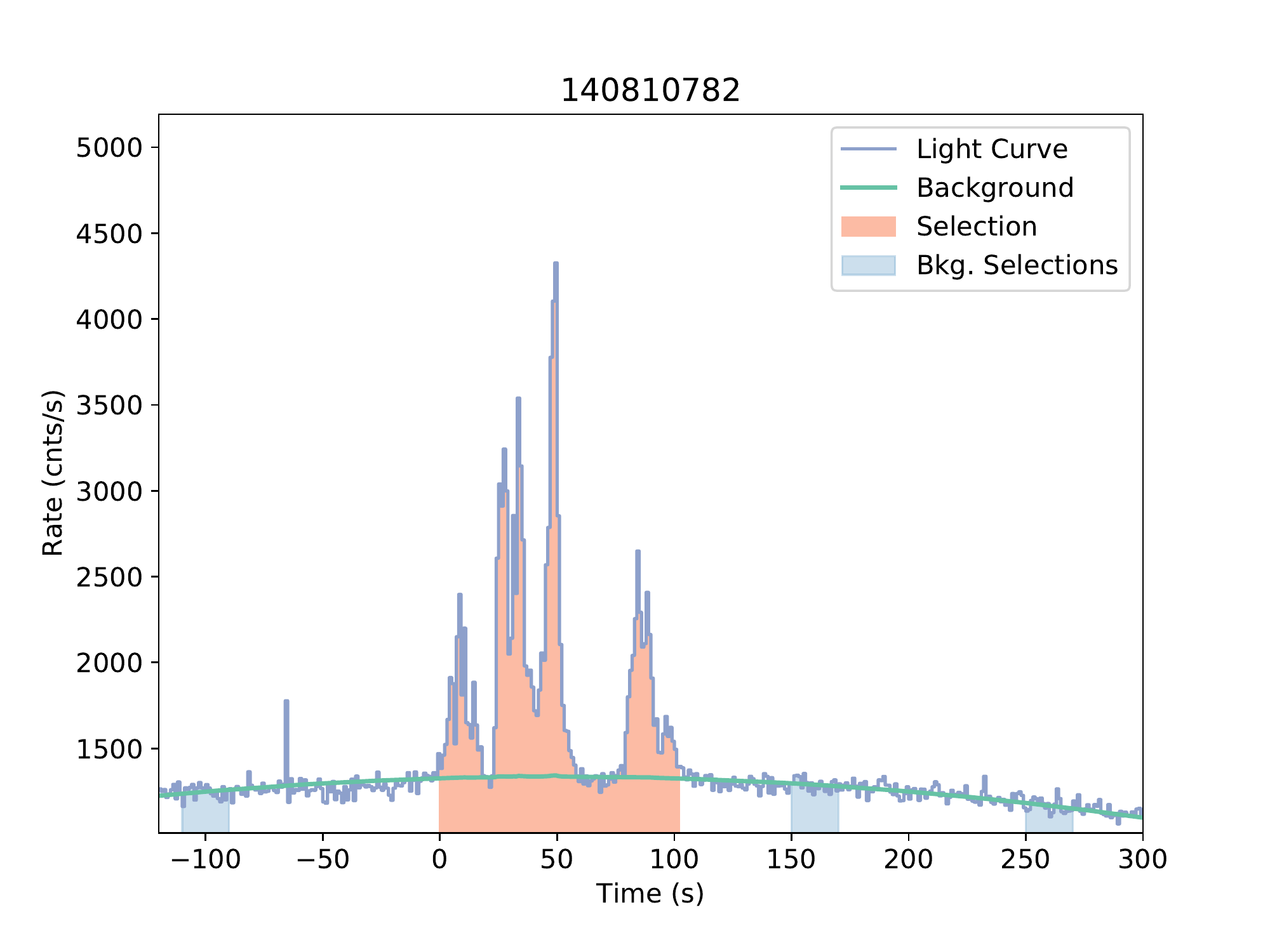}
\includegraphics[angle=0, scale=0.45]{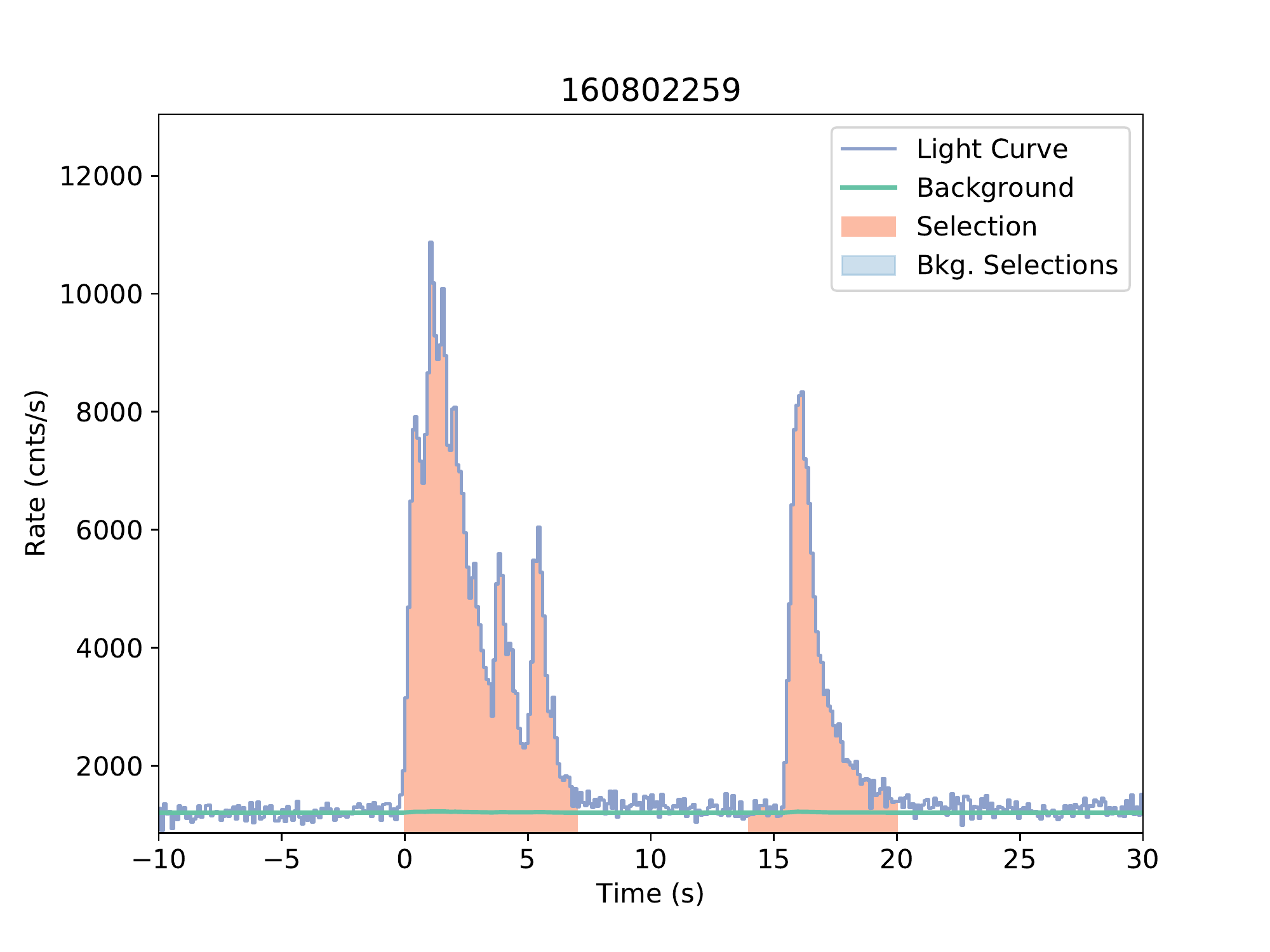}
\includegraphics[angle=0, scale=0.45]{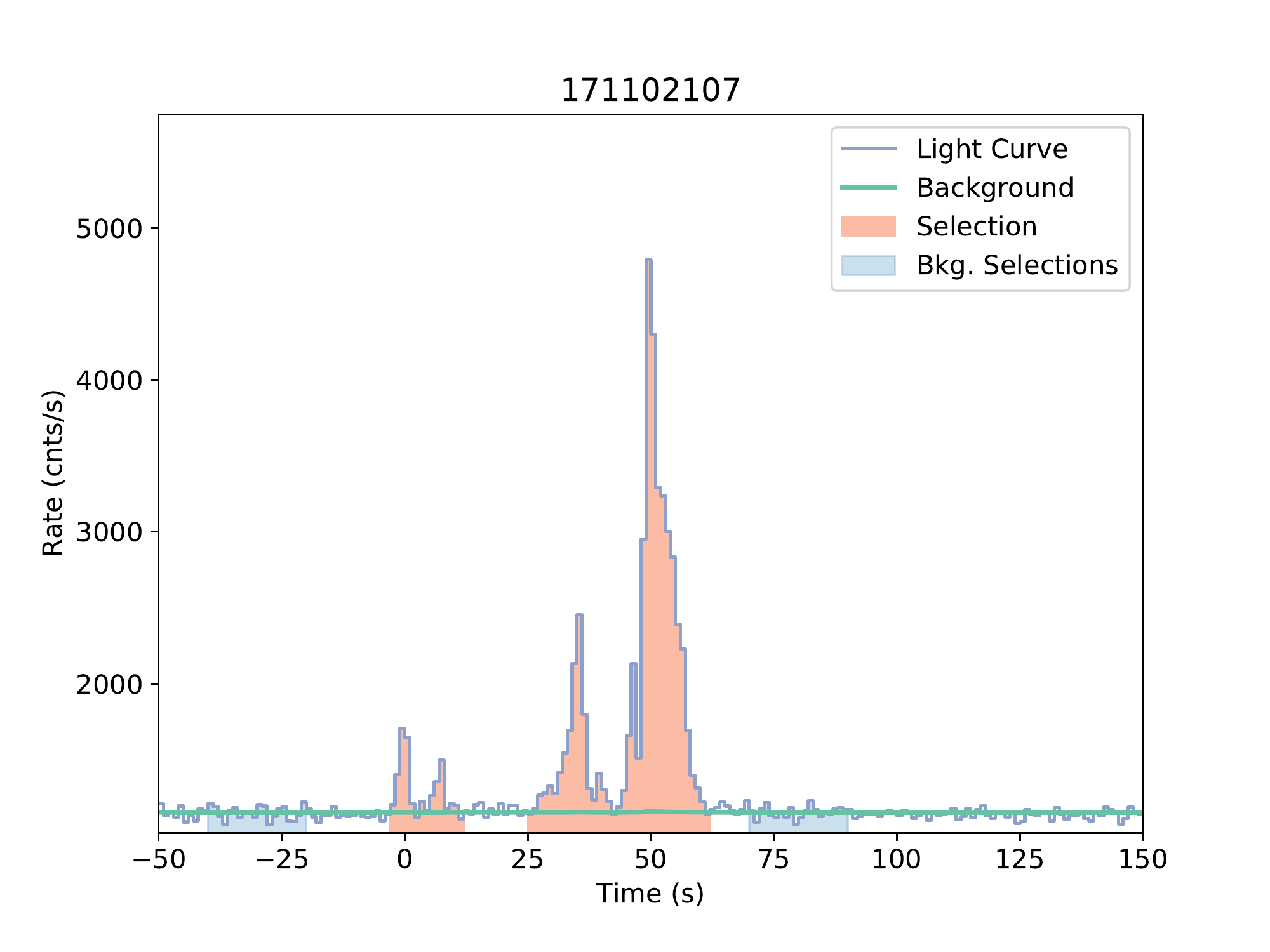}
\caption{Examples of the light curve of the prompt emission, which exhibit clear multi-pulses/sub-bursts in our sample, along with the source selection, the background selections, and the best background fits.}\label{LCs}
\end{figure*}

\begin{figure*}
\includegraphics[angle=0, scale=0.45]{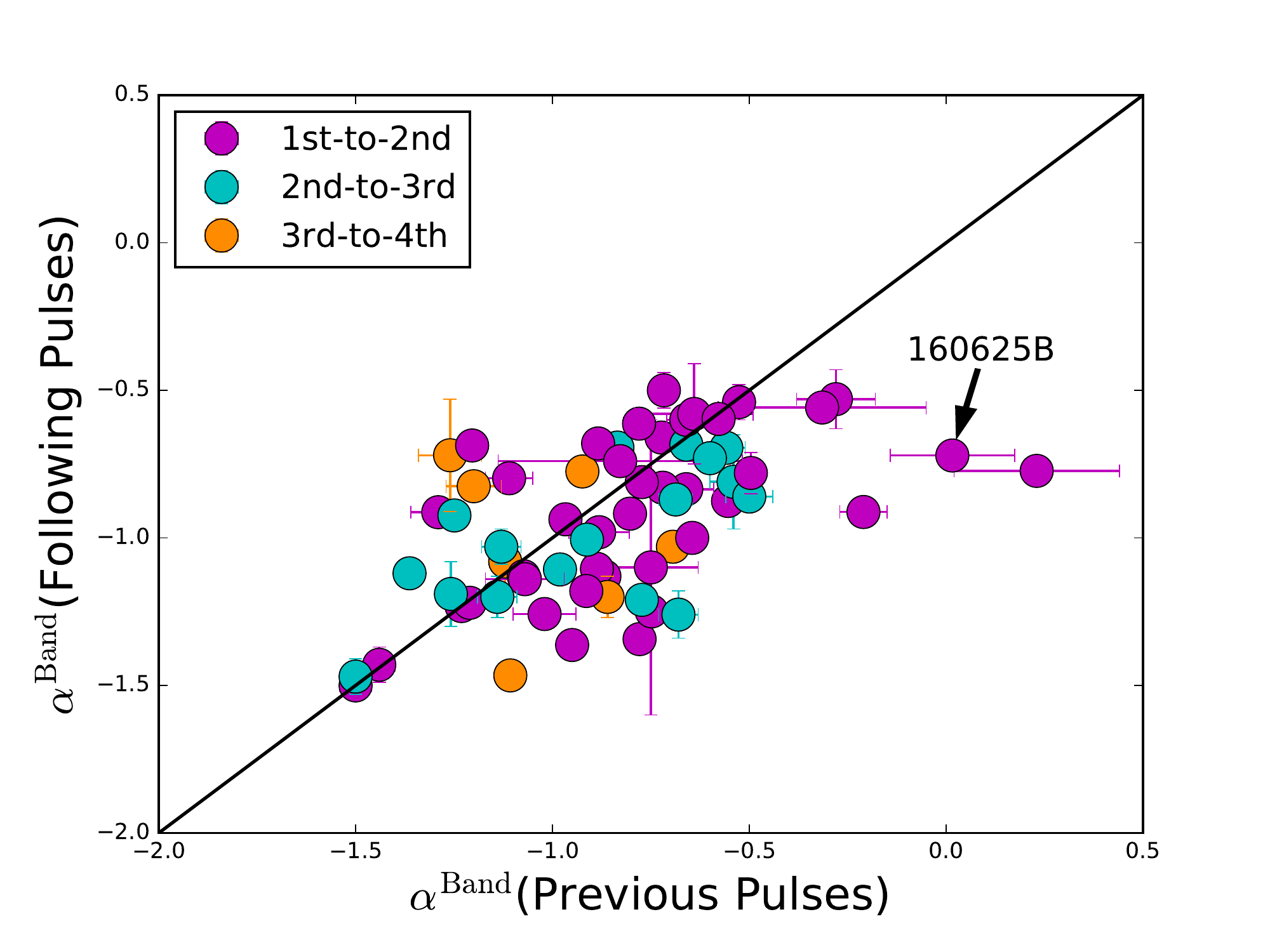}
\includegraphics[angle=0, scale=0.45]{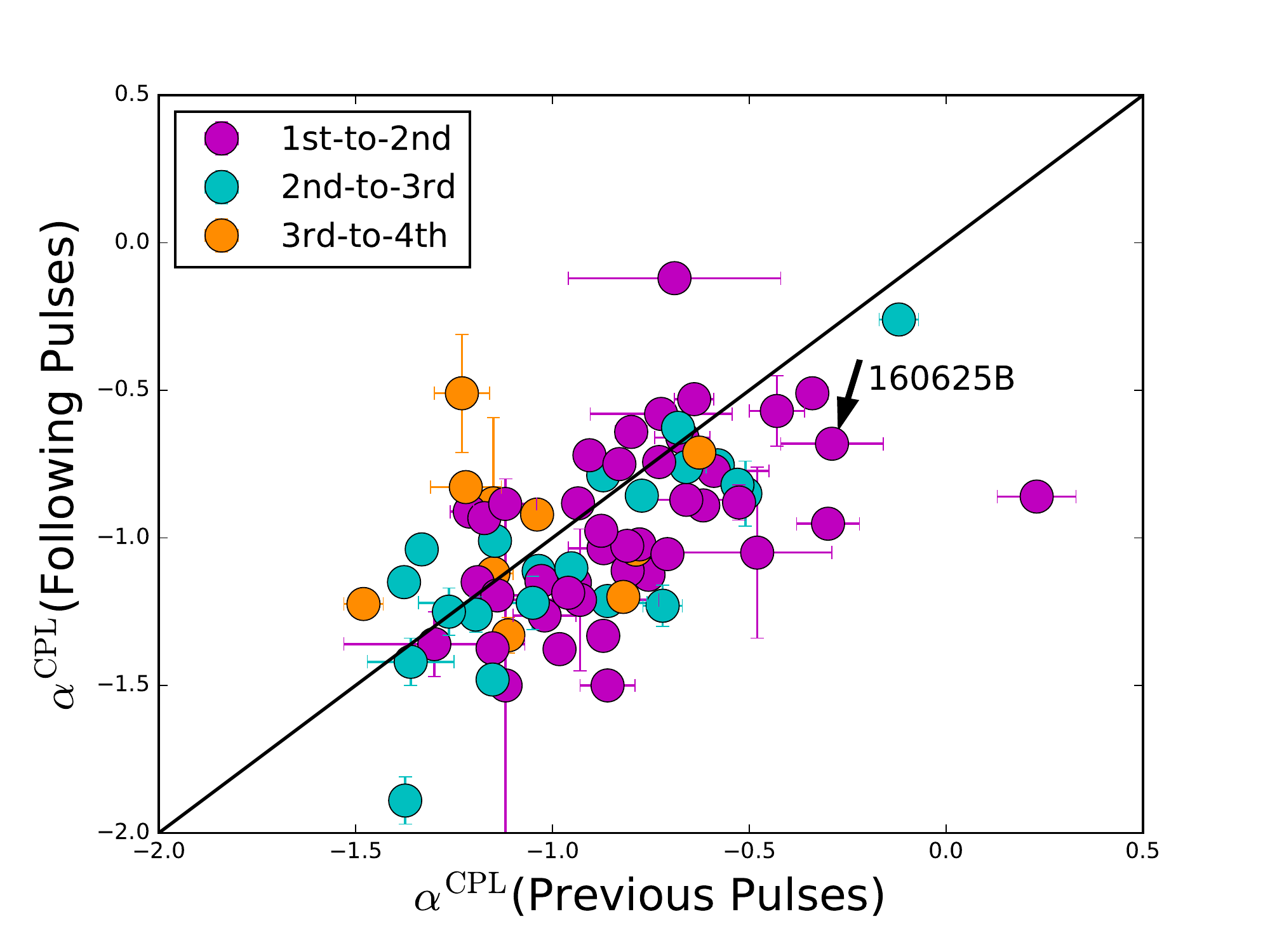}
\includegraphics[angle=0, scale=0.45]{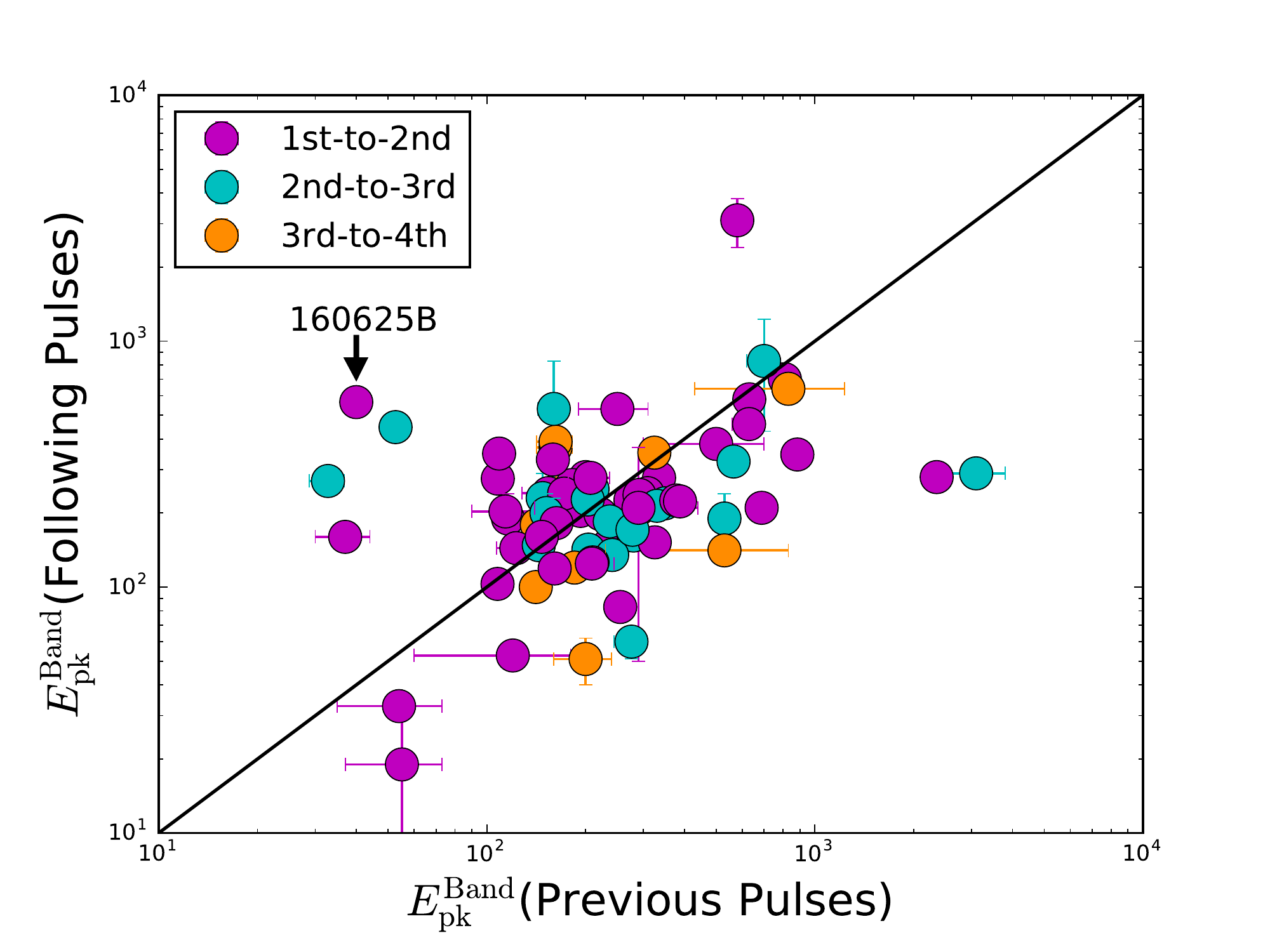}
\includegraphics[angle=0, scale=0.45]{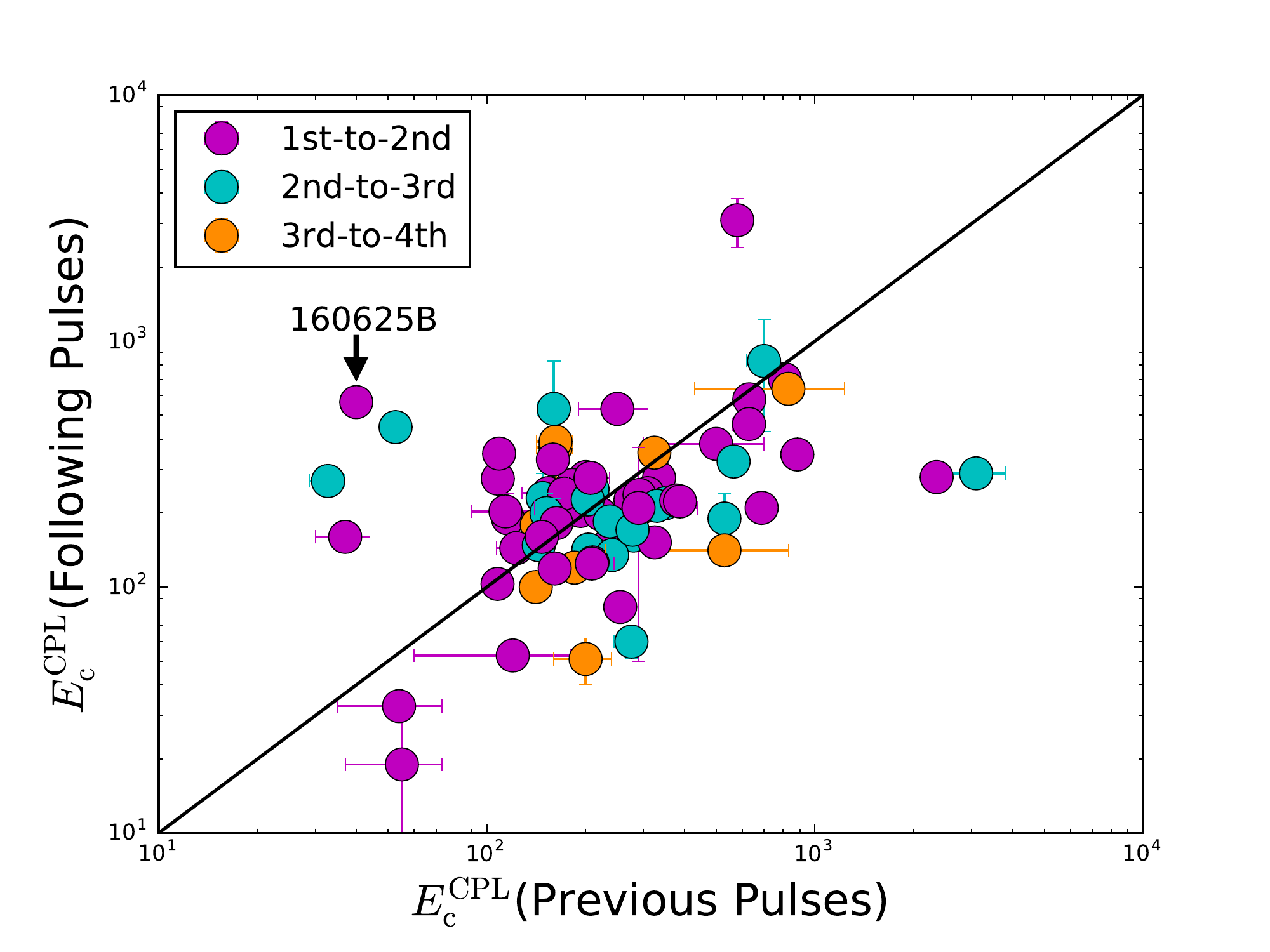}
\caption{Relations of $\alpha$ (top panels) and $E_{\rm pk(c)}$ (bottom panels) for different pulses/sub-bursts: 1st to 2nd (pink), 2nd to 3rd (cyan), and 3rd to 4th (orange), all of which are based on the time-integrated spectral fit results. Top left and bottom left panels are for the Band model, and top right and bottom right panels are for the CPL model. GRB 160625B for the case is labeled with the arrows.}\label{SpectralRelations}
\end{figure*}

\begin{figure*}
\includegraphics[angle=0, scale=0.45]{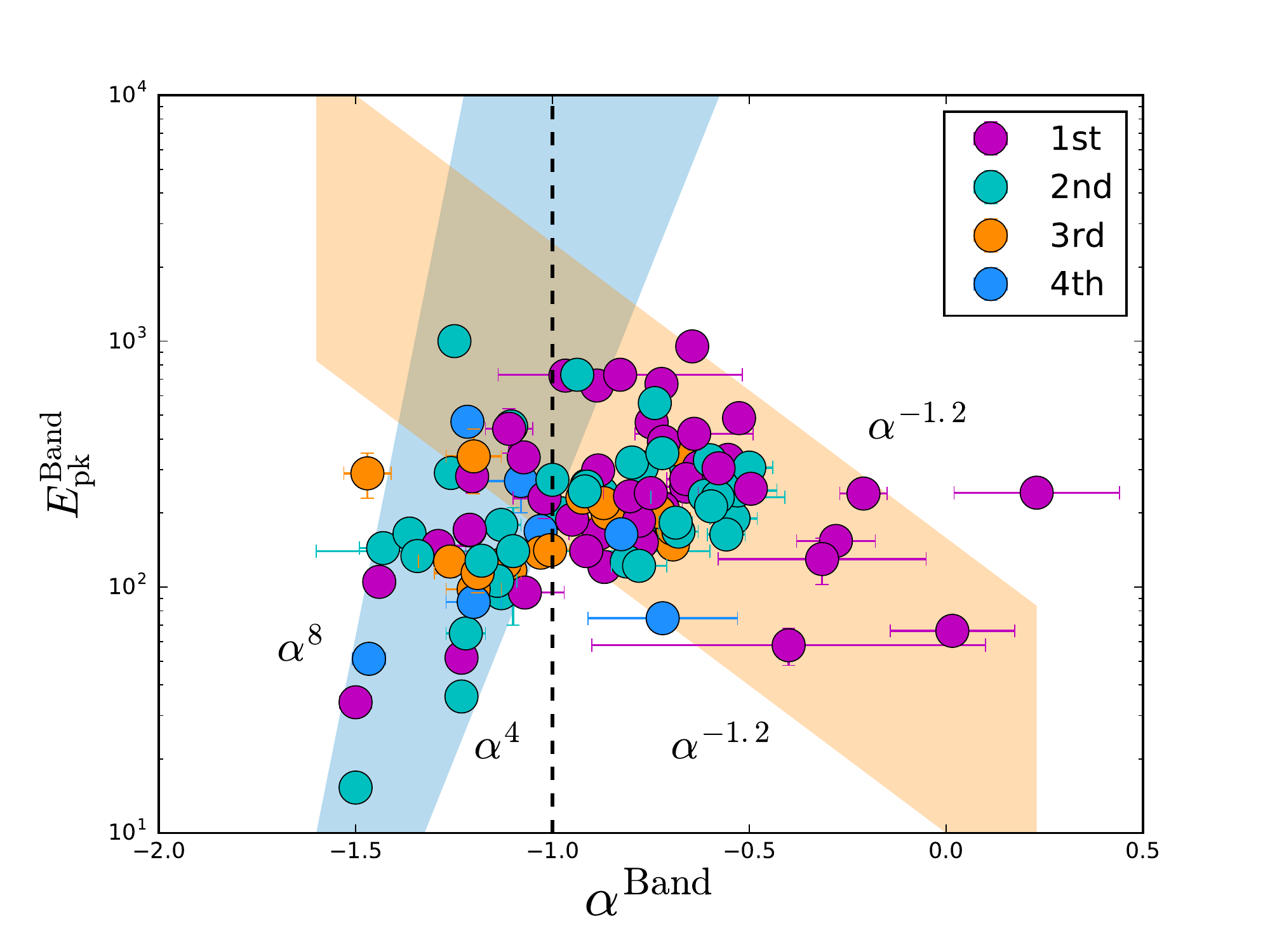}
\includegraphics[angle=0, scale=0.45]{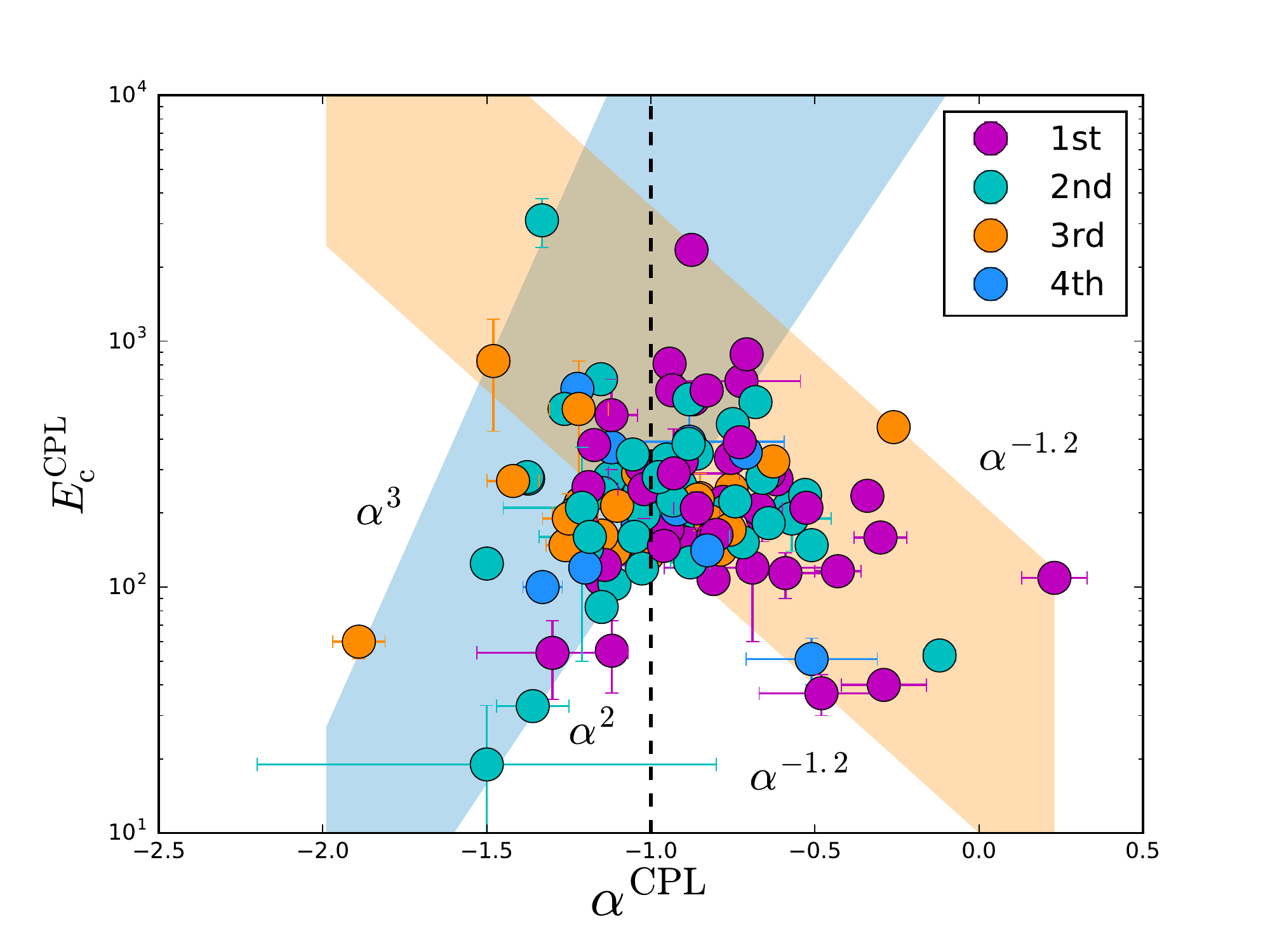}
\caption{$E_{\rm pk}$-$\alpha$ relation for different pulses/sub-bursts: 1st (pink), 2nd (cyan), 3rd (orange), and 4th (blue), which are based on the time-integrated spectral fitting results. One vertical dashed line represents the typical value of $\alpha$=-1.0. The left panel is for the Band model, and the right panel is for the CPL model.}\label{EpAlpha}
\end{figure*}

\begin{figure*}
\centering
\includegraphics[angle=0, scale=1.0]{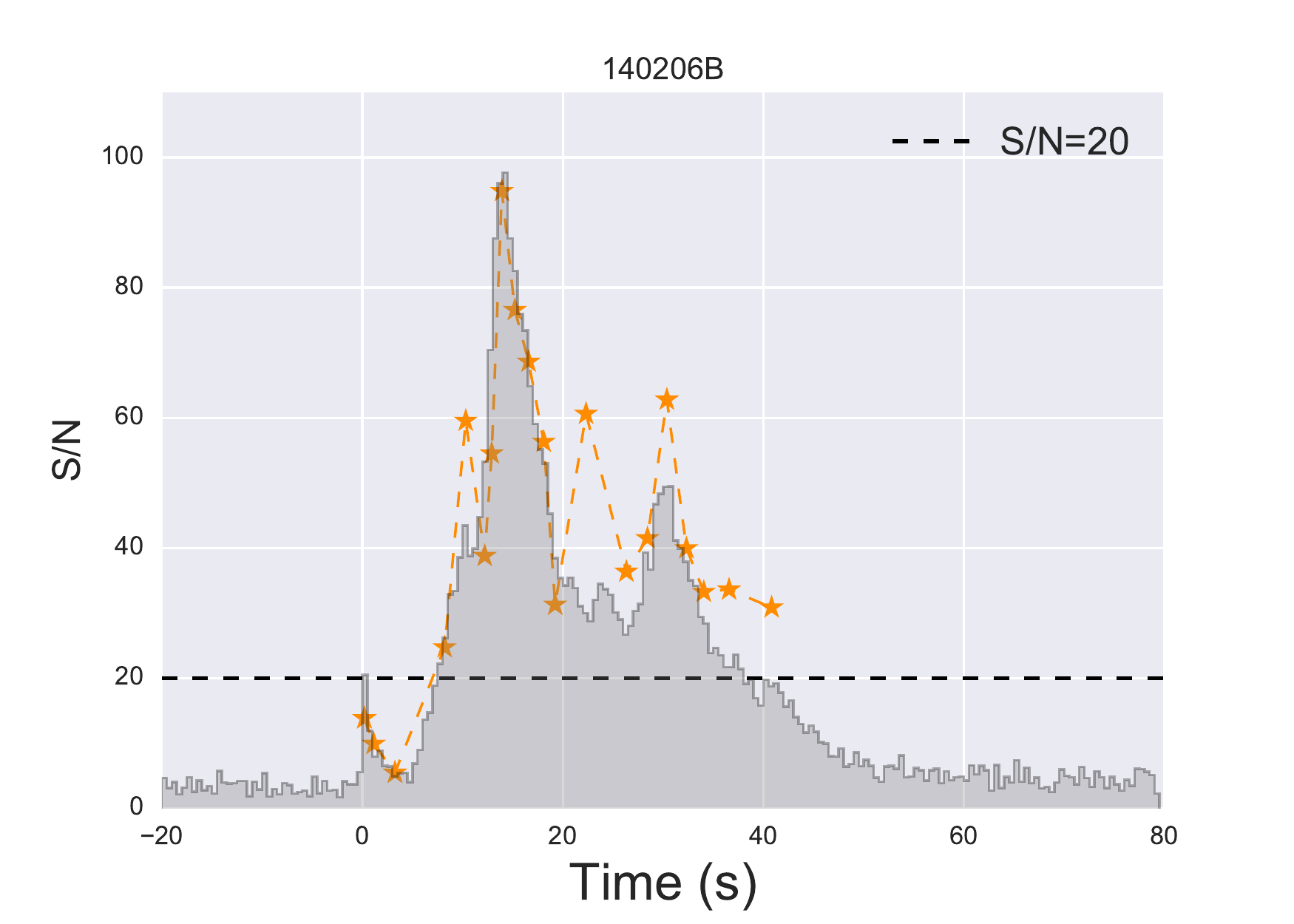}
\caption{Light curve of the prompt emission of GRB 140206B (the right-hand y-axis), along with temporal evolution of the S/N (yellow stars and the left-hand y-axis). The horizontal dashed line represents an S/N value of 20.}\label{SNRTime}
\end{figure*}

\begin{figure*}
\centering
\includegraphics[angle=0, scale=1.0]{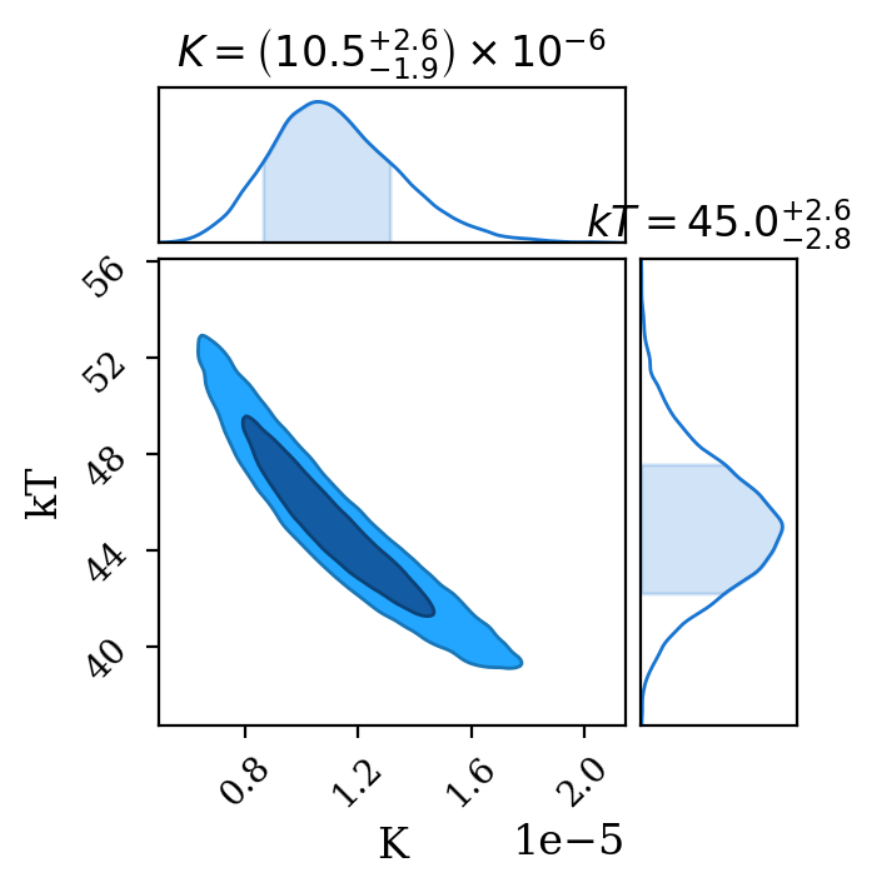}
\caption{Example fit to the data in one time bin (between 0.46s and 1.81s) of GRB 140206B using the BB model.}\label{MCFitBB}
\end{figure*}

\begin{figure*}
\includegraphics[angle=0, scale=0.60]{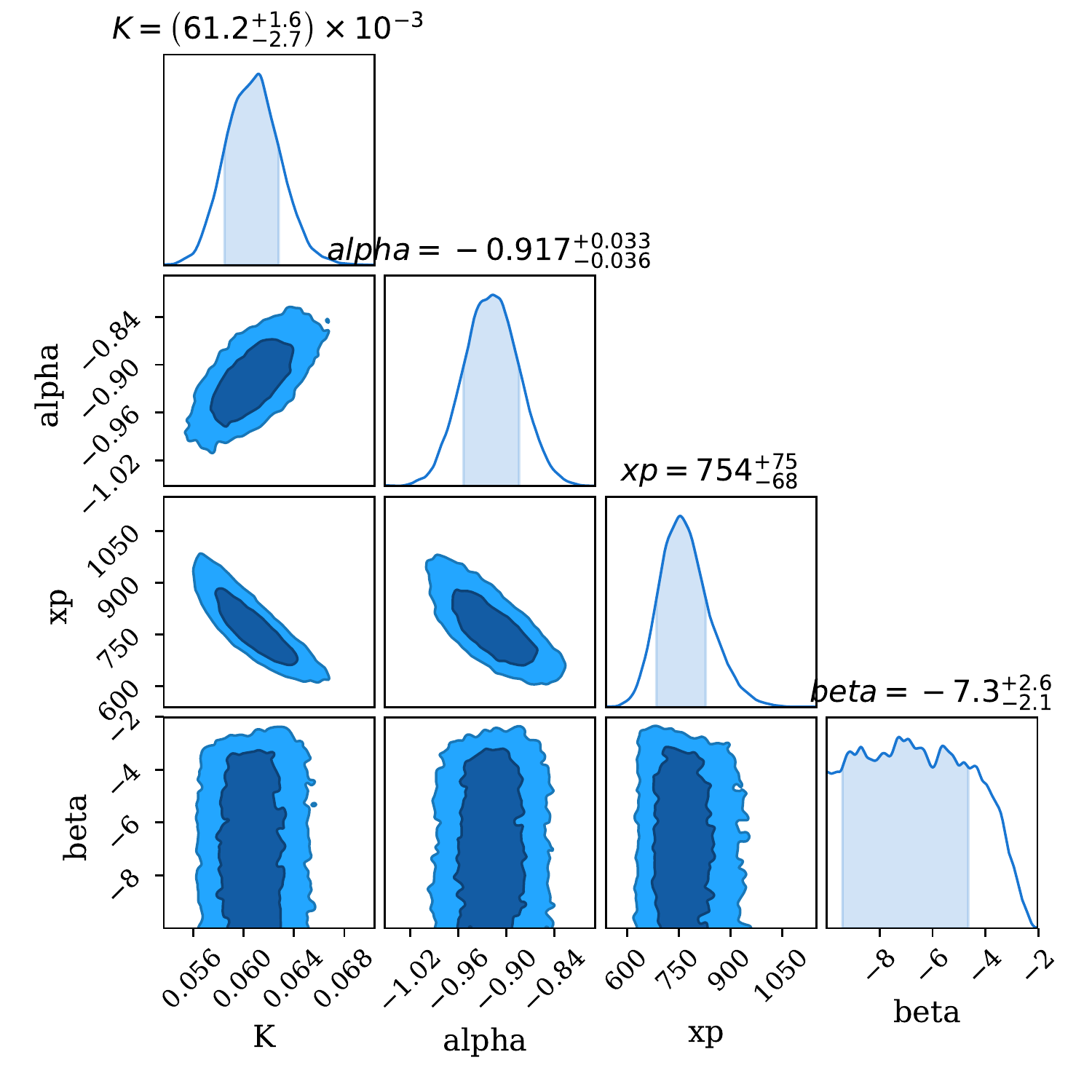}
\includegraphics[angle=0, scale=0.75]{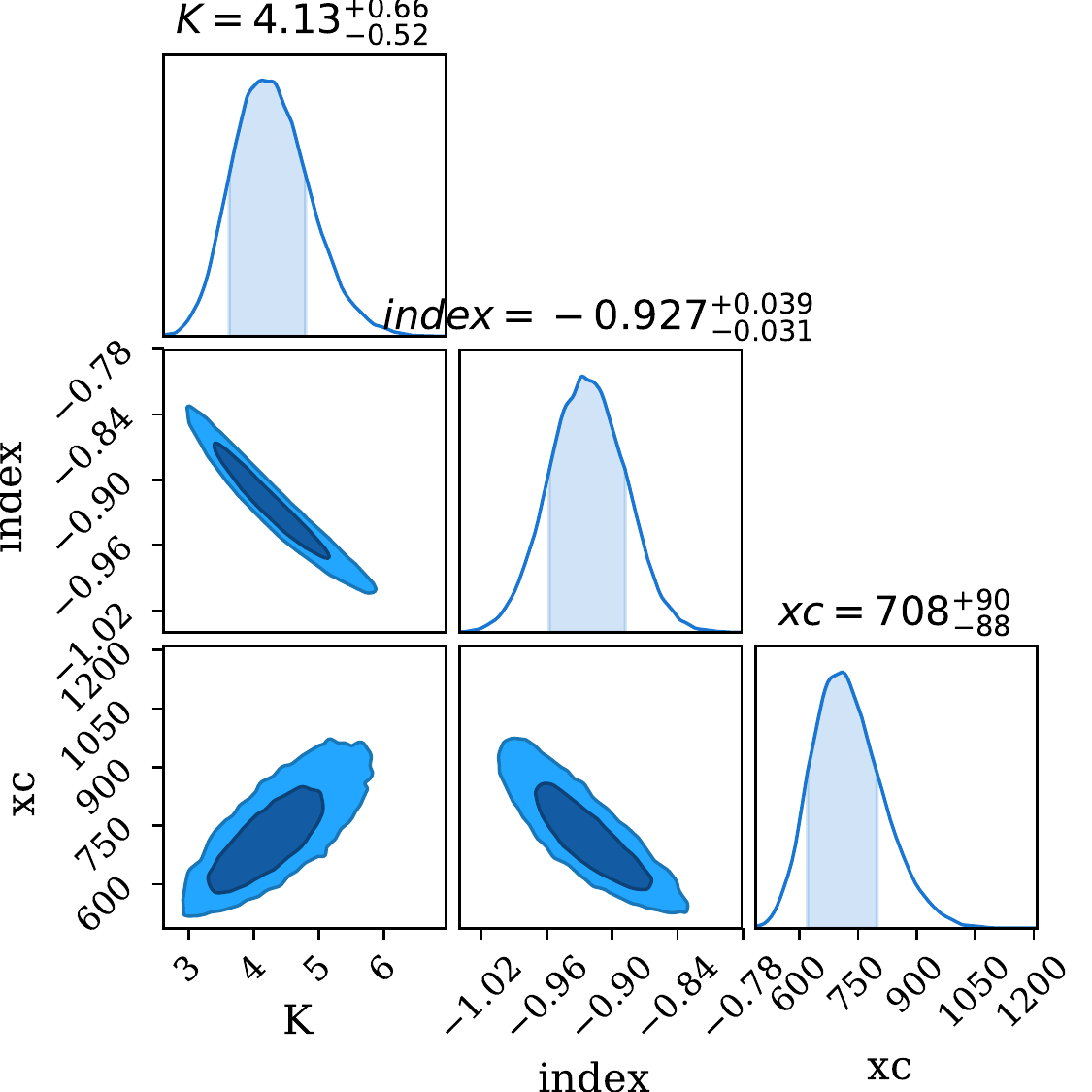}
\caption{Example fit to the data in one time bin (between 11.84s and 12.57s) of GRB 140206B using the Band model and the CPL model, respectively. Left panel: the Band model with an unconstarined parameter $\beta$; right panel: the CPL model.}\label{MCFitBandCPL1}
\end{figure*}

\begin{figure*}
\includegraphics[angle=0, scale=0.60]{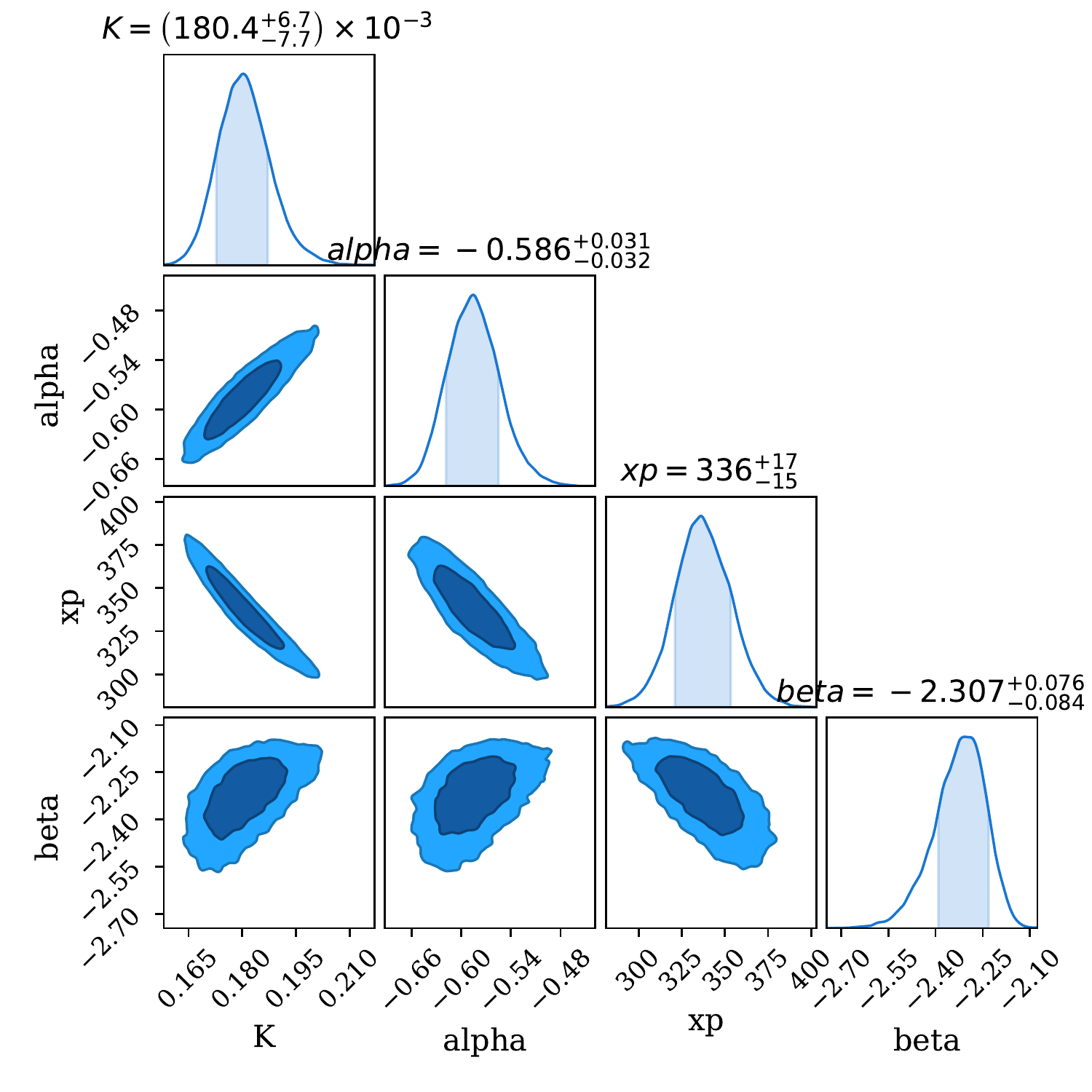}
\includegraphics[angle=0, scale=0.75]{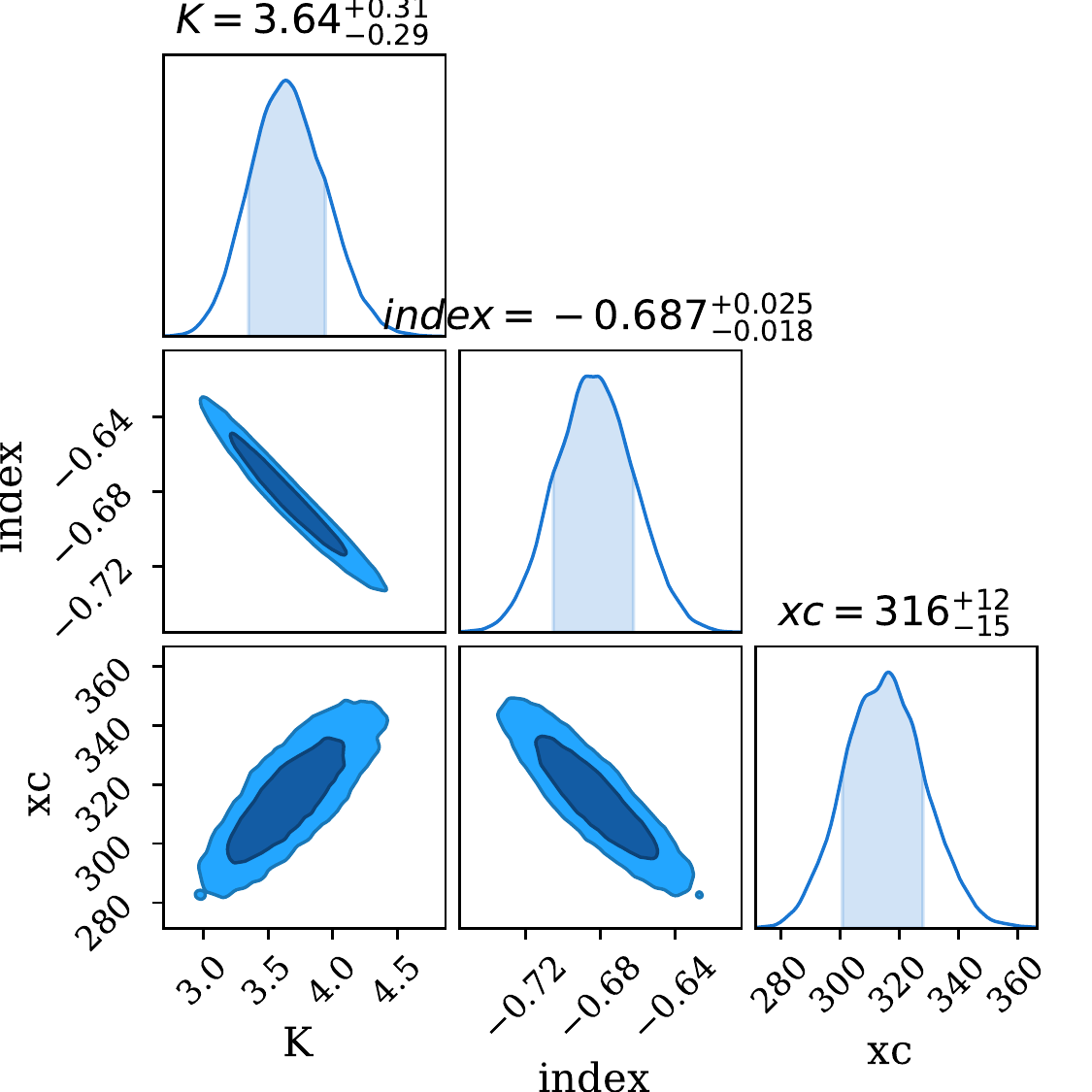}
\caption{Example fit to the data in one time bin (between 13.24 and 14.60s) of GRB 140206B using the Band model and the CPL model, respectively. Left panel: the Band model with a well-constarined parameter $\beta$; right panel: the CPL model.}\label{MCFitBandCPL2}
\end{figure*}

\begin{figure*}
\includegraphics[angle=0, scale=0.45]{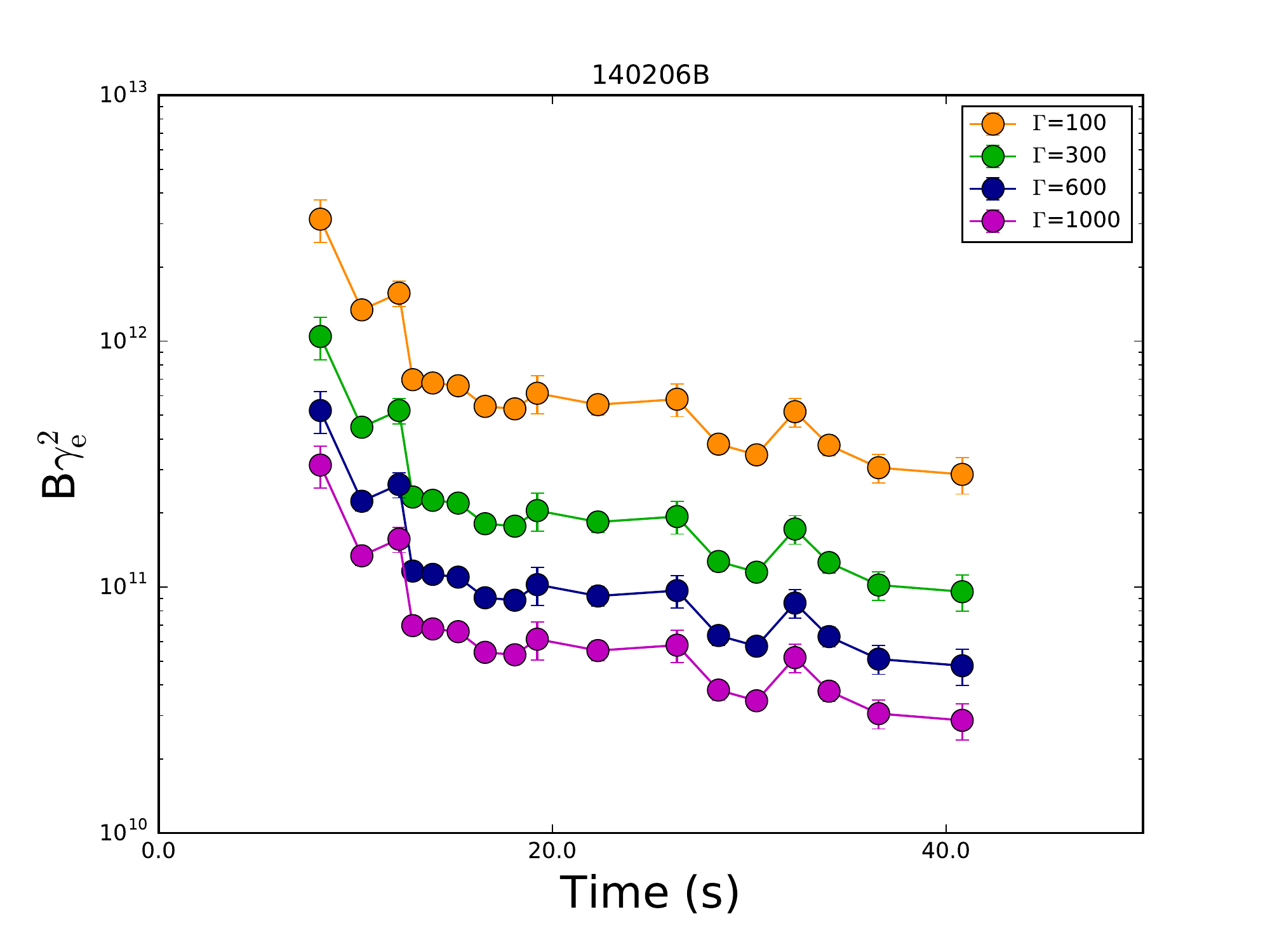}
\includegraphics[angle=0, scale=0.45]{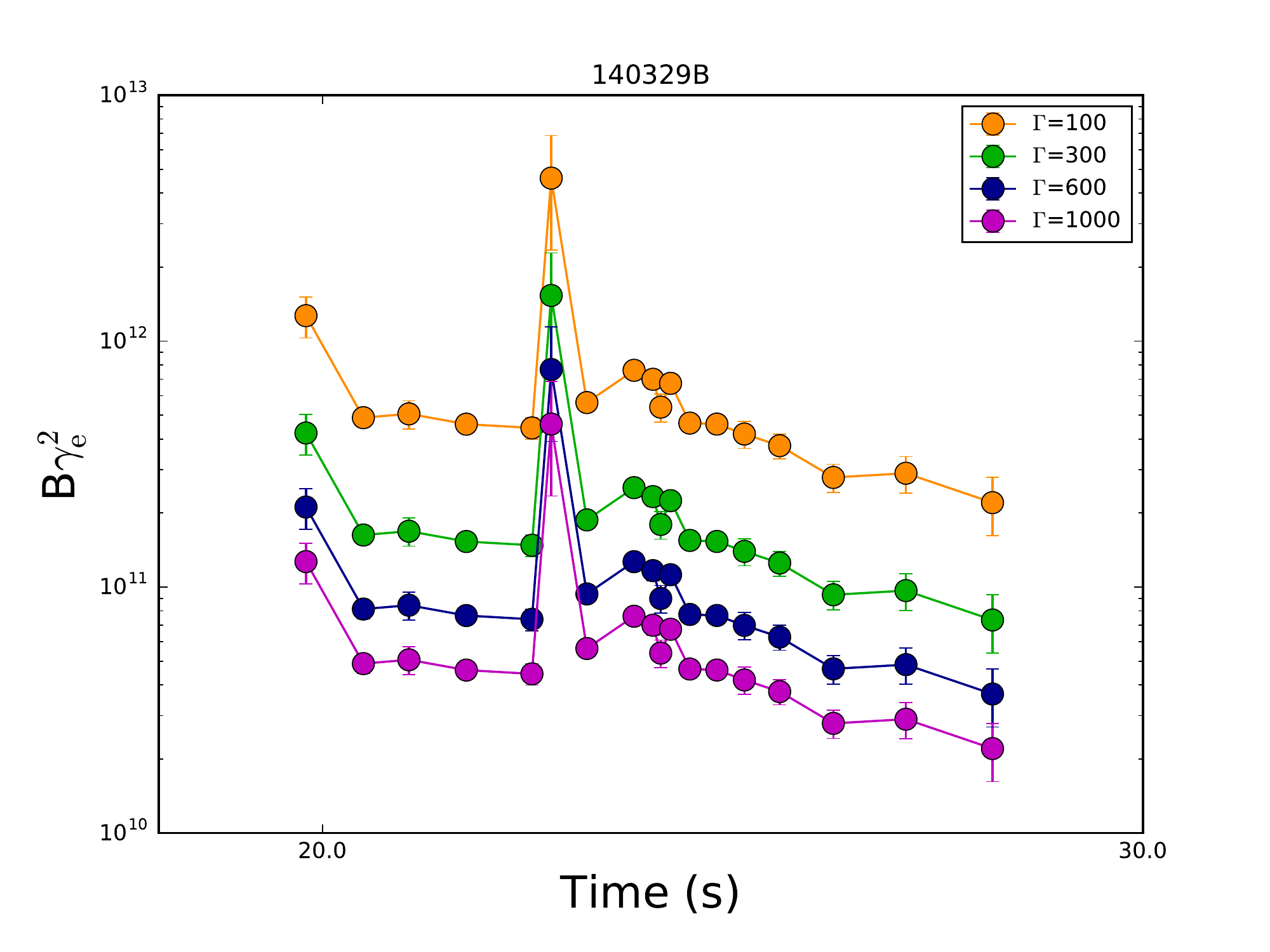}
\includegraphics[angle=0, scale=0.45]{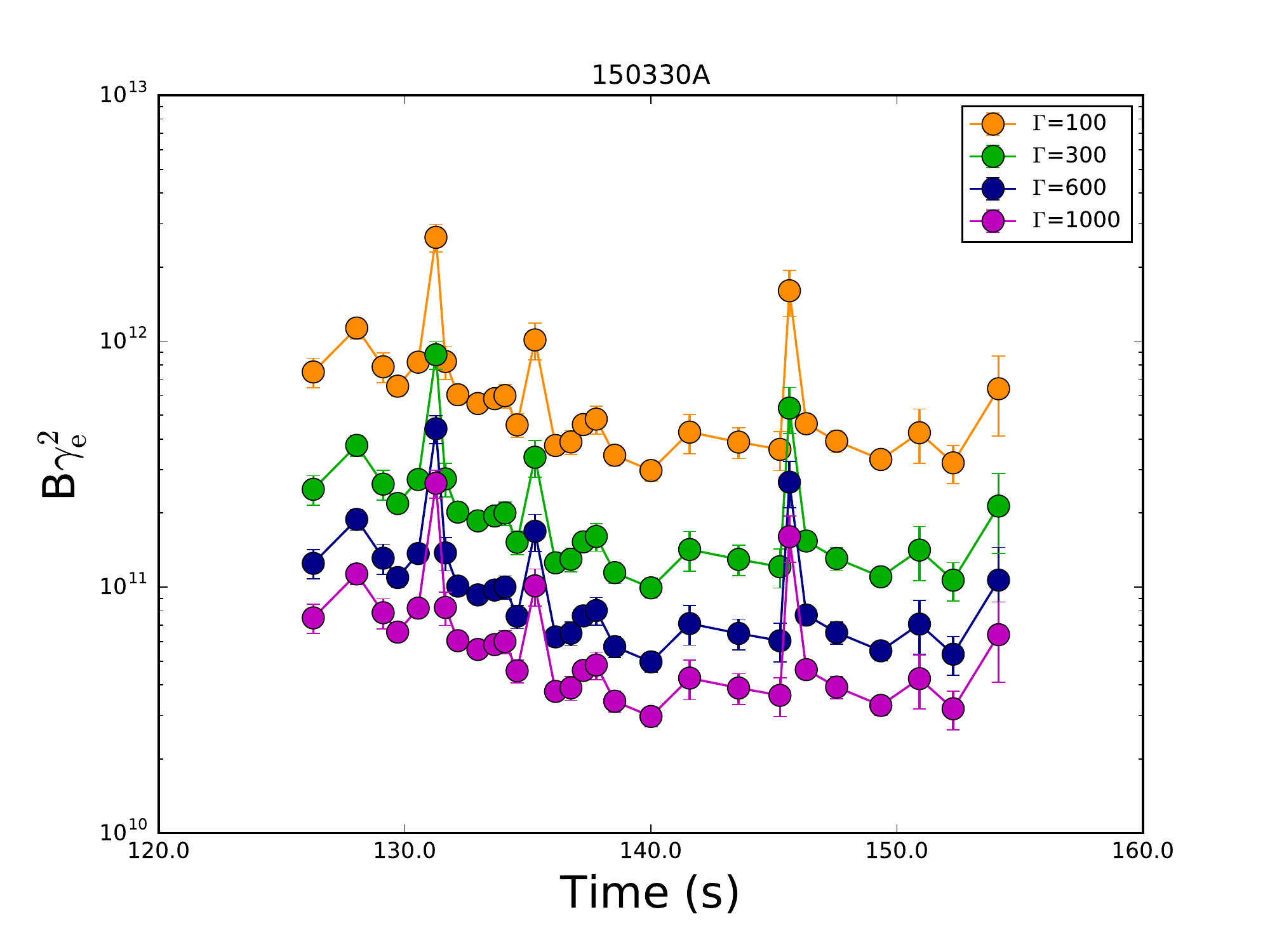}
\includegraphics[angle=0, scale=0.45]{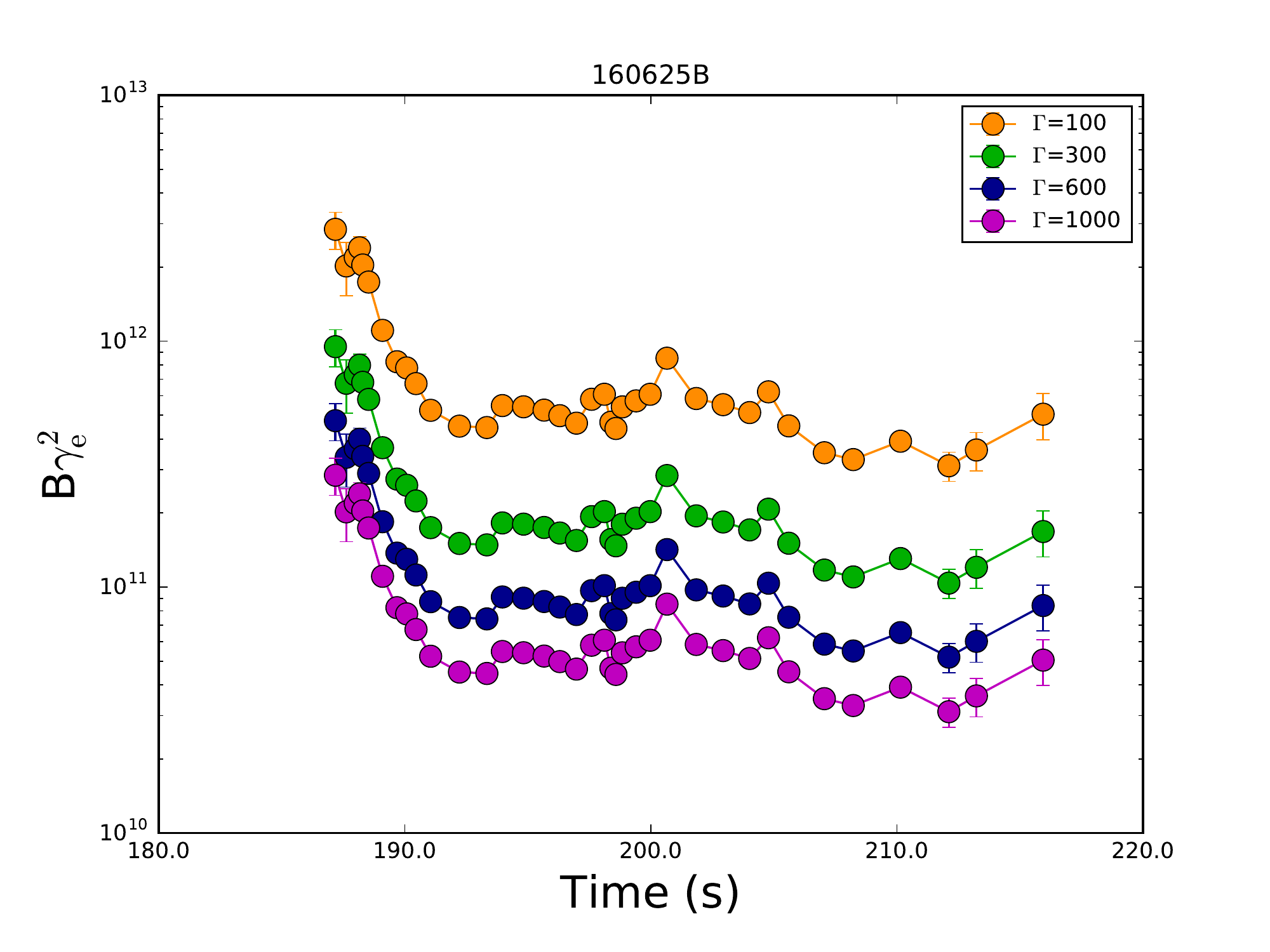}
\includegraphics[angle=0, scale=0.45]{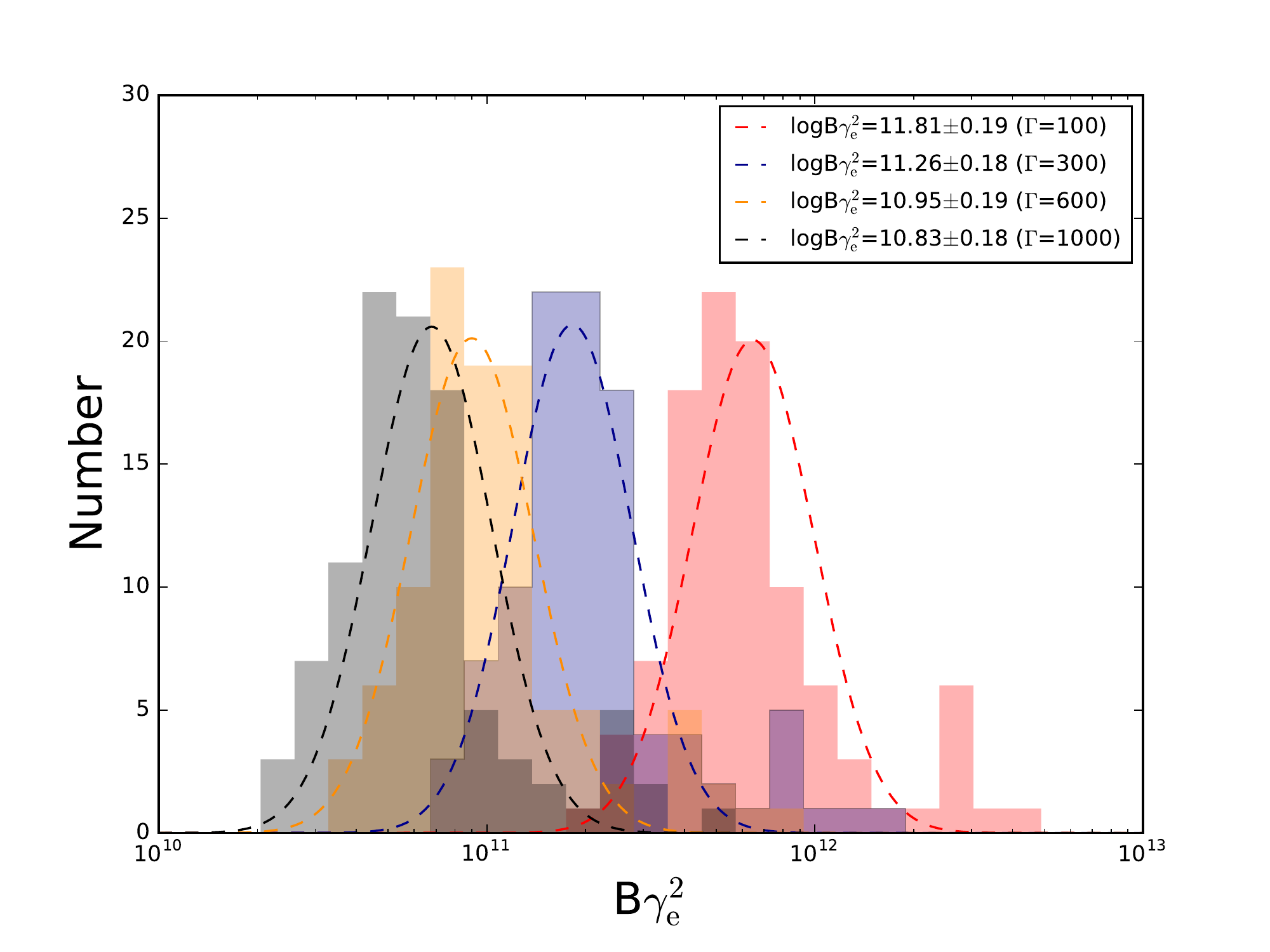}
\caption{Temporal evolution of $B\gamma^{2}_{e}$ for various typical $\Gamma$ values; different colors represent different $\Gamma$ values. Distributions of $B\gamma^{2}_{e}$ are based on these four typical $\Gamma$ values: $\Gamma$=100 (orange), $\Gamma$=300 (green), $\Gamma$=600 (blue), and $\Gamma$=1000 (pink). The best Gaussian fitting gives log $B\gamma^{2}_{e}$=11.81$\pm$0.19 for $\Gamma$=100, log $B\gamma^{2}_{e}$=11.26$\pm$0.18 for $\Gamma$=300, log $B\gamma^{2}_{e}$=10.95$\pm$0.19 for $\Gamma$=600, and log $B\gamma^{2}_{e}$=10.83$\pm$0.18 for $\Gamma$=1000.}\label{Bgammae}
\end{figure*}

In this Appendix, we provide additional figures, including model comparison with different criteria (Figure \ref{BIC}), examples of the GBM light curve (Figure \ref{LCs}), parameter relations for difference pulses/sub-bursts (Figures \ref{SpectralRelations} and \ref{EpAlpha}), temporal evolution of S/N of GRB 140206B (Figure \ref{SNRTime}), examples of the MCMC fit using the different spectral models (Figures \ref{MCFitBB}--\ref{MCFitBandCPL2}), and the temporal evolution of B$\gamma^{2}_{e}$ for various typical $\Gamma$ values (Figure \ref{Bgammae}).

\clearpage

\clearpage

\clearpage

\clearpage

\clearpage

\clearpage

\clearpage

\clearpage

\clearpage

\end{document}